\documentclass{article}
\usepackage[T1]{fontenc}
\usepackage[utf8]{inputenc}
\usepackage{graphics}
\usepackage{graphicx}
\usepackage{longtable}
\usepackage{color}
\usepackage{pslatex}
\usepackage{hyperref}
\usepackage{sverb}

\usepackage{mathtools}
\usepackage{subfig}
\usepackage{wasysym}

\begin{document}

\begin{titlepage}
 
\begin{flushright} 
{ IFJPAN-IV-2016-4 
} 
\end{flushright}
 
\vskip 20 mm
\begin{center}
{\bf\huge  Production of $\tau$ lepton pairs  with high $p_T$ jets at the LHC and  the  {\texttt{TauSpinner}}
reweighting algorithm
}
\end{center}
\vskip 10 mm

\begin{center}
   {\bf J. Kalinowski$^{a}$, W. Kotlarski$^{a,b}$, E. Richter-W\c{a}s$^{c}$ and Z. W\c{a}s$^{d}$  }\\
   \vskip 3 mm
       {\em $^a$ Faculty of Physics, University of Warsaw, Pasteura 5, 02-093 Warsaw,  Poland} \\
       {\em $^b$ Institut f\"ur Kern- und Teilchenphysik, Technische Univesit\"at Dresden, 01069 Dresden, Germany} \\
       {\em $^c$ Institute of Physics, Jagellonian University, Lojasiewicza 11, 30-348 Cracow, Poland} \\
       {\em $^d$ Institute of Nuclear Physics, PAN, Krak\'ow, ul. Radzikowskiego 152, Poland}\\ 
\end{center}
\vspace{1.0 cm}
\begin{center}
{\bf   ABSTRACT  }
\end{center}
The purpose of the {\tt TauSpinner} algorithm is to provide a tool that allows to 
modify the physics model 
of the Monte Carlo generated samples due to the changed assumptions of event production dynamics, 
but without the need of re-generating events. To each event {\tt TauSpinner} attributes the weights.
In this way, for example,  the spin effects of $\tau$-lepton production 
or decay are modified, or the effect of the changes in the production mechanism 
are introduced according to a new physics model.
Such an approach is useful, because there is no need to repeat the 
detector response simulation with each variant of the physics model considered.
In addition, since  only the event weights differ for the
 models, samples are correlated 
and statistical error of the 
modification is proportional to the reweighting only.

We document  the extension of the {\tt TauSpinner} 
algorithm  to ($2 \to 4$) processes in which  the matrix elements for the parton-parton scattering amplitudes into 
a $\tau$-lepton pair and two outgoing partons are used. 
The method is based on tree-level matrix elements with complete helicity information for the Standard Model 
processes, including the Higgs boson production. For this purpose automatically generated codes by {\tt MadGraph5}
have been adapted. Consistency tests of the implemented matrix elements, reweighting algorithm and numerical results
are presented.  

For the sensitive observable, namely the averaged $\tau$ lepton polarisation, we 
perform a systematic comparison 
between ($2\to 2$) and  ($2\to 4$) matrix elements used to calculate the spin weight in 
$p p \to \tau \tau jj$ events.
We show, that for events with $\tau$-lepton pair close to the Z-boson peak, 
the $\tau$-lepton polarisation calculated 
using  ($2\to 4$) matrix elements is very close to the one calculated using ($2\to 2$)  Born process only. 
For the $m_{\tau \tau}$ masses above the Z-boson peak, the effect from including  ($2\to 4$) matrix 
elements is also marginal, 
however when taking into account 
only subprocesses $qq, q\bar q  \to \tau \tau jj$,  it
can lead to a 10\% difference on the predicted  $\tau$-lepton polarisation. On the other hand, we have found that 
the appropriate choice of electroweak scheme  can have significant impact. 
We show that the modification of the electroweak or strong interaction 
initialization (including change of the electroweak schemes or analytic form of scale dependence for $\alpha_S$) can be performed 
with the re-weighting technique as well. 

The new version of {\tt TauSpinner ver.2.0.0 } presented here, allows also to introduce 
non-standard couplings for the Higgs boson 
and study their effects in the vector-boson-fusion processes by exploiting  the spin 
correlations of $\tau$-lepton pair 
decay products. The discussion of physics  effects is however relegated to forthcoming publications.

\vskip 1 cm


\vspace{0.2 cm}
 
\vspace{0.1 cm}
\vfill
{\small
\begin{flushleft}
{   IFJPAN-IV-2016-4
\\ April 2016
}
\end{flushleft}
}
 
\vspace*{1mm}
\footnoterule
\noindent
{
}
\end{titlepage}

\clearpage
\section{Introduction}

With the data collected so far by LHC experiments, there was not much interest to explore
physics  of $\tau$-lepton decays, with the exception of exploiting $\tau$ leptons 
in searches for rare or Standard-Model-forbidden decay channels, see eg. \cite{Aaij:2013fia}. However,  $\tau$-lepton signatures can provide  
a powerful tools  in many areas, like  studies of hard processes characteristics,  measurements of properties of  Higgs boson(s) 
\cite{AtlasHtautau:2015,CMSHtautau:2015}, or in  searches for New Physics \cite{Aad:2015tin,Aad:2015pfa,Aad:2014yka}.

The $\tau$ leptons cannot be observed directly due to their short life-time. All decay products are observed,  with the exception 
of  $\nu$'s. There are more than 20 different $\tau$ decay channels, each of them leading to a somewhat
distinct signature. This makes a preparation of observables involving $\tau$ decays laborious.  
However, such efforts can be rewarding, because $\tau$-lepton spin polarization can be measured directly, 
contrary to the case of electron or muon signatures, giving better insight into the nature of its production
mechanism, e.g. the properties of resonances decaying to $\tau$ leptons.   This is the main motivation for developing {\tt TauSpinner}, 
an algorithm to simplify the task of exploring the $\tau$ physics potential, 
which could be used for evaluation/modification of event
samples including $\tau$ decays.

In the first release, the program algorithms were focused on longitudinal 
spin effects only \cite{Czyczula:2012ny}. Already  {\tt TauSpinner ver.1.1} handled these effects with 
the help of the appropriate spin weight attributed to each event. In this way, spin effects could be introduced, or removed, 
from the sample. With time, variety of extensions were introduced.  
Since Ref. \cite{Banerjee:2012ez}, a second weight was introduced which allows 
to  manipulate the production process by adding additional contributions or completely 
replacing the production process with
an alternative one, including for example an exchange of a new intermediate 
particle.  Ref. \cite{Kaczmarska:2014eoa} brought a possibility of modifying transverse spin 
effects in the cascade $\tau$ decays of  intermediate Higgs boson. Later,  Ref.~\cite{Przedzinski:2014pla} 
enabled the transverse spin effects for the case of $\tau$ leptons produced in Drell-Yan 
processes to be studied  as well.

With time, technical options or important precision improvements were 
introduced too.  
In \cite{Kaczmarska:2014eoa}, an option  to attribute helicity states 
to  $\tau$-leptons was introduced.
One should keep in mind, that because of quantum entanglement, the assignment of a definite helicity 
state to intermediate $\tau$'s is necessarily subject to an approximation.  However, for spin   weight calculation, the 
 complete spin density matrix is taken into account and in general, approximation is not used.
With  later publication \cite{Przedzinski:2014pla}, one-loop electroweak (EW) corrections also 
became available for the Drell-Yan parton- process $q \bar q \to Z/\gamma^* \to \tau \tau$.

Let us mention another technical option. 
Initially the program was expected to work for samples, where spin effects are either 
taken into account 
in full, or are absent. One can  however configure {\tt TauSpinner} algorithm to work 
on generated samples where only part of  spin effects is taken into account 
(only some components of the density matrix
used)  and to correct them to 
full spin effects.

Until now,  for calculations of spin weights, {\tt TauSpinner} algorithm was always using 
the  Born-level ($2 \to 2$) scattering
amplitudes convoluted with the corresponding parton distribution functions (PDFs). Kinematic 
configurations of the incoming/outgoing partons 
were reconstructed from the four-momenta of outgoing $\tau$ leptons and incoming protons 
(using c.m. collision energy), and 
somewhat elaborated kinematical transformations were used for calculating an effective scattering angle 
of the assumed Born process.

{The validity and precision of this approximation became of a concern, especially for configurations 
with high momentum transfers in 
the t-channel and  for outgoing particles with high transverse momentum ($p_T$) that  accompany decay products of the 
electroweak bosons. In such cases,   more elaborated description of the production process dynamics is 
needed.  The aim  of the present paper is to describe an improved version 
of {\tt TauSpinner 2.0.0} which now 
includes hard processes featuring tree-level parton matrix elements for production 
of a $\tau$-lepton pair and two jets. Numerical test, and some results of physics interest, will be also presented.}

 The paper is organized as follows: 
In Section~\ref{sec:theoryBasis} 
we recall assumptions used for  the Monte Carlo reweighting techniques, in particular for the modeling 
of kinematic distributions in the 
multi-dimensional phase-space. We then define the master formula used by {\tt TauSpinner} for modeling 
spin correlations of $\tau$-lepton decay products in events with different topologies in 
proton-proton collisions. 
Section~\ref{sec:MEexplain}  documents details of the tree-level matrix elements used  for the 
calculation of weights 
in $pp \to \tau \tau\ jj$ events. The implemented functionality is based on automatically 
produced {\tt FORTRAN} code from {\tt MadGraph5} package \cite{Alwall:2014hca} for processes 
of the Drell-Yan--type and of the Standard Model
Higgs boson production in vector boson fusion (VBF) processes, which have been 
later manually modified and adapted. Numerical effects of different choices for electroweak 
and QCD interactions initialization are presented in the last two subsections.
We classify parton level processes into groups, which are then used in the following 
Section~\ref{sec:MEtests} for technical tests.
We explain details of the modification which we have introduced to the initialization of 
{\tt MadGraph5} generated amplitudes 
and emphasize the necessity of using the effective $\sin^2 \theta_W^{eff}$ for the 
calculation of the coupling constants to correctly model 
the measured spin asymmetries in the Drell-Yan process. This is even more important for 
a correct generation of angular distributions of leptons in the decay frame of intermediate 
$Z$ bosons.  Then we discuss combinatorial and CP symmetries that allow us  to reduce 
the number of  parton subprocesses for which distinct codes of spin amplitudes are needed. 
(Appendix~\ref{app:HowToUse} is devoted to describe technical details of the introduced extension 
of  {\tt TauSpinner}.) 
Numerical results shown in Section~\ref{sec:MEpheno}  are divided into three parts. 
The first one is devoted to the evaluation of systematic
biases present if  the  ($2 \to 2$) variant of {\tt TauSpinner} is used for spin effects (or for the 
matrix element weights)
in  $pp \to \tau\ \tau\ jj$ processes. 
Next, we present numerical consequences 
of the choice of the electroweak scheme, in particular:
(i) in the $\tau\ \tau\ jj$ production, (ii) in the calculation of the spin correlation 
matrix used for the  generation 
of $\tau$ decays, for the observable distributions. Section~\ref{sec:summary}, closes the paper. 
Somewhat lengthy collection of tests are relegated to Appendices B and C.

In the  present paper we concentrate on physics oriented  aspects of new implementations.  
All technical details 
and a description of available options, resulting not only   from the present work 
but also from the previous publications on {\tt TauSpinner}, will be collected in a  forthcoming publication. 
The most important points for  technical aspects of the program use are nonetheless presented in Appendix A. 
Benchmark outputs from the programs are relegated to the project web page \cite{WebPageSoinner2j}. 

\section{Theoretical basis} \label{sec:theoryBasis}
Before we start the discussion of  new implementations in the {\tt TauSpinner} and 
present numerical results, let 
us shortly recall the basis of the approach being used. 
For the Monte Carlo techniques of calculating integrals or simulating
series of events, to be well established in the mathematical formalism, 
one has to define the phase-space and the function  one is going to integrate.
One can parametrize the integral in the following form
\begin{equation}
G= \int_{0}^{1} \prod_{j=1}^{n} d\hat x_j \;\; {g}(\hat x_1,\hat x_2,...,\hat x_n)
=\lim_{N \to \infty } \frac{1}{N} \sum_{i=1}^N {g}(\ \hat x_1^{\ i},\ \hat x_2^{\ i},...,\ \hat x_n^{\ i}\ ), \label{eq:MC}
\end{equation} 
where on the right-hand side, the sum runs over $n$ dimensional 
vectors $\ \hat x_j^{\ i}$ of random numbers (each $\hat x_j^{\ i}$ in the $[0,1]$ range) which define the point
in the hypercube of coordinates. The $N$ denotes number of events used. 

The function $g$ consists of several components: the phase-space Jacobian resulting from the use
of $\hat x_j$ coordinates for the phase-space parametrization; the matrix element
squared calculated for a given process at prepared phase-space-point; 
and finally the acceptance function which is
zero outside the desired integration region.
Uniformly distributed random numbers $\hat x_j$ 
are used as Monte Carlo integration variables in formula (\ref{eq:MC}). The average value of $g$,
calculated over the event sample, gives the value of integral $G$.
In practical applications a lot of refinements are necessary to assure 
acceptable speed of calculation and numerical stability.  From a single sample of events several 
observables can 
be obtained simultaneously, {\it e.g.} in the form of differential distributions (histograms).

For the convenience of calculating multi-dimensional observables one introduces 
rejection techniques. 
The event $i$  (constructed from random-number variables $\hat x_j^{\ i}$)
 is accepted if an  additional randomly generated number
 is smaller than \\ ${g}(\hat x_1^{\ i},\hat x_2^{\ i},...,\hat x_n^{\ i})/g_{max}$;   
otherwise the 
event is rejected. The result of the integral is then equal to 
$g_{max} \times \frac{n_{accepted}}{n_{generated}}$. Statistical error of this estimate can 
be calculated  using standard textbook Monte Carlo methods.  
In such a method, one has to assure that for the allowed 
$\hat x_j$ range, the condition $0 \le g \le g_{max}$ holds. 
  The accepted events are distributed according to  $dG$  and 
can be used as a starting 
sample for the next step of the generation of weighted (or weight 1) events.

The principle goal of the  {\tt TauSpinner} program is to un-do, modify or supersede the 
discussed above rejection. 
Let us assume
that the sample of events, for which the program will be used, are 
distributed accordingly, {\it  with all details}, to the  known production mechanism described by the  formula
\begin{equation}
d \sigma = \sum_{i,j,k,l} f_i(x_1)f_j(x_2) d x_1  d x_2 
\frac{1}{\Phi_{flux}}d\Omega(p_1,p_2; \;p_3,p_4,p_{\tau^+},p_{\tau^-}) |M_{i,j,k,l}(p_1,p_2,p_3,p_4)|^2. \label{eq:main}
\end{equation}
In Eq.~(\ref{eq:main}), the $\sum_{i,j,k,l}$ extends over all possible 
configurations of incoming and outgoing partons for the processes of $i(p_1)\ j(p_2) \to k(p_3)\ l(p_4) \ \tau^+\tau^-$.
The $p_i$ stand for the 4-momenta of incoming/outgoing partons, 
$ x_1$ and $x_2$ stand for energy fractions of the beams carried by the incoming partons,
parton distribution functions are denoted as $f_i(x_1)$, $f_j(x_2)$ 
respectively for the first and the second incoming proton. The parton-level flux factor
is denoted as $\Phi_{flux}$ and the phase-space volume element as $d\Omega$. 
Finally the parton-level matrix element $M_{i,j,k,l}$ completes the formula. 
Obviously parton distributions (PDFs) are dependent on parton flavour configurations.
In Eq.~(\ref{eq:main}) the $\tau$ decay phase-space and 
the corresponding matrix elements are omitted. Even though it amounts to
semi-factorization, exploited by {\tt TauSpinner} algorithms, we omit for now also 
the discussion of $\tau$-spin correlation matrix.  They are not essential for the clarification of requirements  needed for  {\tt TauSpinner} algorithms.

For the calculation of {\tt TauSpinner} weights in the case of  replacing  one 
production mechanism  $A$ with another one $B$, one has to take into account not 
only differences in the matrix elements and PDFs but also, potentially, 
in $\Phi_{flux}$ and $d\Omega$. Thus the respective weight\footnote{In actual application to 
a sample  of 
experimental events the assumption that events are distributed 
accordingly to Eq.~(\ref{eq:main}), i.e.\ with head-on collision of incoming partons, may not 
hold.  As a result, the reweighing procedure of $A$ to $B$ according to Eq.~(\ref{eq:WT})  
will not   anymore be mathematically rigorous.
Section \ref{sec:PS}  is devoted to tests for this important issue.} 
 is calculated as follows
\begin{eqnarray}
 wt_{prod}^{A \rightarrow B}= 
\frac{ \sum_{i,j,k,l} f_i^B(x_1)f_j^B(x_2)  |M^B_{i,j,k,l}(p_1,p_2,p_3,p_4)|^2  \frac{1}{\Phi_{flux}}d\Omega(p_1,p_2; \;p_3.p_4,p_{\tau^+},p_{\tau^-})}
     { \sum_{i,j,k,l} f_i^A(x_1)f_j^A(x_2)  |M^A_{i,j,k,l}(p_1,p_2,p_3,p_4)|^2  \frac{1}{\Phi_{flux}}d\Omega(p_1,p_2; \;p_3.p_4,p_{\tau^+},p_{\tau^-})}\label{eq:WT}
\end{eqnarray}
Although the factors 
${\Phi_{flux}}$ and
$d\Omega$ may cancel between the numerator and denominator 
in the case when all incoming and outgoing partons are considered 
to be  massless, they 
still may differ due to symmetry factors which are different for identical or distinct  flavours 
of  partons.

\section{Physics and matrix elements of ($2\to 4$) processes.}
\label{sec:MEexplain}

The physics processes of interest are the Standard Model processes in $pp$ collision
with two opposite-sign $\tau$ leptons and 2 jets (quarks or gluons) in the final state\footnote{Here as jets we understand outgoing partons.}. 
Such processes are described at the tree level by ($2 \to 4$) matrix elements, with 
intermediate states being single or double $Z, W, \gamma^*, H$ or fermion exchange in the s- or t-channel. 
Depending on the initial state, tree-level matrix elements are of the order of 
$\alpha_S \alpha_{EW}$ or $\alpha_{EW}^2$, involving sometimes triple $WWZ$ couplings. 
More details are given in Table~\ref{tab:processes}.
We will limit our implementation to the tree-level only, but with the emphasis 
on  controlling the spin configurations.

\subsection{Incorporating {\tt MadGraph} generated code into TauSpinner} 
\label{sub:incorpo}
There are automated programs for generating codes of spin amplitudes calculation. In the development 
of {\tt TauSpinner} we have used  {\tt MadGraph5}~\cite{Alwall:2011uj}. Let us recall some details
of this step of the program development to explain the adopted procedure,  which may be useful in future for introducing 
anomalous couplings or new physics models. 

The {\tt FORTRAN} code for calculating matrix elements squared ($ME^2$) is generated 
using {\tt MadGraph5} with the following commands:
\begin{itemize}
\item[a)] \texttt{import model sm-ckm}
\item[b)] with default definition of "multiparticles"\\ 
\texttt{p = g u c d s u\~{} c\~{} d\~{} s\~}   \\
\texttt{j = g u c d s u\~{} c\~{} d\~{} s\~}
\item[c)] for the Higgs signal processes\\
\texttt{generate p p > j  j h, h > ta+ ta- }
\item[d)] for the Drell-Yan--type SM background processes\\
\texttt{generate p p > j j ta+ ta- / h QED=4}
\item[e)] and print the output using \\
\texttt{output standalone "directory name".}
\end{itemize}
Setting the parameter {\tt QED=4} enforces generation of diagrams up to 4th order in the electroweak couplings.
Other settings are initialized as in the default version of the  {\tt MadGraph5} setup. 
The generated codes for the individual subprocesses are then grouped together into subroutines, 
depending
on the  flavour of initial state partons, and named accordingly. For example, 
\begin{verbatim}        
SUBROUTINE  UDX(P,I3,I4,H1,H2,KEY,ANS)
\end{verbatim}
corresponds to processes initiated by $u\bar d$ partons.  {\tt X} after the letter {\tt U,D,S} and {\tt C} means the antiquark, 
i.e. {\tt UXCX} corresponds to processes initiated by $\bar u\bar c$, while {\tt GUX} -- processes initiated by $g\bar u$.  
The input variables are: real matrix {\tt P(0:3,6)} for four-momenta of incoming and outgoing
particles, integers {\tt I3,I4} for the Particle Data Group (PDG) identifiers for final parton flavours, 
integers {\tt H1,H2} stand for  
outgoing $\tau$ helicity states; integer {\tt KEY}  
selects the requested matrix element for the SM background ({\tt KEY=0}), the SM Higgs boson%
\footnote{The {\tt KEY} $>$ 1 is reserved for non-standard scenarios, then the code discussed in the present Section is not necessarily used.} ({\tt KEY=1}),
{\tt ANS} returns the calculated value of the matrix-element squared. 
According to the value of {\tt I3,I4,KEY}  the corresponding subroutine generated 
by {\tt  MadGraph5} is called\footnote{ Note that 
by convention,  setting {\tt I3 = 0} and {\tt I4 = 0} returns the matrix element squared  summed 
over all possible final 
state partons; as a default this option is not used, and  the corresponding sum is performed 
explicitly in the  code.}.
The {\tt TauSpinner} user usually will not access to the {\tt KEY} variable.

Before integrating these subroutines into the {\tt TauSpinner} program, a number of modifications have 
been done for the following reasons:
\begin{itemize}
\item[a)] Since {\tt MadGraph5} by default sums and averages over spins of incoming and outgoing particles,
 while we are interested in $\tau$ spin states,  
the generated codes have to be modified to keep track of the $\tau$ polarization; 
\item[b)] Moreover, since  the subroutines and internal functions  generated by {\tt MadGraph5} have the same 
names for all subprocesses {\tt SMATRIX(P,ANS)}, the names had to be changed to be unique for each subprocess.
 To be more specific, for the Higgs signal subprocess 
$u\bar d\to c\bar d \, h,\, h\to\tau^+\tau^-$ the 
generated subroutine name is changed to 
{\tt  UDX\_CDX\_H(P,H1,H2,ANS)}, 
while for 
the background $u\bar d\to c\bar d \,\tau^+\tau^-$ process, the generated subroutine name is 
changed to
{\tt UDX\_CDX\_noH(P,H1,H2,ANS)}.
\end{itemize}
For other processes and internal functions
similar convention is used, see Table \ref{tab:processes}.
Note that for example for the processes with $c s$ quarks in the initial state, exchange of $W's$ is allowed,
the final states cannot include gluons and the only allowed final states are:
$cs, cd, us, ud$.  After taking into account permutation of incoming and outgoing partons and CP symmetric states
this gives in total $4 \times 4\times 2 = 32$ non-zero contributions to the sum of Eq. (\ref{eq:main}).
This is the case both for Drell-Yan--type background and Higgs-boson production processes.
For the remaining processes the codes listed in Table~\ref{tab:processes} are also used  with the help of $CP$ symmetry
or re-ordering of partons. 

At the parton level  each of the incoming or outgoing parton can be one of flavours: 
$ \bar b\ \bar c\ \bar s\ \bar u\ \bar d\  g\ d\ u\ s\ c\ b$, with Particle Data Group (PDG) identifiers: 
 -5, -4, -3, -2, -1, 21, 1, 2, 3, 4, 5 respectively.
For processes with two incoming partons,  two outgoing $\tau$ leptons and two outgoing patrons 
that gives
$11^4$ possibilities, most of them with the zero contribution, and many available one from another by relations following from  CP symmetries 
and/or  permutations of incoming and/or outgoing partons. 
Grouped by the type of initial state partons,  the 
subroutines listed in Table~\ref{tab:processes} are currently limited to the first
two flavour families. The matrix elements for processes involving $b$-quarks are not yet 
implemented\footnote{The matrix elements with $b$ quarks are set to zero in our
default installation. However,  the program has already been set up so that  the 
user-provided codes featuring $b$-quark processes can be activated by a C++ pointer at 
any moment using the {\tt TauSpinner::set\_vbfdistrModif()} method,  
see Appendix \ref{app:HowToUse} for details.}.

Also, for practical purposes, for  a pair of final-state parton flavours $k \ne l$, the 
{\tt MadGraph5} generated codes have been obtained 
for a definite ordering
$(k,l)$, but not for  $(l,k)$, to reduce the number of generated configurations.
When  {\tt TauSpinner} is invoked, 
the flavour configuration of outgoing partons is unknown and it takes into account 
both possibilities: thus a compensating factor  $\frac{1+\delta_{ij}}{2} $ has to be introduced.
This is because of the organization of the sum in Eq.~(\ref{eq:WT}).

\begin{table}
 \caption{List of implemented processes for calculating matrix element squared grouped into
 categories, which differ by flavours of incoming partons. For each category,
{\tt FORTRAN} files with 
implemented subroutines for calculating the matrix element square, grouped by the 
 flavour of incoming partons, are given in the second column. Examples of processes in each category are 
given in the last column. 
Partially redundant codes for some of the processes are used for tests only, this is the case of amplitudes
stored in files {\tt UCX.f} and {\tt CUX.f}, the amplitudes of these two files can be obtained from each other
by CP symmetry.}
\label{tab:processes}
 \vspace{2mm}
  \begin{center}
  \begin{tabular}{|l|l|l|}
  \hline\hline 
 Category of               &  Corresponding FORTRAN files    & Processes       \\ 
 Matrix Elements           &                                 &                 \\ 
  \hline\hline
   {  (1)}                  & GG.f                             &$g g \to \sum_{f} q_f \bar q_f$ \\
\hline
   {  (2) }                 & GD.f, GU.f     & $g q_f (\bar q_f) \to g q_f (\bar q_f)$\\
\hline
   {  (3) }                 & DD.f,   UD.f,  UU.f,     & $q_{f_1} \ q_{f_2}\ (\bar q_{f_1} \ \bar q_{f_2}) \to q_{f_1} \ q_{f_2} (\bar q_{f_1} \ \bar q_{f_2})$   \\
                           & CC.f, CS.f,            & \\ 
                           & DC.f, DS.f, SS.f  CD.f,             &  \\
                           & CU.f, SD.f, SU.f, US.f             &   \\
\hline   
   {  (4) }                &  DDX.f, UDX.f,  UUX.f      & $ q_{f_1} \ \bar q_{f_2}\ (\bar q_{f_1} \ \bar q_{f_2}) \to q_{f_1} \ \bar q_{f_2} (\bar q_{f_1} \ \bar q_{f_2})$ \\
                           & CCX.f, CSX.f, DCX.f,  DSX.f,             & $q_{f_1} \ \bar q_{f_2}\ (\bar q_{f_1} \ \bar q_{f_2}) \to gg$              \\ 
                           & SCX.f, SSX.f,  UCX.f, USX.f,          &    \\
                           & CDX.f, CUX.f, SDX.f, SUX.f            &  \\
\hline
\end{tabular}
\end{center}
\end{table}

\subsection{Topologies and the dynamical structure of subprocesses}\label{sec:topologies}

\begin{figure}
\begin{center}
\hspace*{\fill}
\subfloat[]{\includegraphics[width=0.20\textwidth]{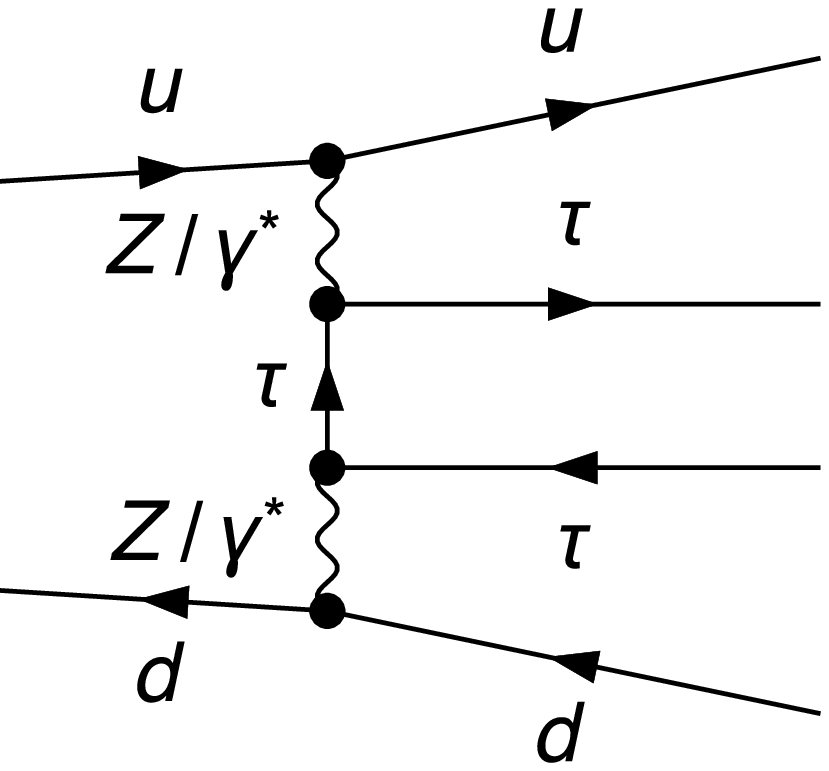}}
\hfill
\subfloat[]{\includegraphics[width=0.20\textwidth]{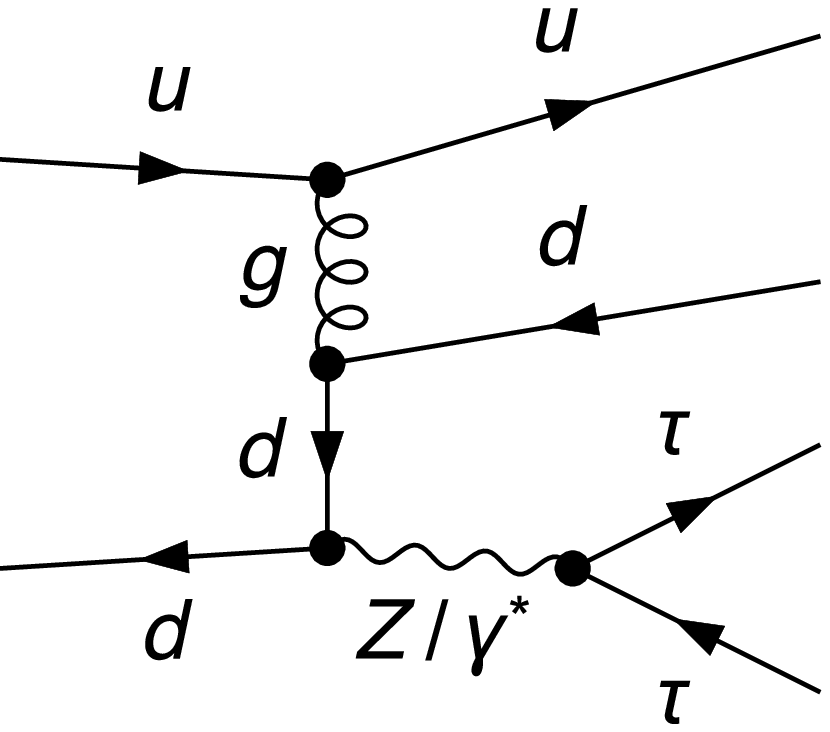}}
\hfill
\subfloat[]{\includegraphics[width=0.20\textwidth]{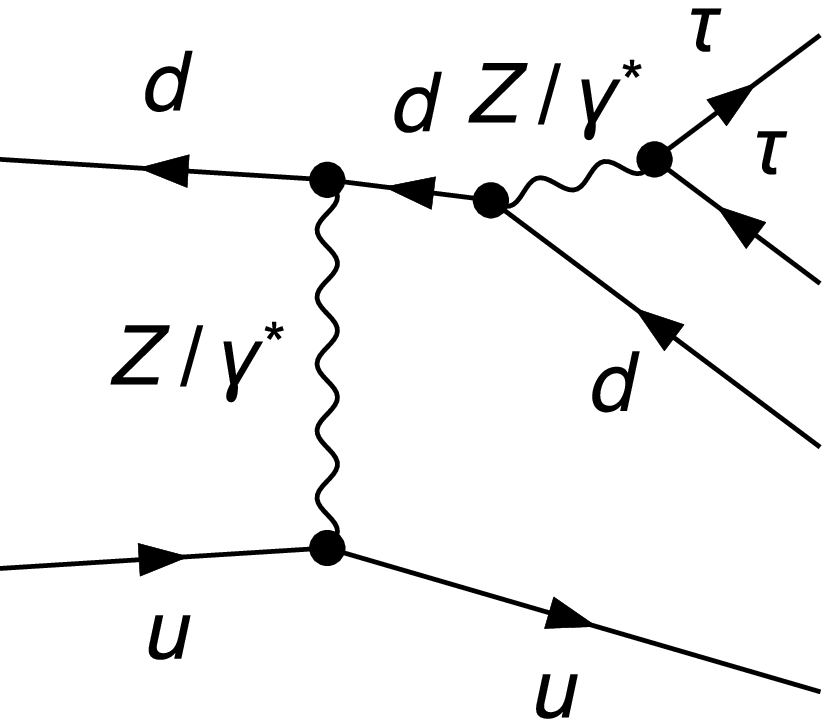}}
\hfill
\subfloat[]{\includegraphics[width=0.20\textwidth]{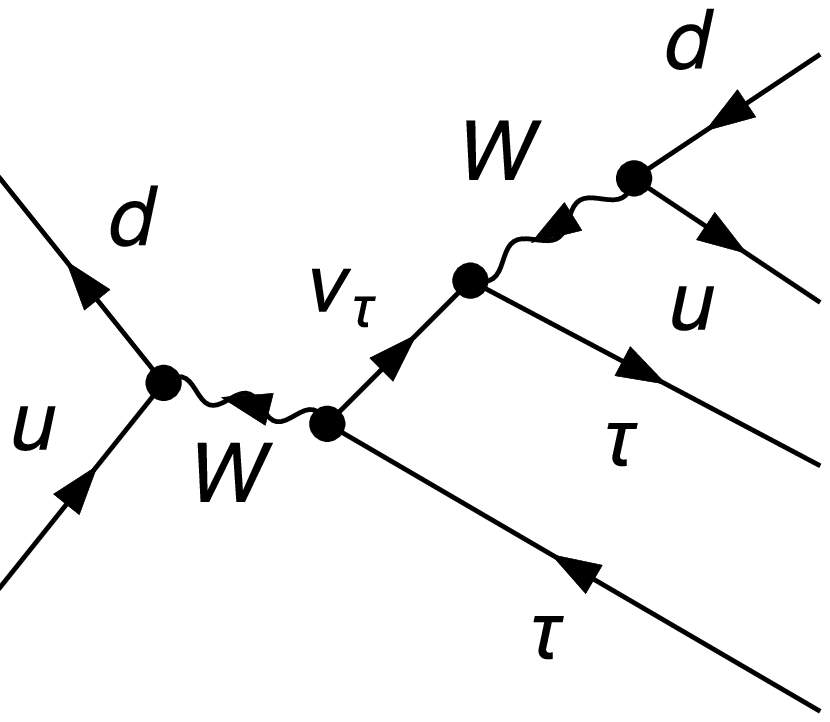}}
\hspace*{\fill}\\
\hspace*{\fill}
\subfloat[]{\includegraphics[width=0.20\textwidth]{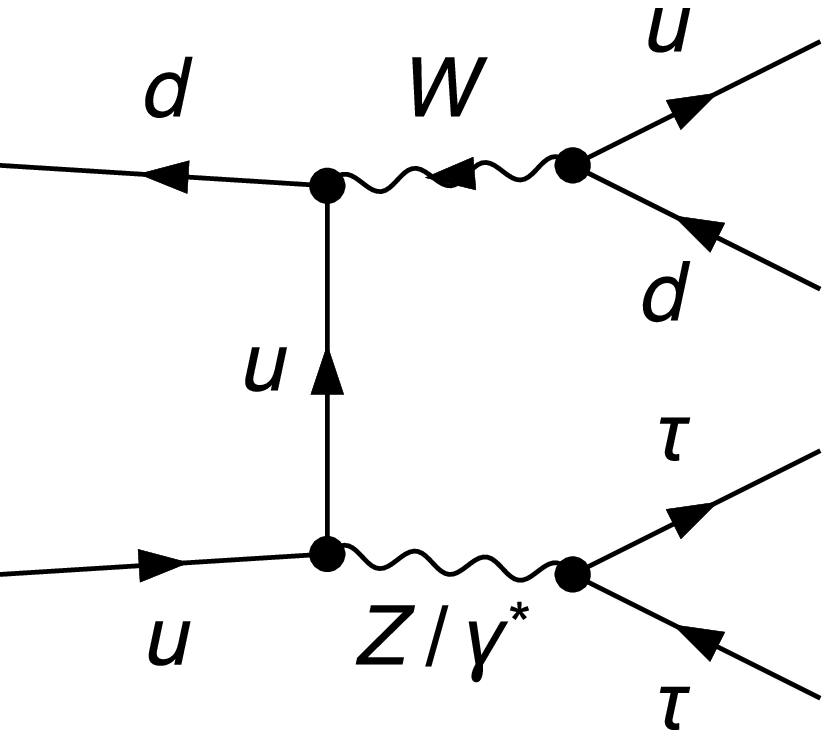}}
\hfill
\subfloat[]{\includegraphics[width=0.20\textwidth]{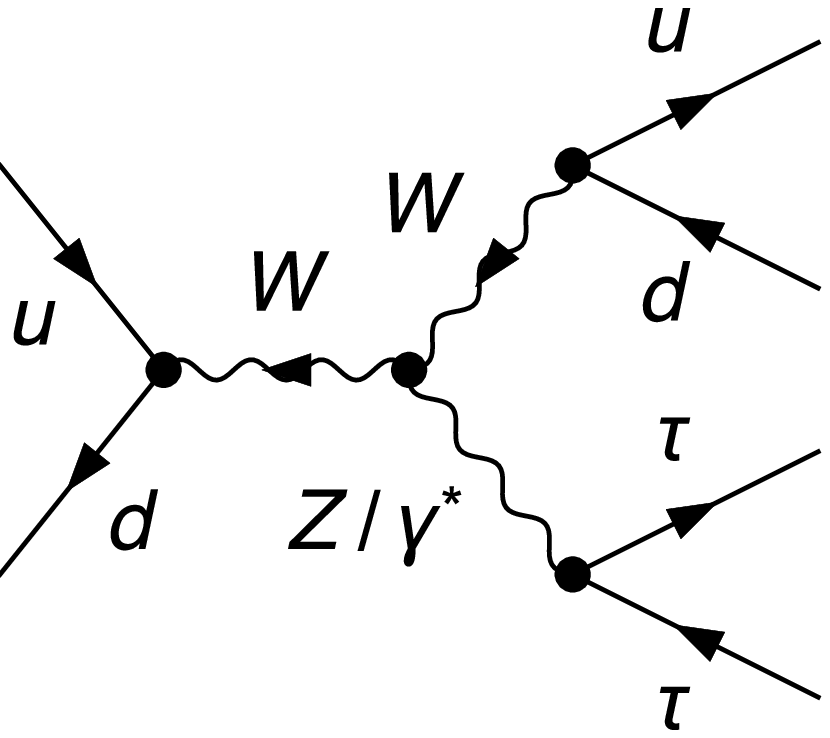}}
\hfill
\subfloat[]{\includegraphics[width=0.20\textwidth]{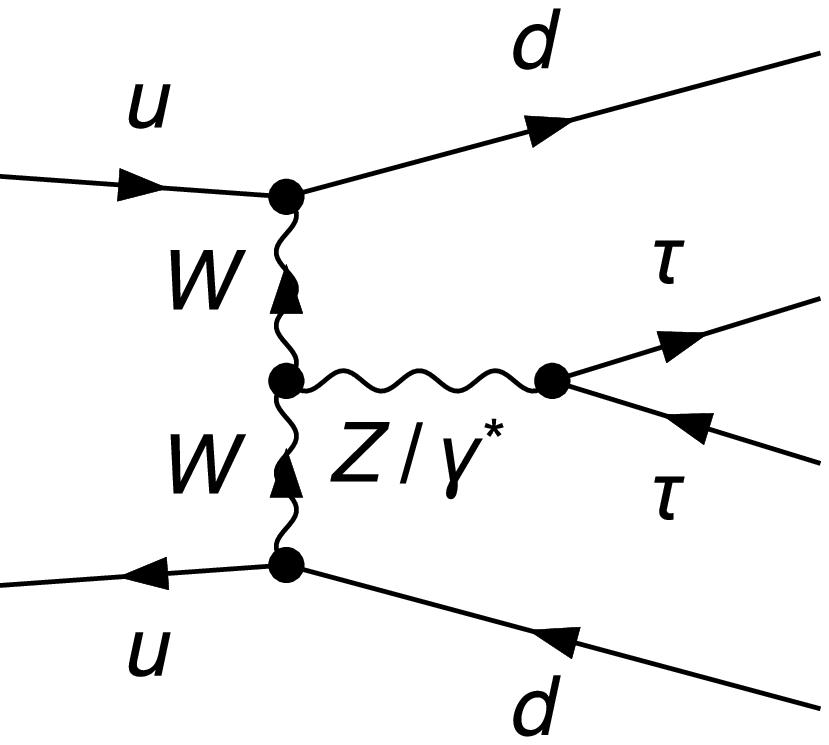}}
\hspace*{\fill}
\end{center}
\caption{ Typical topologies of diagrams contributing to the Drell-Yan--type SM process in $u \bar d \to \tau^+\tau^- u \bar d$: 
multi-pheripheral (a), double-t (b), t-cascade (c), s-cascade (d), 
double-s (e), mercedes (f) and fusion (g) type of diagrams.
\label{fig:diagram_noh}}
\begin{center}
\hspace*{\fill}
\subfloat[]{\includegraphics[width=0.20\textwidth]{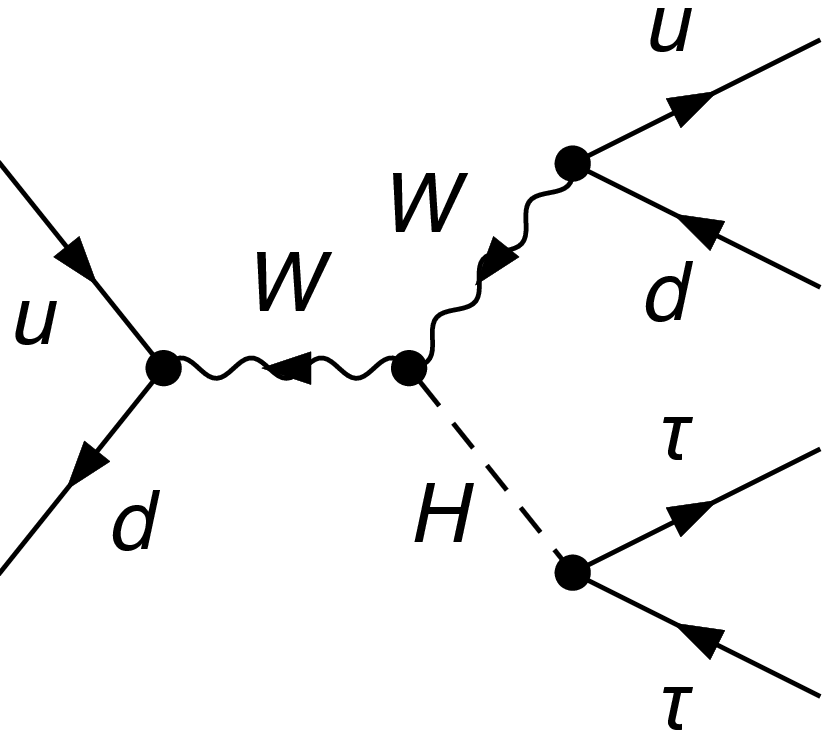}}
\hfill
\subfloat[]{\includegraphics[width=0.20\textwidth]{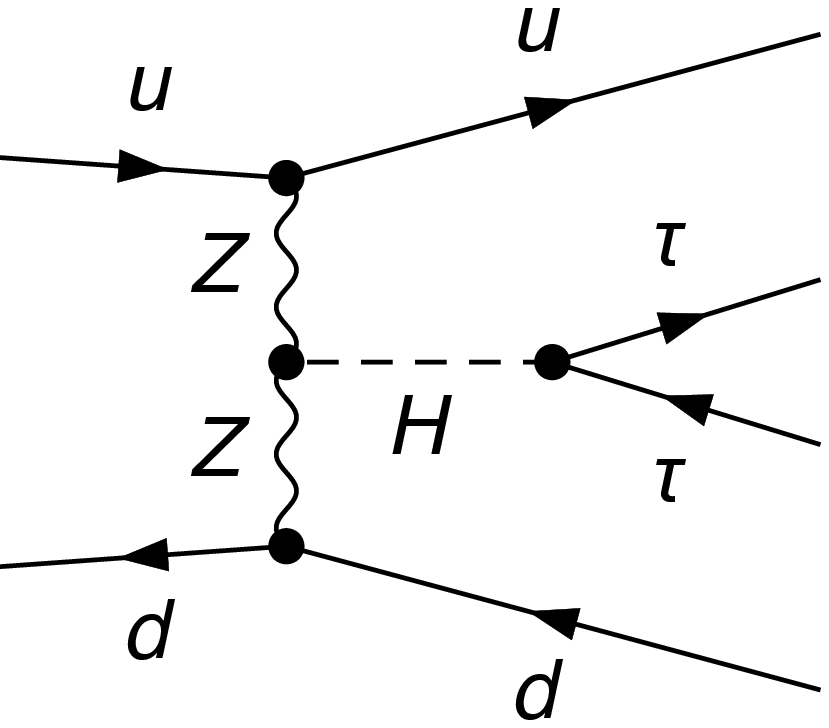}}
\hspace*{\fill}
\end{center}
\caption{Topologies of diagrams contributing to the Higgs production process $u \bar d \to H(\to \tau^+\tau^-) u \bar d$: vector boson fusion (a),  
Higgs-strahlung (b).
In general, depending on the flavour of incoming partons, mediating boson could be $W$ or $Z$. 
\label{fig:diagram_h}}
\end{figure}

The number of contributing subprocesses is very large. 
For the case of  the non-Higgs Drell-Yan--type background processes, in which the $\tau$-pair 
originates either from the vector boson decay 
(including also cascade decays) or from multi-peripheral vector-boson fusion processes, 
{\tt MadGraph5} generates 82 subprocesses with partons belonging to the first two 
generations of quarks, or gluons. 
Subprocesses in which all partons are of the same flavour (like $u\bar u\to u\bar u\tau^+\tau_-$)  
receive contributions from 64 Feynman diagrams, subprocesses with two pairs of flavours -- 
either 43 diagrams (if one pair is of up-type and the other down-type, like 
$u\bar u\to s\bar s \tau^+\tau^-$) or 32 diagrams (if both pairs are either down- or up-type, 
like $u\bar u\to c\bar c\tau^+\tau^-$), subprocesses with three or four different flavours -- 11 diagrams 
(like $us\to ud\tau^+\tau^-$), and subprocesses with two quarks and two gluons -- 16 diagrams.
 As far as the dynamical structure of the amplitudes is concerned, there are all together 
seven different topologies of Feynman diagrams,   
with representatives shown  in Fig.~\ref{fig:diagram_noh}. Which of them contribute 
to a given subprocess depends on flavours of incoming and outgoing partons.  
Irrespectively of their origin, in all processes the polarizations of $\tau$ leptons are 
strongly correlated due to the helicity-conserving couplings to the vector bosons.
The spin correlations of the produced $\tau$ pair depend on the 
relative size of the subprocesses with vector and pseudo-vector couplings contributing 
to the given final state configuration.
For example, in the case of $ q\ \bar q \to \tau^+\tau^- q \bar q$ , see Fig.~\ref{fig:diagram_noh},
diagram (d)  contributes with 100\% polarised $\tau$'s since they couple directly 
to $W^{\pm}$. 
In diagram (g), the polarisation of $Z/\gamma^*$ is different than in 
the Born-like production   because $Z/\gamma^*$ decaying to $\tau^+\tau^-$ originates from the $WWZ/\gamma^*$ 
vertex. This 
 leads to a distinct polarisation  of $\tau$ leptons. 

For the Higgs signal processes the $\tau$ pairs originate from the Higgs boson decay, 
as imposed at the generation level, and the number of subprocesses is reduced to 67. Each 
subprocess receives contributions from at most two Feynman diagrams, since with massless 
quarks of the first two generations, the Higgs boson can originate either from the vector boson 
fusion or from Higgs-strahlung diagrams, as illustrated  in Fig.~\ref{fig:diagram_h}. Depending
on the flavour configuration of incoming partons, mediating boson is $W$ or $Z$, which leads
to almost 10 GeV shift between resonance invariant mass of the outgoing pair of jets in case of 
Higgs-strahlung process.
The helicity-flipping scalar coupling to the Higgs boson results in the opposite spin 
correlation as compared to  the case of the Drell-Yan--process. The individual $\tau$ 
polarization is absent. 


Concerning the analytic structure of the differential cross sections, it is determined by 
topologies of contributing diagrams to a particular subprocess. 
For example,  $s$-channel propagators will result in a resonance enhancement, while 
the $t$-channel ones may lead to collinear  or soft singularities (in the limit $m_W^2/s\ll 1$, 
$m_Z^2/s\ll 1$) regulated either 
by the phase space cuts or by the virtuality of the attached boson line.  Understanding differences 
in analytic structures of subprocesses will turn important when discussing tests of reweighing 
technique of {\tt TauSpinner} in Subsection ~\ref{sec:TestsMEevents}.

Technically speaking,   the sums in Eq.~(\ref{eq:MC}) or (\ref{eq:WT}) defining the production weights
used in {\tt TauSpinner} consist of $9^4$ ($11^4$ if b-quarks are allowed)  elements, which are 
potentially distinct
and require their own subroutines for the matrix element calculation.
Since most of the elements are equal zero, or some  matrix elements  are related to others  
by permutation of partons  and/or  CP symmetries, special interfacing procedure is 
prepared to exploit those relations.
It reduces significantly the computation time and size of the program code.
Details are given in Appendix~\ref{app:HowToUse}.

\subsection{EW scheme and parameters}
\label{sec:EWdetails}

In early  versions of {\tt TauSpinner } 
the electroweak interactions 
were embedded into an effective ($2\to 2$) Born process for $q \bar q \to \tau^+\tau^-$.
Its analytic form is given by Eqs.~(3)-(5) and Table 2 of Ref. \cite{Pierzchala:2001gc}.
The adopted  scheme is fully compatible with the one of {\tt Tauola universal interface }
\cite{Davidson:2010rw}. 
It is using the lowest order ME for the $q \bar q \to Z/\gamma^* \to \tau \tau$ 
process, however with the effective value for the $ \sin^2\theta_W^{eff} $ and running Z-boson  
width. Such a choice corresponds  to a partial resummation 
of higher order electroweak effects, exactly as it was adopted at the time of precision tests 
of the Standard Model at LEP \cite{ALEPH:2005ab},   with the remaining loop weak corrections  at the 
per mille level.

However,  since the effects of $WW$ boxes can be numerically significant for $\tau$ lepton 
pairs of large virtuality or  large invariant mass, there is an option  to include genuine 
weak loop effects into {\tt TauSpinner} 
effective Born, already since its version 1.4.0 of June 2014 as well. It can be done for {\tt TauSpinner} in a manner similar to 
{\tt Tauola universal interface} \cite{Davidson:2010rw} because this process is implemented in both codes in the same way.

Table~\ref{tab:EWdetails} compares numerical values of the input parameters for the 
 ($2 \to 2$) and ($2 \to 4$) processes. Variants of initialization for $(2 \to 4)$  processes are
explained in Table~\ref{tab:EWschemes}. It is worth to point out that by using  over-constrained set of parameters ({\tt EWSH=4}): 
$\alpha_{QED}(M_{Z}), M_{Z}, \sin^2\theta_W^{eff}, M_W, G_F$ essential effects of the loop corrections
 are taken into account, 
providing the results for $\tau$-lepton polarisation close to LEP measurements \cite{ALEPH:2005ab}.  
 As the parameters are not independent, this can lead to problems if the input values are not consistent,  
especially when  applied to processes  other than ($2\to 2$),  which is the main focus of our paper. 
  
Let us now turn to the details of electroweak schemes used for the  matrix elements of 
the ($2\to 4$) hard subprocesses entering the $pp\to \tau \tau\ j j$. The code from {\tt MadGraph5} 
has its own initialisation module consistent with the so called $G_{F}$ 
scheme, which uses $G_F$, $\alpha_{QED}$ and $m_Z$ as input parameters, see~Table~\ref{tab:EWdetails}
(and {\tt EWSH=1} scheme in Table~\ref{tab:EWschemes}).
As such, it uses tree-level (equivalent to on-shell) definition of the weak mixing angle  
$ \sin^2\theta_{W} = 1 -{M_W^2}/{M_Z^2}=0.222246$. This  is far from the measured value
from the $Z$ boson couplings to fermions; also the constant width of $Z$-boson is used.  Since the 
$\tau$-lepton polarization is very sensitive to the value of the  mixing angle, and  for both {\tt Tauola} 
and {\tt TauSpinner} the $\tau$ physics is important target, such a LO implementation in the $G_F$ 
scheme is not sufficiently realistic. This is even more serious issue for the angular distributions of leptons themselves, 
making such a scheme phenomenologically inadequate to any observable that relies on  directions of leptons.
Alternatively, one could adopt the scheme with $G_F$, $m_Z$ and $\sin^2\theta_W^{eff}$ as input parameters 
(and {\tt EWSH=2} scheme in Table~\ref{tab:EWschemes}), but then the predicted 
tree-level $W$-boson mass is away from the measured value which would result in distorted spectra of jets coming
from $W$ decays (and shift in the resonance structure of the matrix element). 
In some regions of the phase space the distortion can reach 40\%. One can also use scheme with  $G_F$, $m_Z$ and  $m_W$, 
 as input parameters (and {\tt EWSH=3} scheme in Table~\ref{tab:EWschemes}), but the on-shell 
definition $ \sin^2\theta_{W} = 1 -{M_W^2}/{M_Z^2}=0.222246$ will lead back to far from measured value of  $ \sin^2\theta_{W}$.
There are two options: either include EW loop corrections simultaneously with QCD corrections, or adopt an effective 
scheme which would allow at tree-level to account correctly for the $\tau$-lepton polarization at 
the $Z$-boson peak and physical $W$-boson mass. Since the former is beyond the scope of the present paper, we take the 
second option.   


To this end, we define an effective scheme with  $\theta_W^{eff}$, in which the effective weak 
mixing angle  $\sin^2\theta_W^{eff} =0.2315$ is used (instead of the on-shell one) together with the 
on-shell boson masses, i.e. as input we take $G_F$,  $m_Z$, $m_W$ and 
$\sin^2\theta_W^{eff}$  ({\tt EWSH=4} scheme in Table~\ref{tab:EWschemes}). 
 Although being in principle flavour dependent, the  value $\sin^2\theta_W^{eff}$ is flavour universal with   an accuracy of order  
0.1\%. Effectively such a procedure amounts to
the inclusion of some of higher order EW corrections to the $Z\tau^+\tau^-$ 
vertex.\footnote{Although by itself the vertex correction is not gauge invariant, it has been shown 
for the case of $e^+e^- \to f \bar f$ that near the $Z$-pole the box contribution, needed to 
cancel gauge dependence, is numerically negligible. See for example Ref.~\cite{Hollik:1992kq}.}
This value is used in all vertices, also in the triple gauge-boson coupling since the $WWZ$  
coupling is essential for the gauge cancellation and it must match the couplings in other Feynman 
diagrams, forming together the gauge invariant part of the whole amplitude. 
 In our case we are not aiming at a careful theoretical study of higher order corrections; 
instead we checked numerically that the introduction of dominant loop corrections to 
$Z\tau^+\tau^-$ vertex through the effective $\sin^2\theta_W^{eff}$ does not lead to 
numerically important consequences for the $WWZ$ vertex.  For example, the
effect of the mismatch of $WWZ$ and $Zf\bar f$ couplings for the  case of $qq \to qq\tau\tau$ 
subprocess is small, see Fig.~\ref{fig:EWModif}, in Section~\ref{sec:WWZtest}. Thus,  we gain consistency with observables, such as $\tau$-polarization 
or $\tau$-directions, which would otherwise be off by $\sim$ 40\% at the expense of breaking EW relations in higher order 
of perturbation theory. Moreover, since {\tt TauSpinner} is used to reweigh events, as given in Eq.~(\ref{eq:WT}),
 the  uncertainties of our procedure should to a large degree cancel out. 

For the purpose of comparison of the predicted  $\tau$-lepton polarisation at the Z-boson peak,  we provide four  initialisation options for the 
($2 \to 4$) matrix elements, the first three motivated by the schemes used in~\cite{Baglio:2014uba}, and the fourth one corresponding to $\theta_W^{eff}$. 
They are specified in Table~\ref{tab:EWschemes}. Scheme labeled $\texttt{EWSH}=4$ is the only one numerically   appropriate for 
use when predicting $\tau$ polarisation, and taking into account configurations with two additional  jets, as 
shown in Section~\ref{sec:MEpheno}.
For technical testing purposes we also introduce a scheme like  {\tt EWSH=4} but with modified  $WWZ$ coupling by $  5 \%$
which we label as  {\tt EWSH=5}.

\begin{table}
 \caption{ Input parameters for initialising couplings calculations for $(2 \to 2)$ in {\tt Tauola} code 
and $(2 \to 4)$  in {\tt MadGraph5} code. Note that in {\tt Tauola} code $\alpha_{QED} =\alpha_{QED}(Q^2=0) $
is used as an input for calculation of the $Z$ couplings as well. This leads, in principle, to an over-all missing factor of 
$(\frac{\alpha_{QED}(Q^2)}{\alpha_{QED}(0)})^2$.
It can be thus 
 dropped off, as long as it cancels out in calculation of  weights, the 
ratios of differential cross-sections. 
The numerical values  of CKM matrix are taken from Ref.~\cite{Alwall:2014hca}.}
\label{tab:EWdetails}
 \vspace{2mm}
  \begin{center}
  \begin{tabular}{|l|r|r||r|r|}
  \hline\hline 
 Type                & {\tt Tauola} code & Input/Calculated & {\tt MadGraph5} code   & Input/Calculated \\ 
                     & (default for $2 \to 2$)&     &  (SM default for $2 \to 4$)                  &                  \\
  \hline\hline
 $ m_{H} $            &   -----            &         &  125.0   GeV                   &  Input\\ 
 $ \Gamma_{H} $       &   -----            &         &  0.0057531 GeV                 &  Input \\ 
 $ m_{Z} $            &  91.1882 GeV       & Input   &  91.1880 GeV                   &  Input  \\ 
 $ \Gamma_{Z} $       &  2.4952 GeV        & Input   &  2.44140 GeV                   &  Input\\ 
 $ m_{W} $            &   -----            &         &  80.4190                       & Calculated\\ 
 $ \Gamma_{W} $       &   ------           &         &  2.04760 GeV                   &  Input\\ 
 $ m_{\tau} $          & 1.77703 GeV        & Input   &  1.77703 GeV                   &  Input\\ 
 $ sin^2\theta_{W} $   & 0.23147           &  Input   &  0.222220                      & Calculated        \\ 
 $ 1/\alpha_{QED}  $   &   137.036          &  Input   &  132.507                      &  Input \\ 
 $ G_{F}  $            &  -----            &          &  1.16639 $10^{-5}$ GeV$^{-2}$    &  Input  \\ 
\hline
\end{tabular}
\end{center}
 \caption{Implemented EW schemes, the recommended EW scheme is {\tt EWSH=4} which gives the 
$\tau$ lepton polarisation on the Z-boson mass peak, in agreement with the measurement at LEP1 \cite{Altarelli:1996gh}, 
and physical $W$ boson mass.
}
\label{tab:EWschemes}
 \vspace{2mm}
  \begin{center}
  \begin{tabular}{|l|r|r|r|r|}
  \hline\hline 
 Type        &   {\tt EWSH=1}            & {\tt EWSH=2}                 &       {\tt EWSH=3}             & {\tt EWSH=4}      \\
             & input: $G_F,\alpha_{QED}, m_Z$ & input: $G_F,sin^2\theta_{W}, m_Z$ & input:  $G_F, m_W, m_Z$   & input: $G_F, m_W, m_Z,\sin^2\theta_W^{eff}$                  \\
  \hline\hline
 $ m_{Z} $            &  91.1882 GeV                 &  91.1882 GeV               &  91.1882 GeV                  &  91.1882 GeV   \\ 
 $ m_{W} $            &  80.4190                     &  79.9407  GeV              &  80.4189 GeV                  & 80.4189 GeV  \\ 
 $ sin^2\theta_{W} $  &  0.222246                    &  0.231470                   &  0.222246                    &  0.231470 \\ 
 $ 1/\alpha_{QED}  $  &  132.5070                    &  128.7538                   &  132.5069                    & 127.2272  \\ 
 $ G_{F}  $           &  1.16639 $10^{-5}$ GeV$^{-2}$  & 1.16639 $10^{-5}$ GeV$^{-2}$ &  1.16639 $10^{-5}$ GeV$^{-2}$   & 1.16639 $10^{-5}$ GeV$^{-2}$  \\ 
\hline
\end{tabular}
\end{center}
\end{table}


\subsection{QCD scales and parton density functions}
\label{app:QCDdetails}

The distribution version of 
\texttt{TauSpinner} is interfaced with \texttt{LHAPDF} v6 library \cite{Buckley:2014ana}.
User has the freedom of choosing renormalization and factorization scales,
within the constraint that $\mu_F=\mu_R$, otherwise minor re-coding is necessary.
To this end we have implemented four predefined choices for the scale $\mu^2$ as should be expected
for our processes:
\begin{center}
\begin{tabular}{ccl}
  {\tt scalePDFOpt=0}&~  & 200 GeV                  \\
  {\tt scalePDFOpt=1} & & $\mu = \sqrt{\hat s}$ \\
  {\tt scalePDFOpt=2}  & & $\mu = \sum m_T$, \quad $m_T^2 = m^2 + p_\perp^2$ \\
  {\tt scalePDFOpt=3} & & $\mu = \sum E_\perp$, \quad $E_\perp = E p_\perp /| \vec{p}|$ \\
\end{tabular}
\end{center}
where sums are taken over final state particles of hard scattering process.
For the $\alpha_s(\mu^2)$  we provide, as a default, a simple choice of the $\mu^2$ dependence, 
following the leading logarithmic formula,
\begin{equation}
\alpha_s(\mu^2)= \frac{\alpha_s(M_Z^2)} { 1 +4 \pi\alpha_s(M_Z^2) (11 - 2N_f/3 )  \ln{\frac{\mu^2}{M_Z^2}}}
\label{eq:alfas}
\end{equation}
 with the starting point $\alpha_s(M_Z^2)=0.118$. The same value of $\alpha_s$ is used for the case of 
the  fixed coupling constant,  that is 
for {\tt scalePDFOpt=0}.

The reweighting procedure of {\tt TauSpinner} itself may be used to study numerically the effects of 
different scale choices, as well as for the electroweak schemes, see the discussion later in Section 
\ref{sec:MEpheno} and Appendix~\ref{sect:Initialisation}.

\section{Tests of implementation of ($2\to 4$) matrix elements.}
\label{sec:MEtests}

\subsection{Tests of matrix elements using fixed kinematical configurations}
\label{sect:fixKine}
For the purpose of testing the consistency of implemented  codes, generated with 
{\tt MadGraph5} and modified  
as explained in Sect. \ref{sub:incorpo},  we have chosen a fixed kinematic configuration 
at the parton level\footnote{This test is  build 
into the {\tt TauSpinner} testing and can be activated with the hard-coded local variable of
{\tt TAUOLA/TauSpinner/src/VBF/vbfdistr.cxx}  by setting {\tt const bool  DEBUG = 1;}.
 Numerical results are collected
on the project web page \cite{WebPageSoinner2j}.}.
For such kinematics we have calculated the matrix element squared for all possible helicity configurations 
of all subprocesses using the codes implemented in {\tt TauSpinner} and checked against 
the numerical values obtained directly 
from {\tt MadGraph5}.  The agreement of at least 6~significant digits has been confirmed.

               
\subsection{Tests of matrix elements using series of generated events}
\label{sec:TestsMEevents}
{As further tests of the internal consistency  of matrix element implementation in 
{\tt TauSpinner} we have used the reweighting procedure by comparing 
a number of kinematic distributions obtained in two different ways: the first one  
obtained directly from events generated for 
a specified parton level process REF (a reference distribution REF), and  
the second one (GEN reweighted) obtained by reweighting with {\tt TauSpinner} events 
generated for a different process GEN. 
These tests have been performed in a few steps as follows.}
\begin{itemize}
\item
{Series of 10 million events each for a number of different processes in $pp \to \tau \tau jj$ 
(with specified flavours of final state jets, or for subprocesses with selected flavours of incoming partons) 
with {\tt MadGraph5\_aMC@NLO} \cite{Alwall:2014hca} v2.3.3 at LO have been generated.}
Samples were generated for $pp$ collisions at the c.m.\ energy of 13 TeV using CTEQ6L1 
PDFs \cite{Pumplin:2002vw} linked through LHAPDF v6 interface.
Renormalization and factorization scales were fixed to $\mu_R = \mu_F =m_Z$. Only very 
loose selection criteria at the generation level were applied: invariant mass of 
the $\tau \tau$ pair was required to be in the range\footnote{Several tests were repeated also 
for the full spectrum, i.e. starting from $m_{\tau \tau} > 10$ GeV}
 $m_{\tau \tau}=60-130$~GeV, and jets to be separated by  $\Delta R_{jj} > 0.1$ and with transverse 
momenta $p_T^{j} > 1$~GeV.  A complete configuration file used for events generation is given in the file
{\tt MadgraphCards.txt} which is included for reference in 
{\tt TAUOLA/TauSpinner/examples/example-VBF/benchfiles} directory. 
\item
The testing program was reading generated events stored in the  {\tt LesHouches Event File} 
format \cite{Alwall:2006yp} filtering 
the ones of a given {\tt ID1, ID2, ID3, ID4} configuration of flavour of incoming/outgoing partons 
corresponding to the process GEN. 
The weight $wt_{ME}$ allowing to transform this subset of events into  the equivalent of reference  REF one,
was calculated as
\begin{equation}
wt_{ME} = \frac{|ME(ID1, ID2, ID3', ID4')|^2}{|ME(ID1, ID2, ID3, ID4)|^2}
\end{equation}
and kinematic distributions of reweighted events (GEN reweighted)  
were compared to distributions of 
the reference process  REF {\tt (ID1, ID2, ID3', ID4')}. 
Note that for this test
to be meaningful one has to select processes with the same initial state partons, so that the 
dependence on  the structure functions cancels out.
A very good agreement between the REF and GEN reweighted distributions 
was found for 10 different kinematic distributions  for several configurations of  
{\tt (ID1, ID2, ID3, ID4, ID3', ID4')}. It has shown a very good
numerical stability, which was not obvious from the beginning as events corresponding  to 
the REF  and GEN  processes may have very  different kinematic distributions due to their 
specific topologies and resonance structures of Feynman diagrams.
\item
In the next step, the tests were repeated, but now reweighting the matrix elements   convoluted 
with the structure functions of the incoming patrons and summing over final states  restricted to 
the selected sub-groups (named respectively $C$ and $D$ of parton level processes). In 
this case the weight is calculated as
\begin{eqnarray}
wt_{prod}^{C \rightarrow D} = 
\frac{ \sum^D_{i,j,k,l} f_i(x_1)f_j(x_2)  |M_{i,j,k,l}(p_1,p_2,p_3,p_4)|^2  \frac{1}{\Phi_{flux}}d\Omega(p_1,p_2; \;p_3.p_4,p_{\tau^+},p_{\tau^-})}
     { \sum^C_{i,j,k,l} f_i(x_1)f_j(x_2)  |M_{i,j,k,l}(p_1,p_2,p_3,p_4)|^2  \frac{1}{\Phi_{flux}}d\Omega(p_1,p_2; \;p_3.p_4,p_{\tau^+},p_{\tau^-})}\label{wtxsec}
\end{eqnarray}
where the notation as for Eq. (\ref{eq:WT}) is used, except that now the $\sum^{C,D}$ mean 
that summation is restricted to processes belonging to sub-groups $C,D$, respectively.
For testing the code implementation for the Drell-Yan process the groups, listed in the first column of Table~\ref{tab:processes},   
 were reweighted, one to another.
\end{itemize}

The reweighting tests performed between sub-groups of processes, and later, between groups of processes listed in 
Table~\ref{tab:processes}, 
 allowed to check relative normalization of amplitudes. 
 Again, a good 
agreement has been found\footnote{ Technical point is worth mentioning: we had to 
 randomize order of final state partons in events generated by {\tt MadGraph5}, as such order
in not imposed in the  matrix elements implemented in {\tt TauSpinner}.
}.
For the tests, the following kinematical distributions were used: \\
$-$ Pseudorapidity of an outgoing parton $j$. \\
$-$  Pseudorapidity gap of outgoing partons. \\
$-$ Rapidity of the $ \tau \tau$ and  $ j j $ systems. \\
$-$  Transverse  momentum of the $ \tau \tau$ and  $ j j $ system. \\
$-$  Invariant mass of the $ \tau \tau$ and  $ j j $ system. \\
$-$ Longitudinal momentum of the  $ \tau \tau$ and  $ \tau \tau j j $.  \\
$-$  Cosine of the azimuthal angle of $\tau$ lepton in the $\tau \tau$ rest frame.

Let us discuss some of these results, shown 
in Fig.~\ref{fig:pseudojj} and  Fig.~\ref{fig:pTtautau} (the complete set of distributions is 
shown in Appendices  \ref{app:XsecTestsDY} and \ref{app:XsecTestsH}).  
In each plot the distribution  REF for the reference process is shown as a black histogram, 
while the red histogram shows the distribution for a different process GEN. Both histograms 
are obtained  directly from the 
{\tt MadGraph5} generated samples of REF and GEN processes, respectively. 
Now the histogram GEN is reweighted using {\tt TauSpinner} and 
the resulting reweighted histogram  is represented  by the 
red points with error bars.  For the test to be successful  
the  red points should follow the black histogram; the ratio of the REF and 
GEN reweighted  distributions is shown in the bottom panel of each figure.

Let us note, that in our tests, we reweight events of substantially  different dynamical structures 
over the multi-dimensional 
phase-space. This may be not evident from the histograms shown in figures, which can 
be both for the REF  and GEN reweighted 
distributions rather regular and similar. Nevertheless, several 
bins of GEN reweighted distributions with small errors can be found to lie 
below the REF distribution, whereas a few above  with large errors. 
This second category of bins is populated by a few events, which originate  from  the flat 
distribution of the GEN process, receiving high weight due to some 
resonance/collinear 
configuration of the REF process. This is a technical difficulty 
for the testing, but is not an issue of the actual use of {\tt TauSpinner} when all 
subprocesses are used together.
To confirm that the observed deviations are not significant statistically
we have reproduced plots from Fig.~\ref{fig:pseudojj} and~Fig.~\ref{fig:pTtautau} for 
four independent series of events. We observed that bins with large error or sequences of few bins with large deviations
 were randomly distributed between these series strongly indicating that observed deviations are of statistical origin.
As primarily we are not interested in use of 
implemented code to reweight between the groups of parton level processes, for checking general correctness of its 
implementation it was sufficient to use four statistically independent samples only.  In practical applications, 
contributions from all processes will be merged together and weights will become less dispersed.

\begin{figure}
  \begin{center}                               
{
    \includegraphics[width=7.5cm,angle=0]{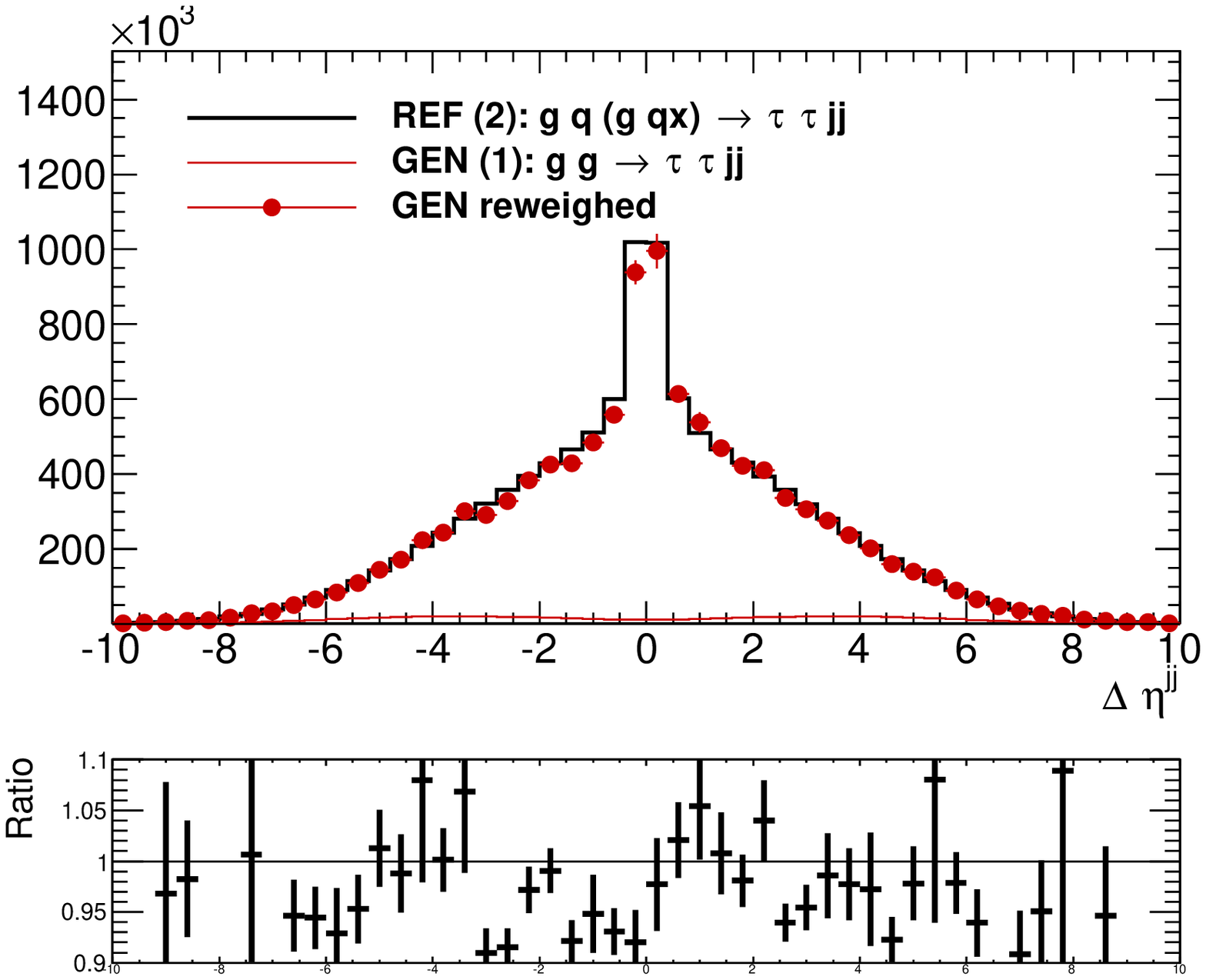}
    \includegraphics[width=7.5cm,angle=0]{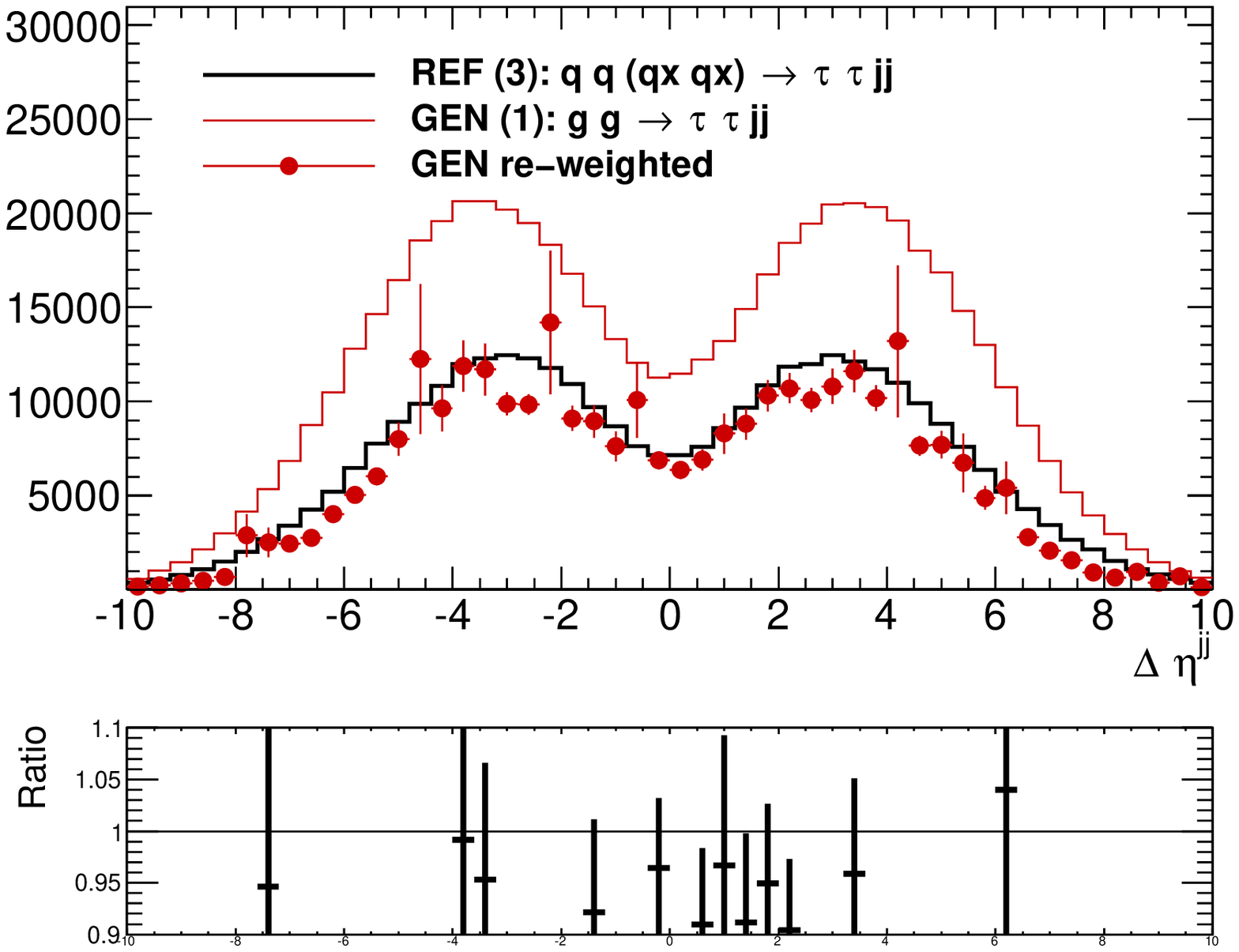}
    \includegraphics[width=7.5cm,angle=0]{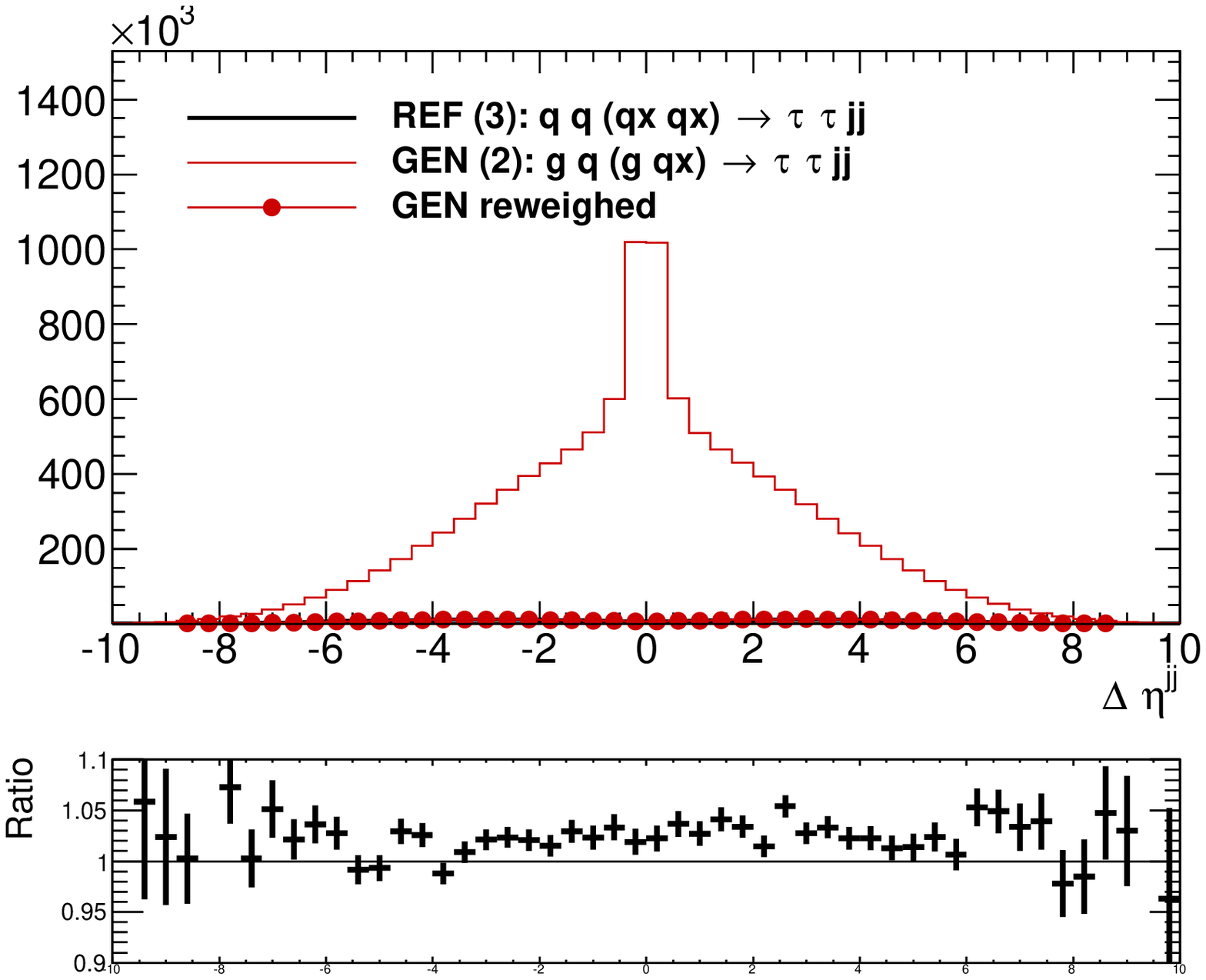}
    \includegraphics[width=7.5cm,angle=0]{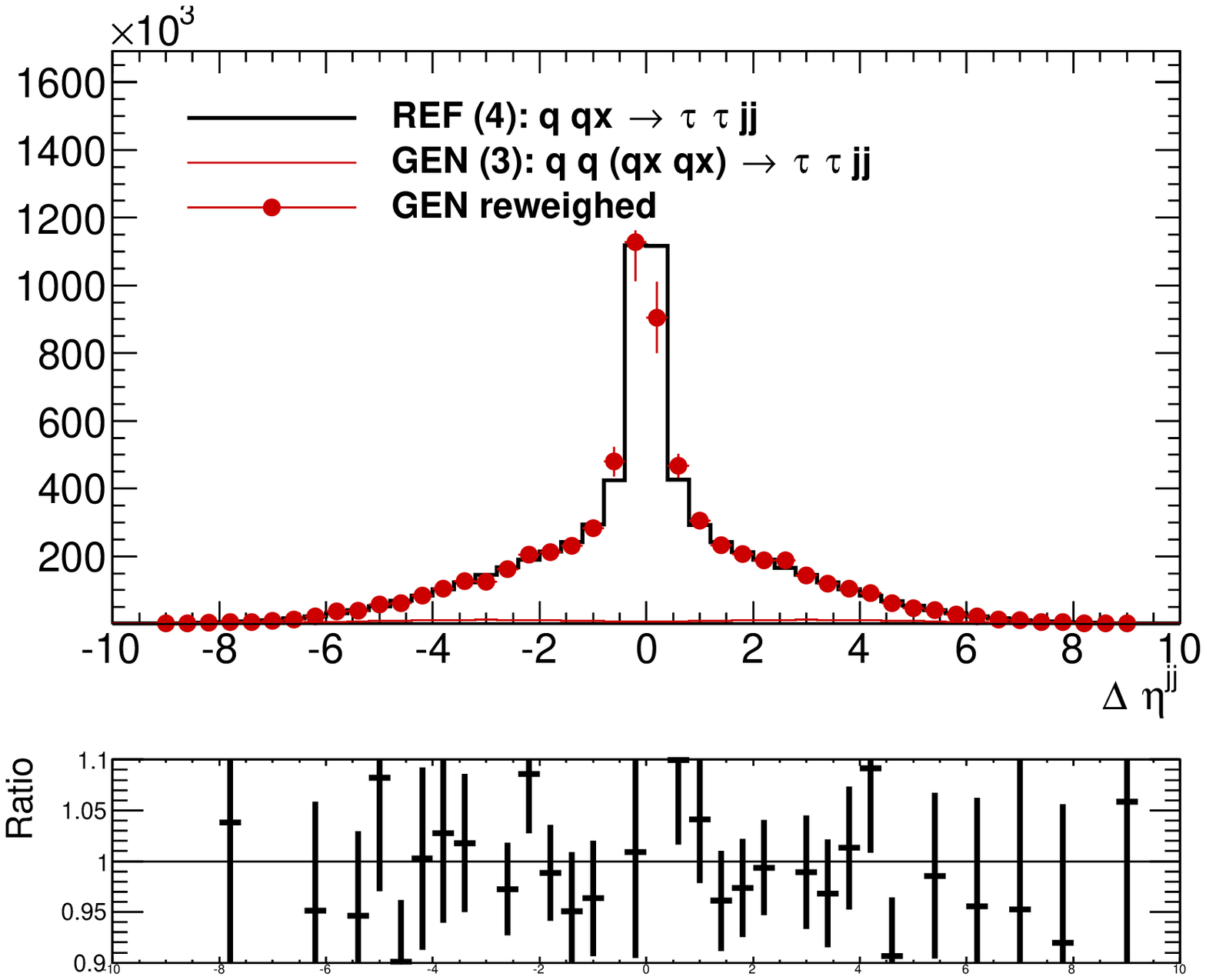}
    \includegraphics[width=7.5cm,angle=0]{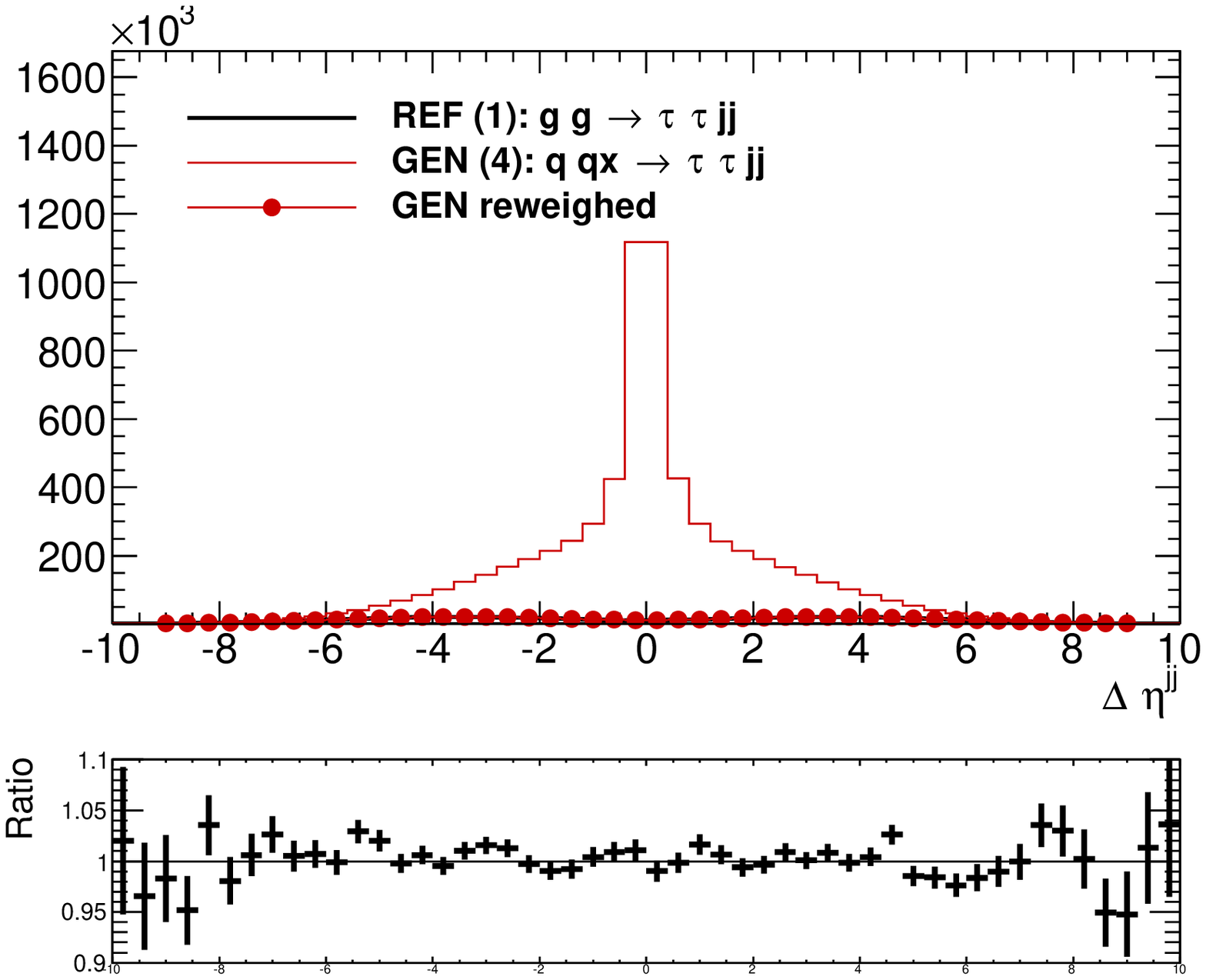}
    \includegraphics[width=7.5cm,angle=0]{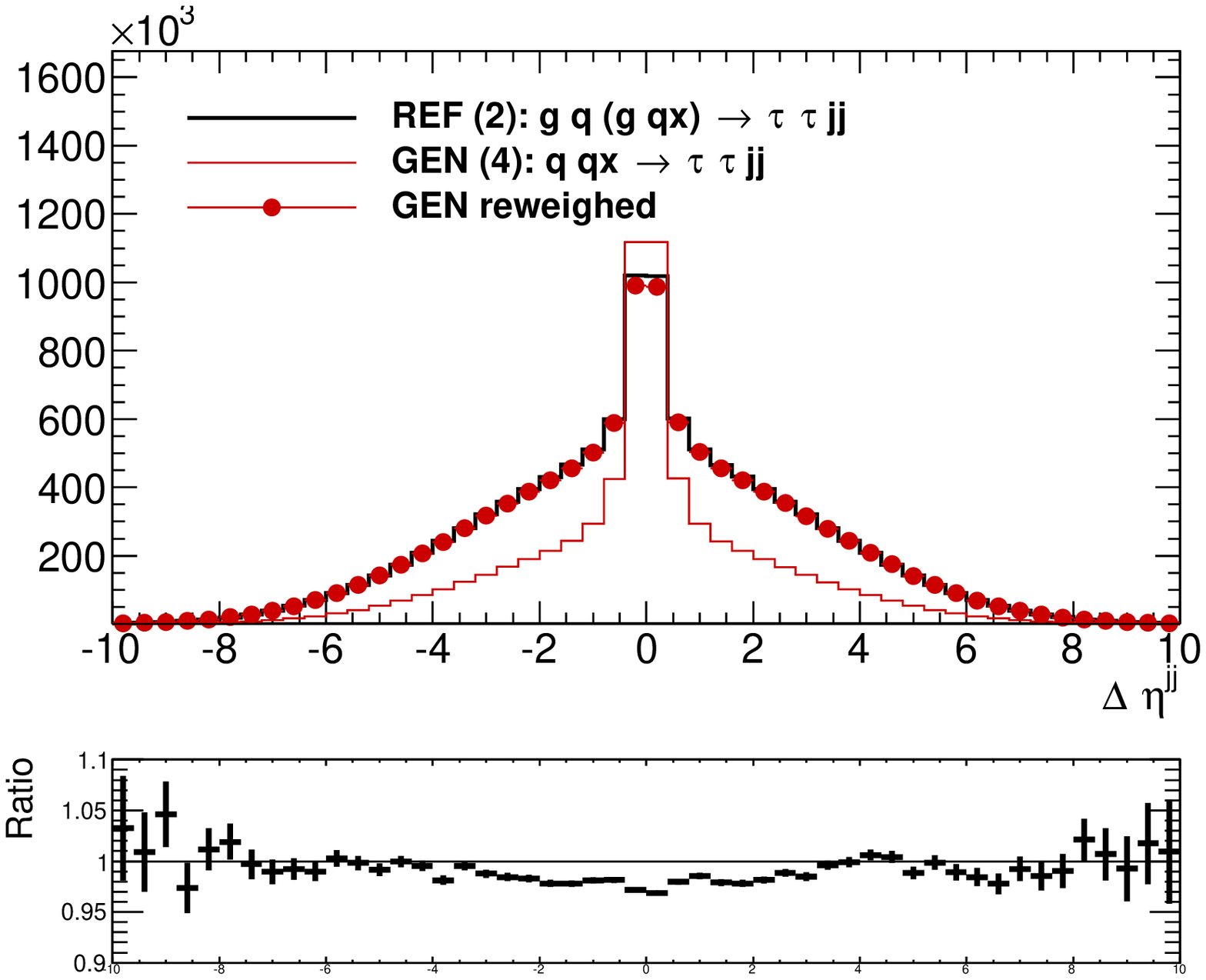}
}
\end{center}
\caption{Shown are distributions of the pseudorapidity gap between outgoing partons 
for the  GEN sub-process (thin red line) and  after its reweighting to the reference one   
(GEN reweighted, red points). Reference distribution REF is shown with 
a black line.  GEN and REF sub-processes are grouped as listed in Table \ref{tab:processes}.  The qx on plots, denote antiquark i.e. $\bar q$.
More plots for other distributions are given
in Appendix~\ref{app:XsecTestsDY}.   
\label{fig:pseudojj} }
\end{figure}

\begin{figure}
  \begin{center}                               
{
    \includegraphics[width=7.5cm,angle=0]{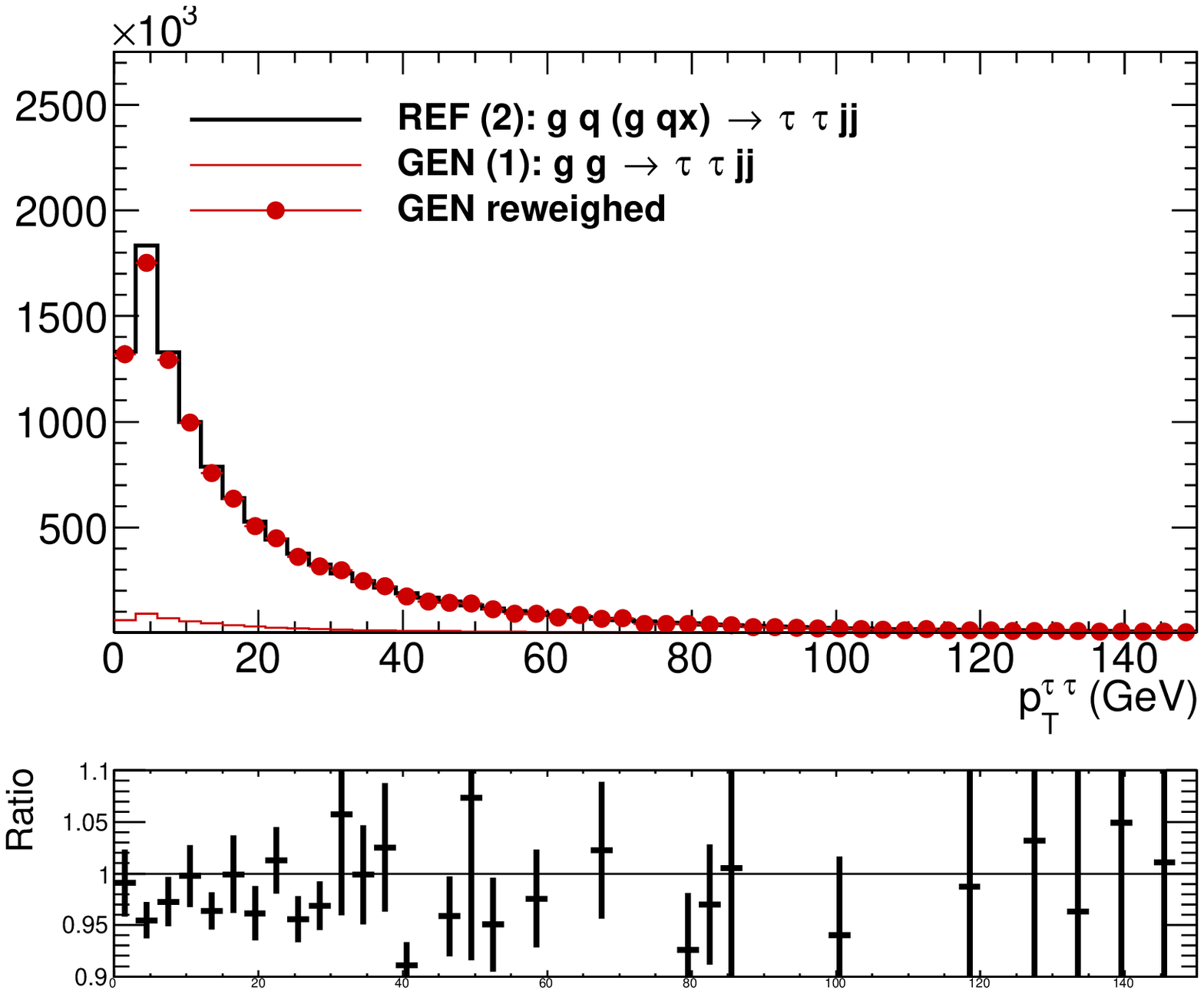}
    \includegraphics[width=7.5cm,angle=0]{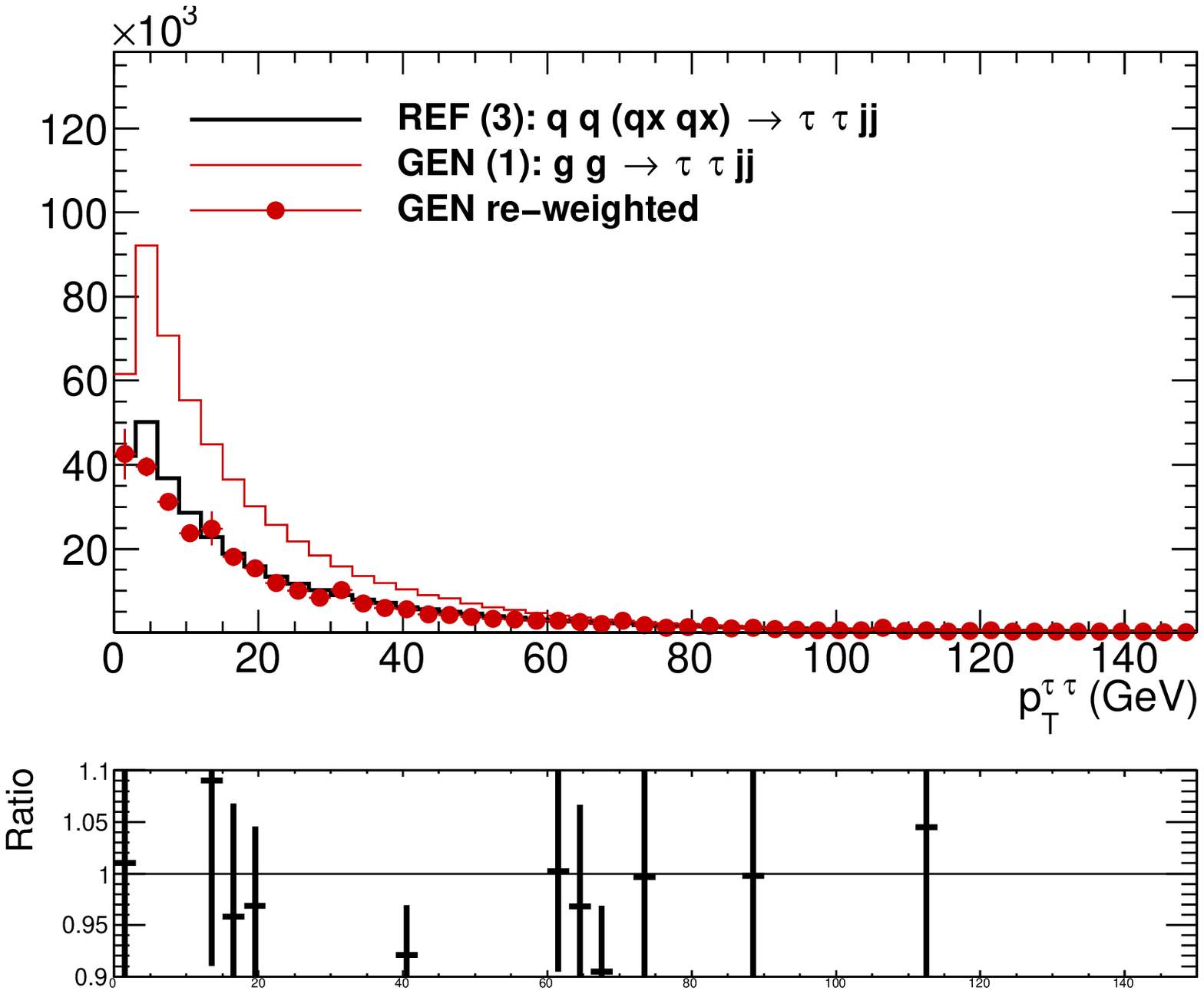}
    \includegraphics[width=7.5cm,angle=0]{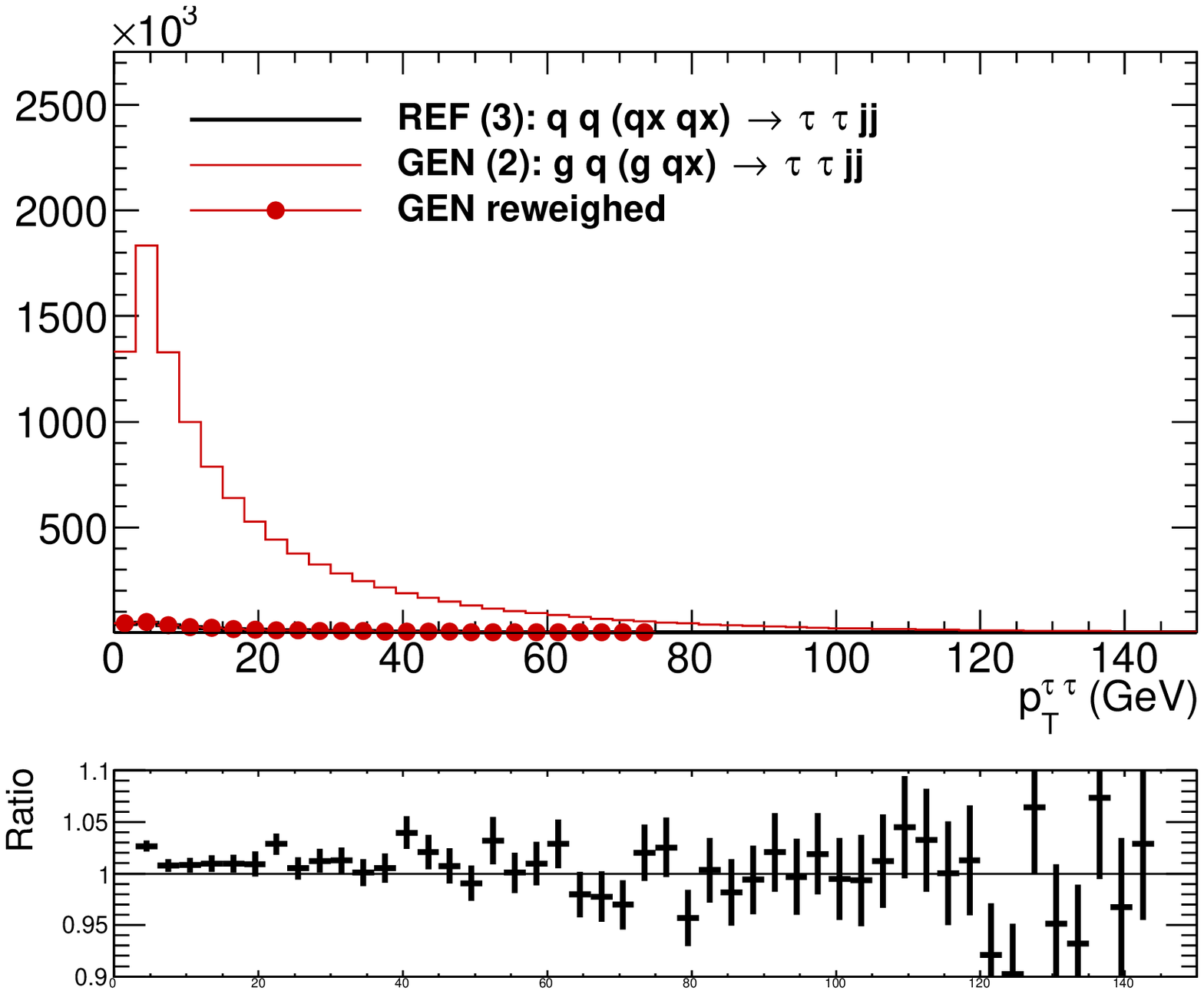}
    \includegraphics[width=7.5cm,angle=0]{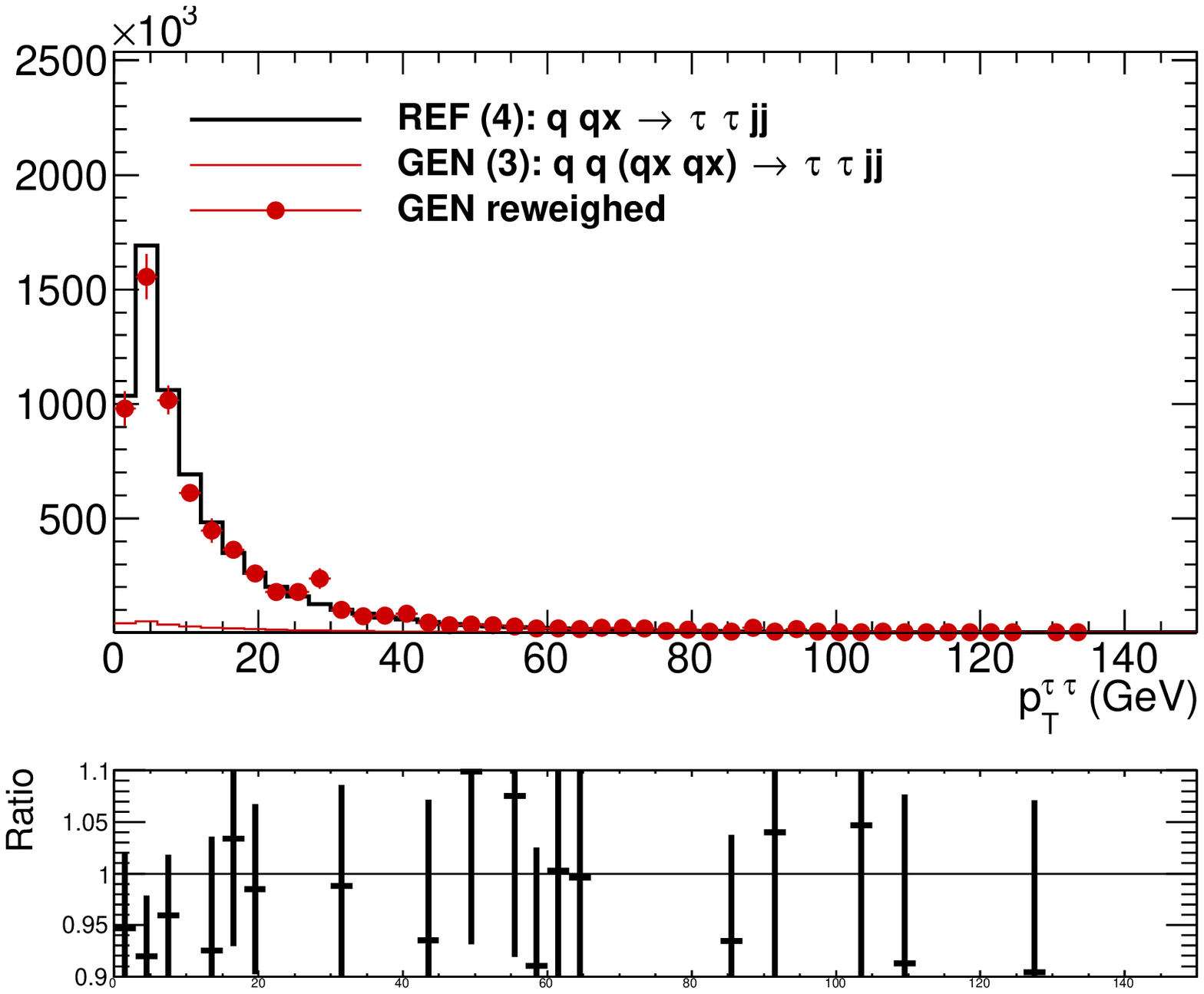}
    \includegraphics[width=7.5cm,angle=0]{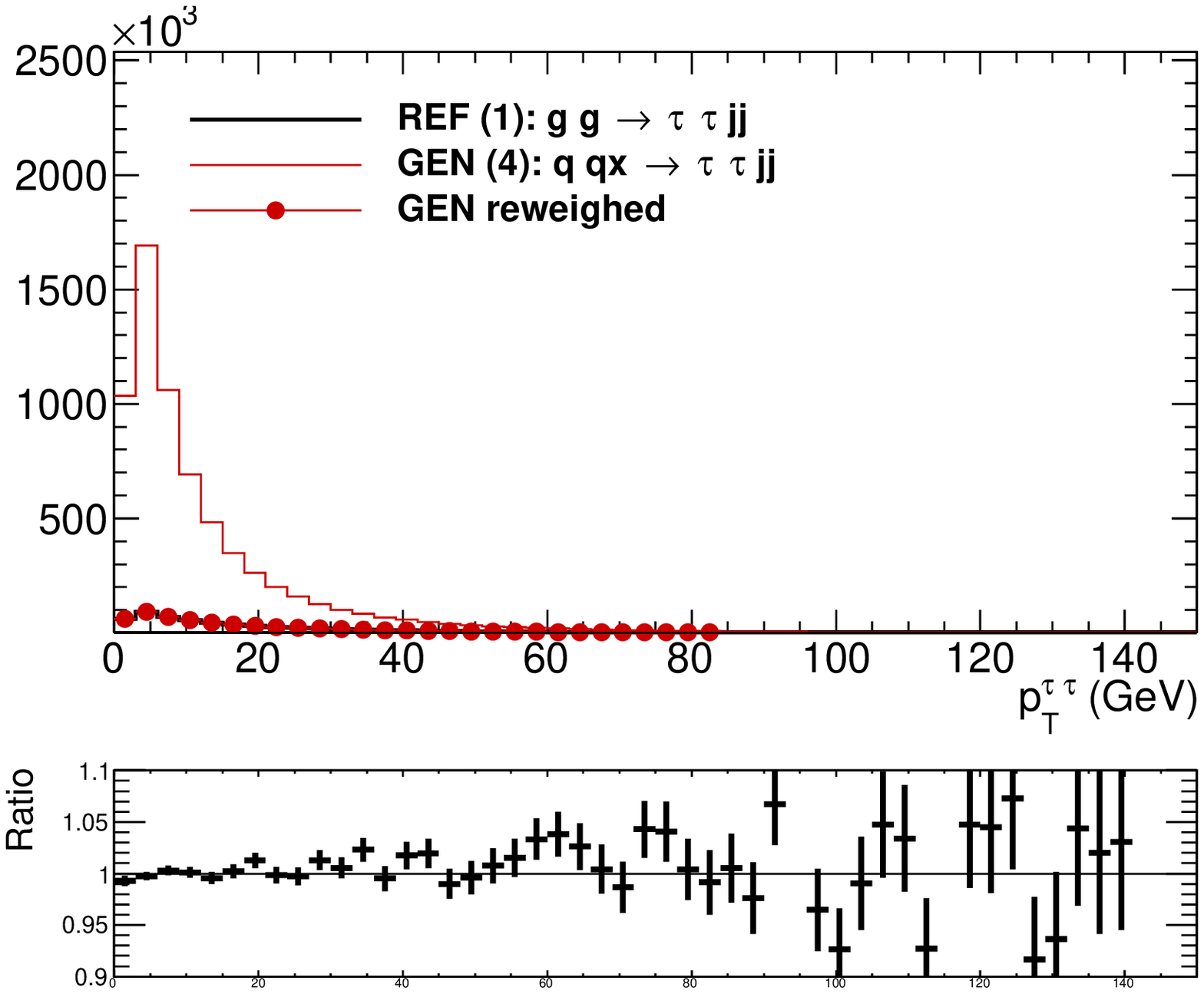}
    \includegraphics[width=7.5cm,angle=0]{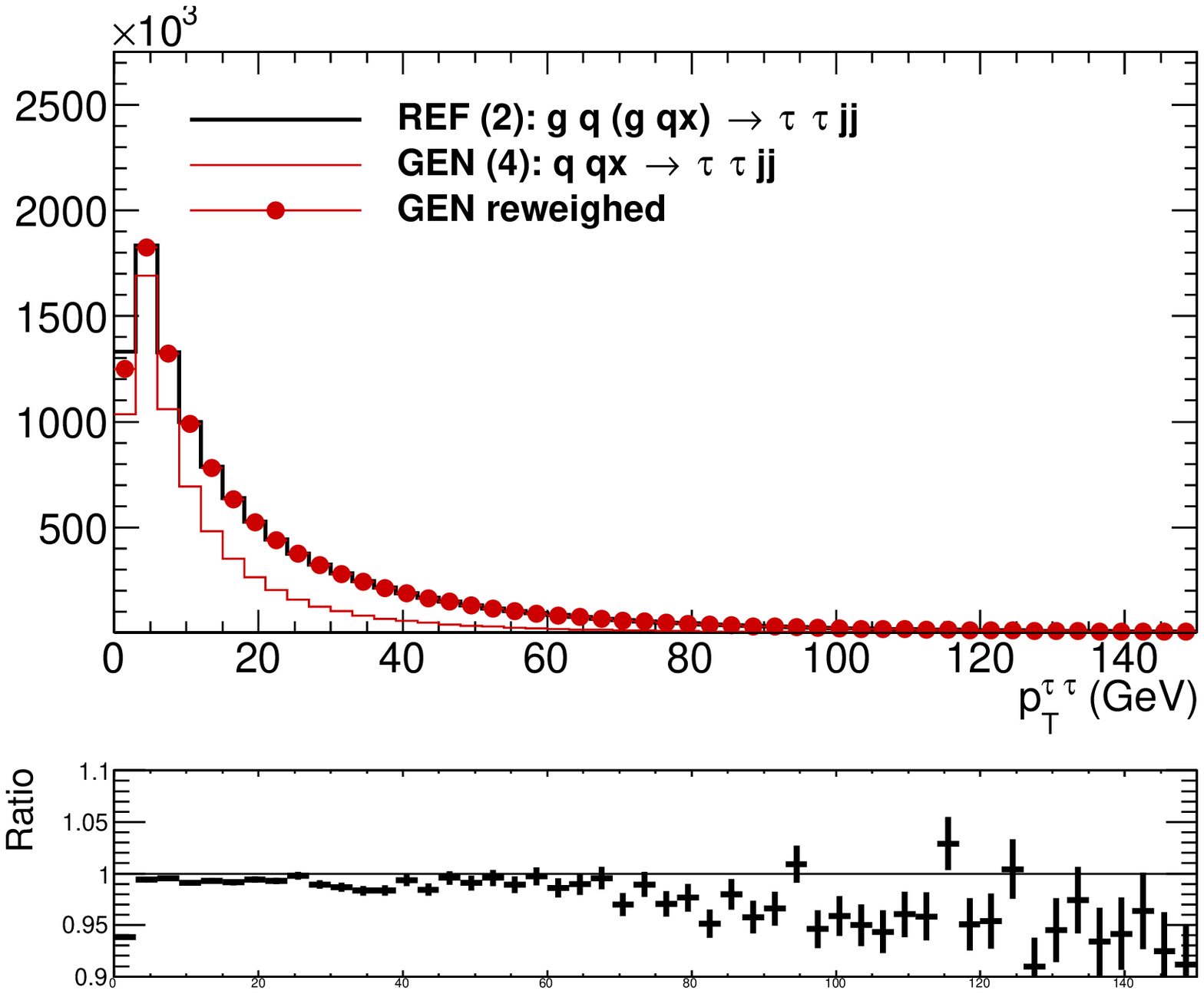}
}
\end{center}
\caption{Shown are distributions of transverse momenta of $\tau$ pairs, $p_T^{\tau \tau}$
with labeling as in Fig.~\ref{fig:pseudojj}.
\label{fig:pTtautau} }
\end{figure}

\begin{figure}
  \begin{center}                               
{
   \includegraphics[width=7.5cm,angle=0]{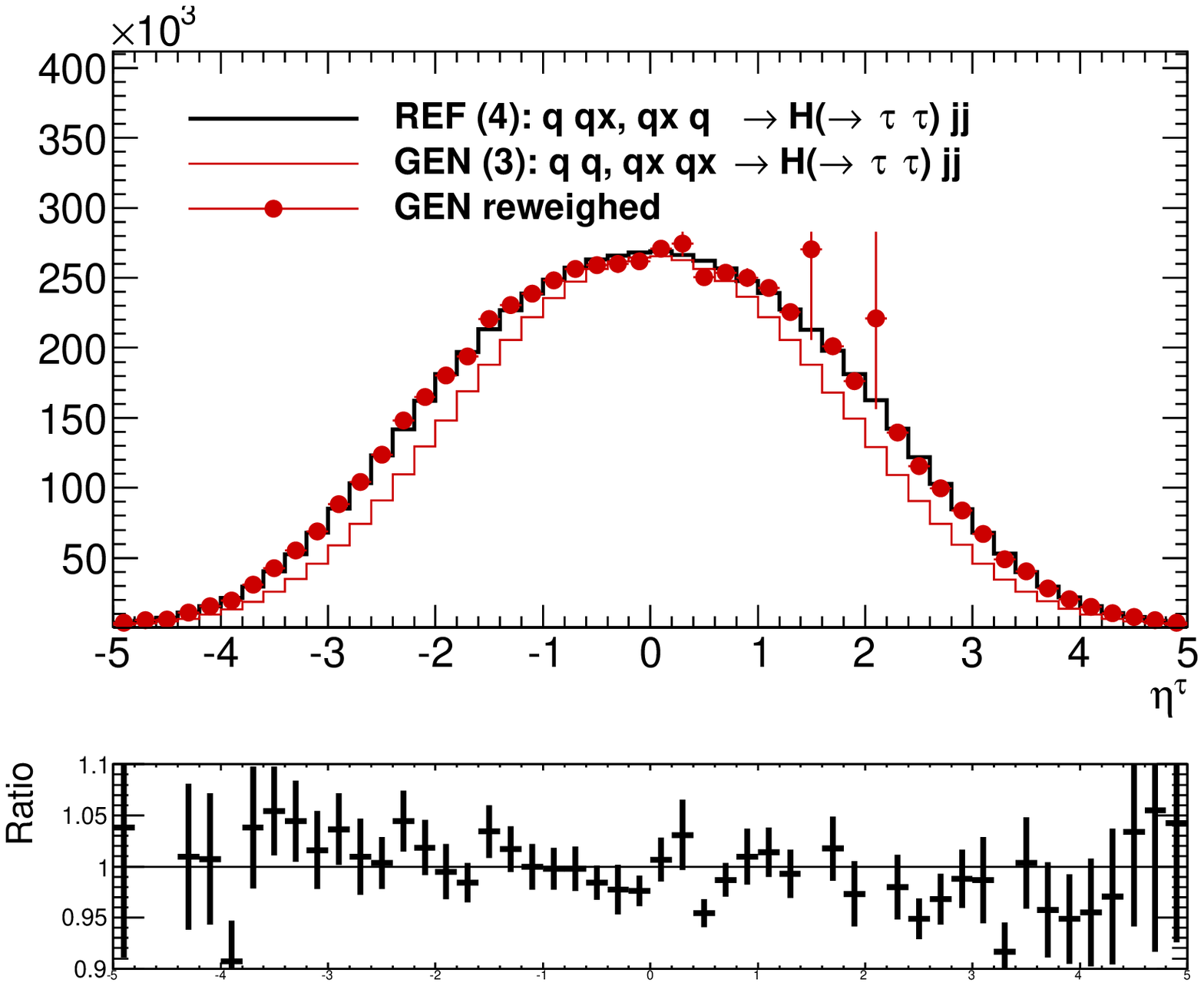}
   \includegraphics[width=7.5cm,angle=0]{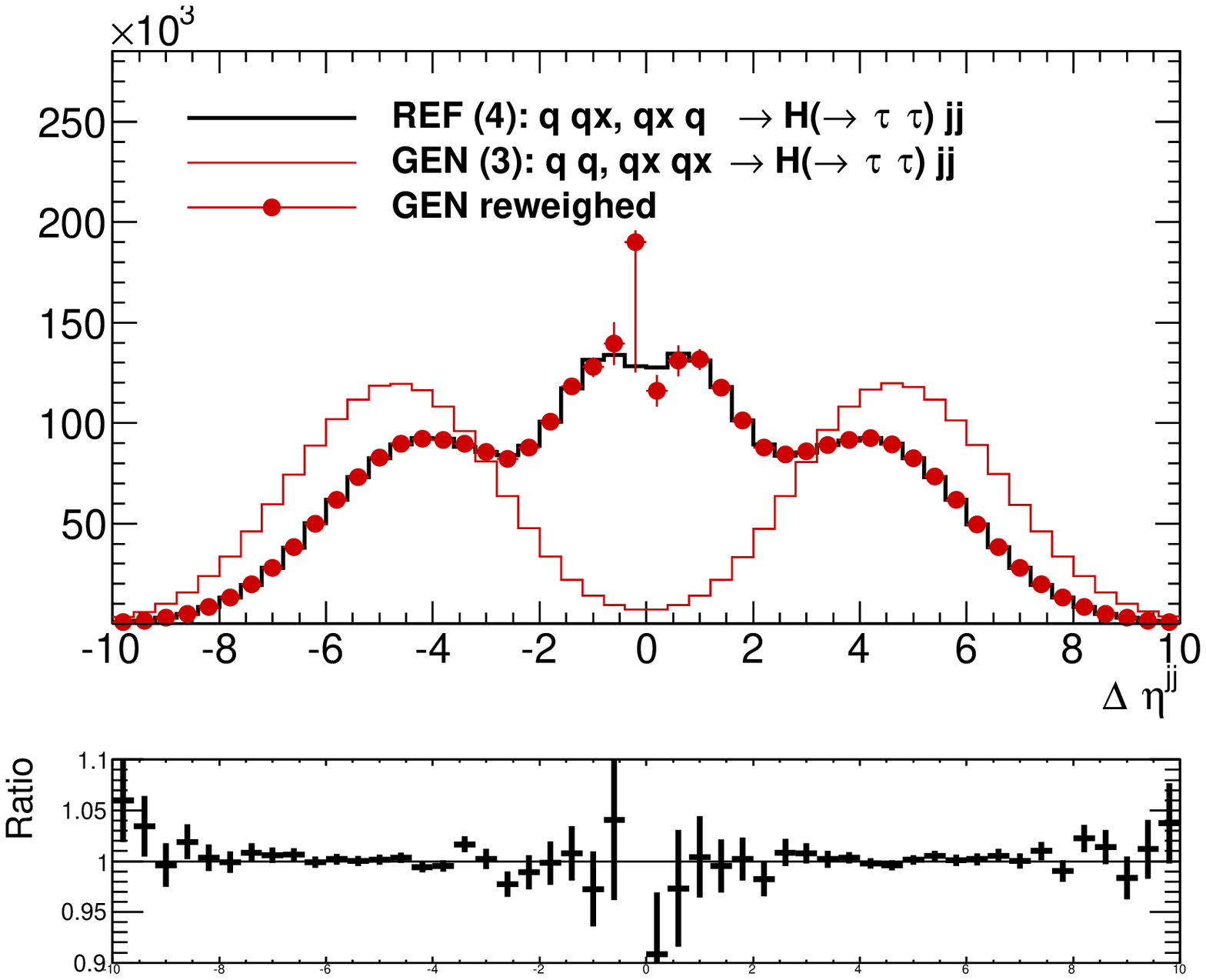}
   \includegraphics[width=7.5cm,angle=0]{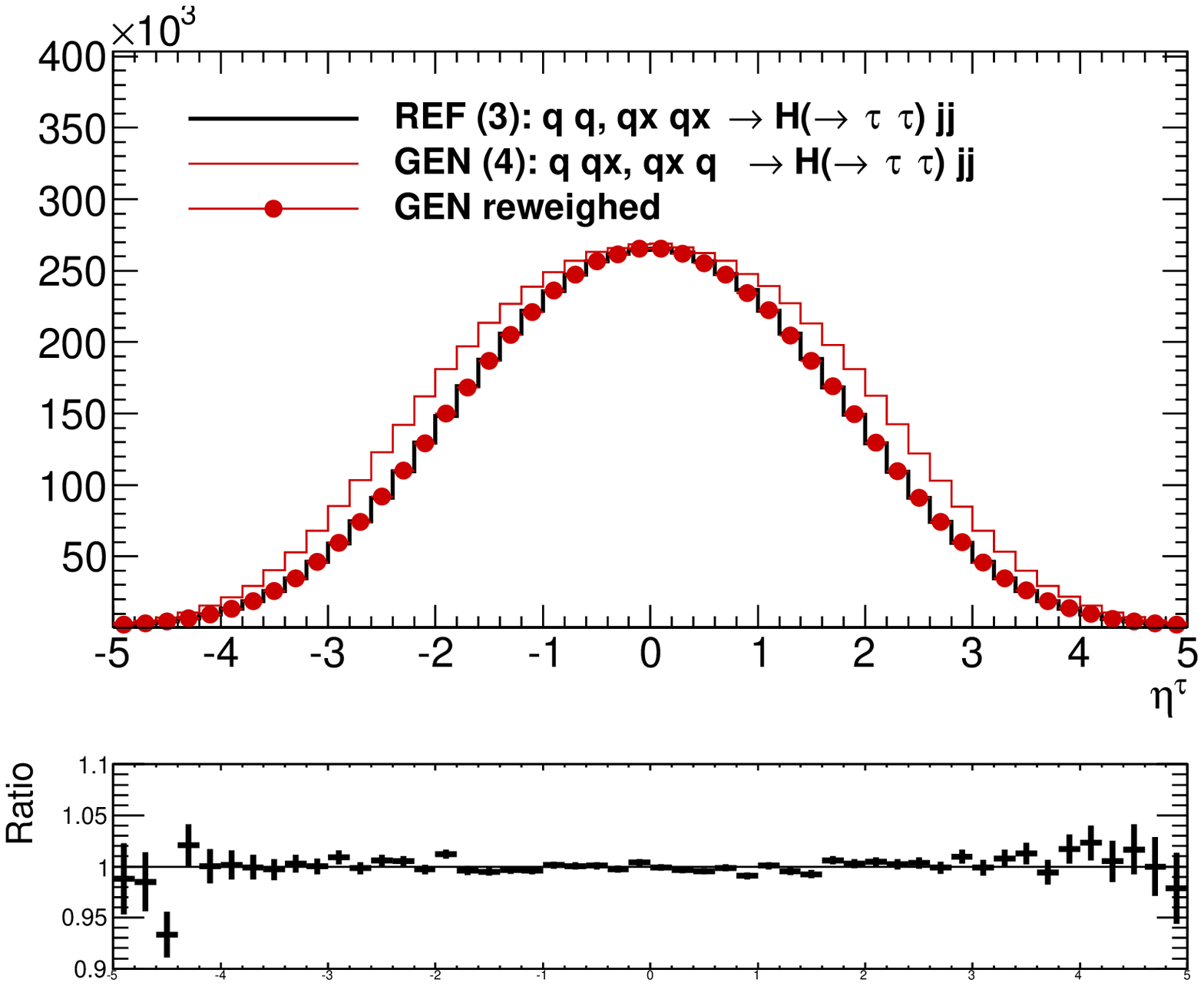}
   \includegraphics[width=7.5cm,angle=0]{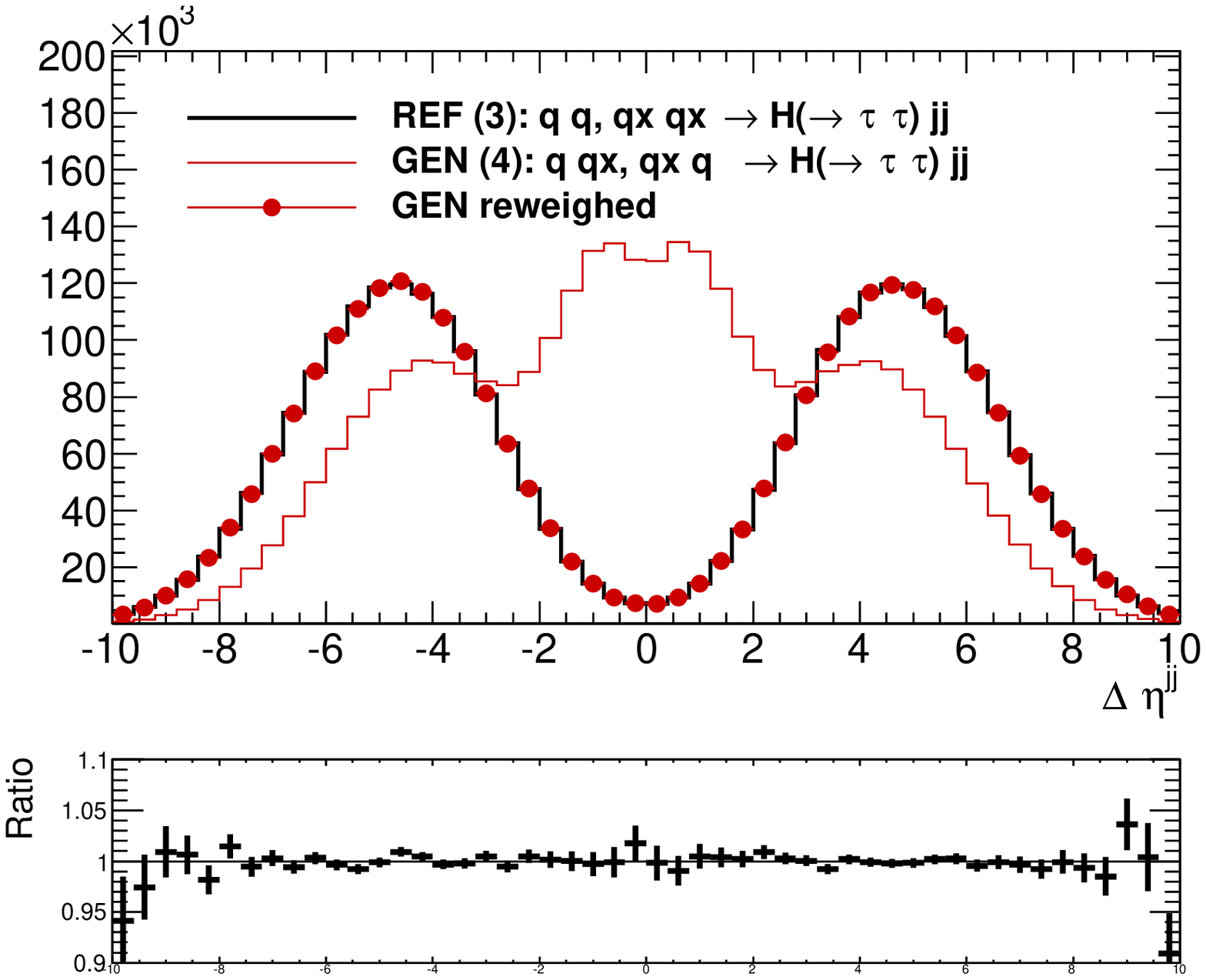}
}
\end{center}
\caption{Shown are the distributions of the jet pseudorapidity (left plots)  and the pseudorapidity gap between outgoing partons 
(right plots) with labeling as in Fig.~\ref{fig:pseudojj} but for the processes of Higgs boson production.
More plots for other distributions are given
in Appendix~\ref{app:XsecTestsH}.
\label{fig:Higgstests} }
\end{figure}

Similar tests have been performed  for the 
Higgs boson production. 
Fig.~\ref{fig:Higgstests} shows the  comparison of generated and reweighed distributions 
for the jet pseudorapidity and for the 
pseudorapidity gap between jets in the  case of $qq$ and $q \bar q$ processes. 
As the resonant structure in the $m_{jj}$ distribution coming from $Z \to q \bar q$ and 
$W \to q \bar q$ is 
different in REF and GEN processes, results of some bins feature unexpectedly large
statistical fluctuations.

Finally, let us stress that simple, but nonetheless, necessary check have been done as well: from the inspection of the control outputs 
we confirmed that the dominant contributions to cross sections are distinct 
for Drell-Yan and Higgs production processes, and that the slopes of energy 
spectra of $\tau$-decay products  are of a proper sign. 
That confirms that our installation is free of possible trivial 
errors in spin implementation. 

\subsection{Tests with hard process + parton shower events}\label{sec:PS}

\begin{figure}
  \begin{center}                               
{
   \includegraphics[width=7.5cm,angle=0]{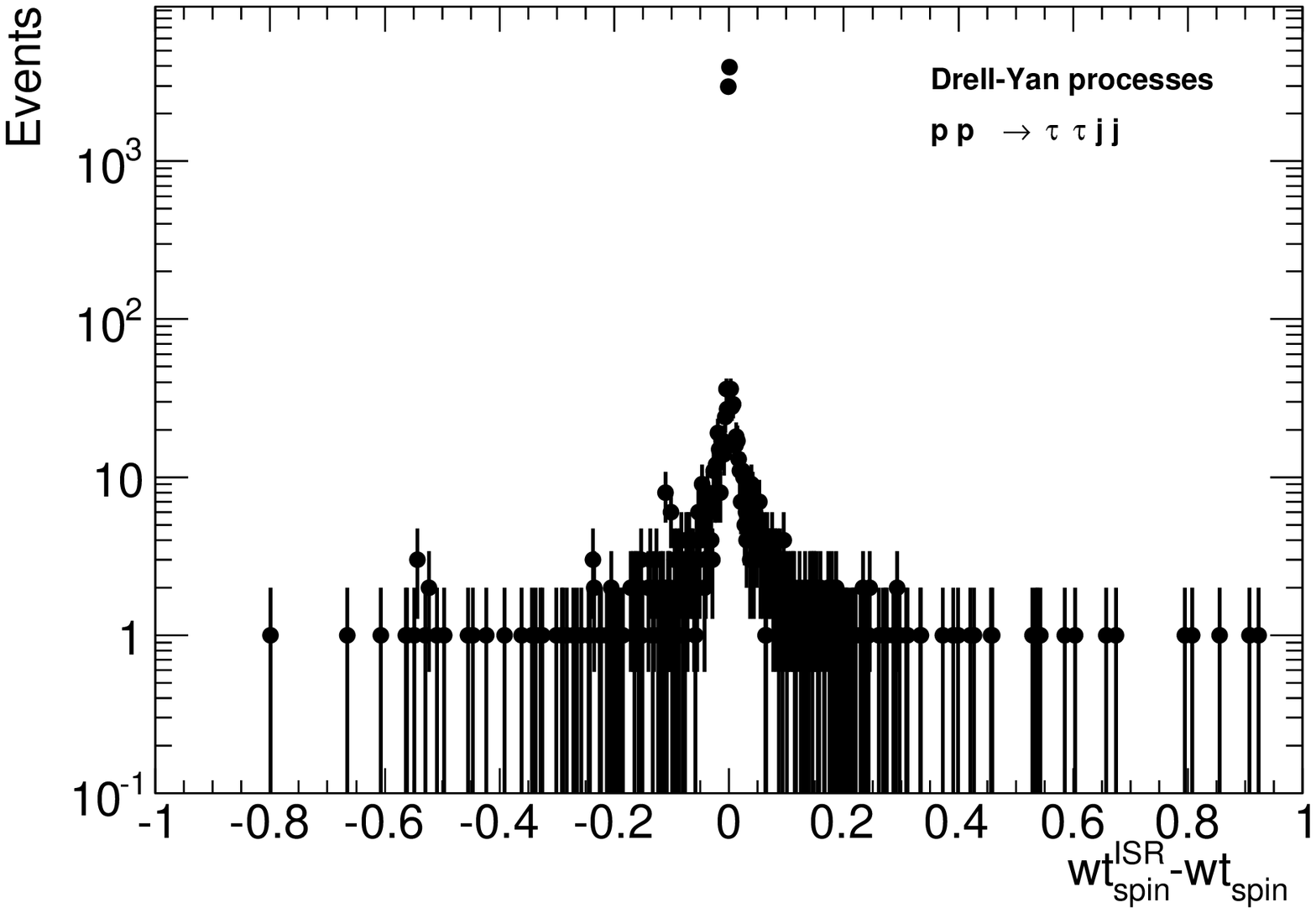}
   \includegraphics[width=7.5cm,angle=0]{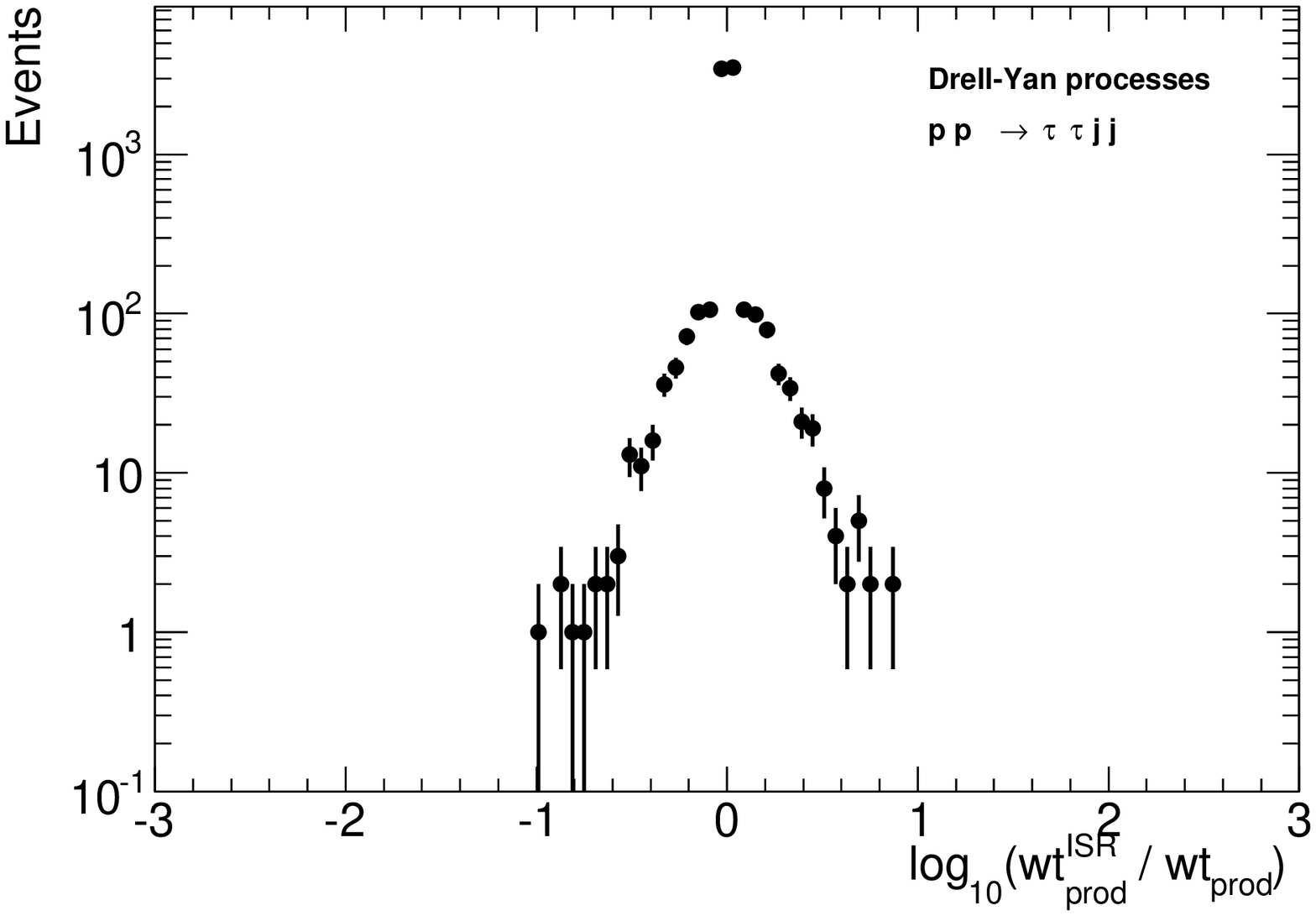}
}
\end{center}
\caption{ Impact on the matrix element calculation of parton shower smearing, as explained in the text.
On the left, the difference of spin weights calculated
with and without ISR parton shower kinematic smearing is shown.
On the right,  the  ratio of matrix element weights calculated for the two cases is shown.
 Sample of 10000 events was used.
\label{fig:WTShower} }
\end{figure}

After technical tests at the hard process level (convoluted with structure functions), we turn to check the algorithm on 
events where the incoming parton momenta can not be assumed to be along the 
beam direction due to the presence
of the parton shower in the initial state (ISR). 
 For that purpose, we have taken events generated with {\tt MadGraph5} and 
 added ISR with the default version of {\tt Pythia 8.2}
 (as described in Ref.~\cite{Sjostrand:2014zea}).  The two statistically correlated samples were constructed 
and  used
by {\tt TauSpinner} for calculation of spin weights ($wt_{spin}$) 
and production weights ($wt_{prod}$). 
Fig. \ref{fig:WTShower} shows the number of events as a function of differences for 
the spin weights calculated  for each event from configurations  with and without ISR parton shower. Similarly, 
shown is the  ratio of $wt_{prod}$ weights calculated for configurations with and without ISR parton shower. 
One can see from Fig. \ref{fig:WTShower} (left plot), that the spin weights for the cases with and without ISR 
are strongly  correlated. Majority of events reside in central bins of the distribution and the difference in weights
is smaller than the bin width.
Also the matrix element weights for the two cases are strongly correlated, 
see Fig.~\ref{fig:WTShower} (right plot). Majority of events reside in central bins. 
We can conclude that, 
 similarly as in the past \cite{Czyczula:2012ny} for the
($2\to 2$) process, the algorithm which is applied to kinematics of the hard process 
particles effectively removes impact of the initial state transverse momentum
 and 
leads to results which are stable with respect to the presence of extra showering.
 This test is of more physical nature, since in such a case Eq. (\ref{eq:main}) does not 
 hold for the distribution of reweighted events 
 and, as a consequence,  
reweighting with Eq.~(\ref{eq:WT}) is featuring an approximation, which we have validated with this test.
Note that adding ISR means only that the system of partons and $\tau$ leptons outgoing from the hard process  underwent (as a whole) a boost and rotation
before calculating matrix elements and PDF's. This justifies the evaluation of $x_1, x_2$, fraction of proton energies carried by the incoming partons in collinear approximation.

\subsection{Tests on the EW schemes and WWZ coupling }
\label{sec:WWZtest}

\begin{figure}
  \begin{center}                               
{
   \includegraphics[width=7.5cm,angle=0]{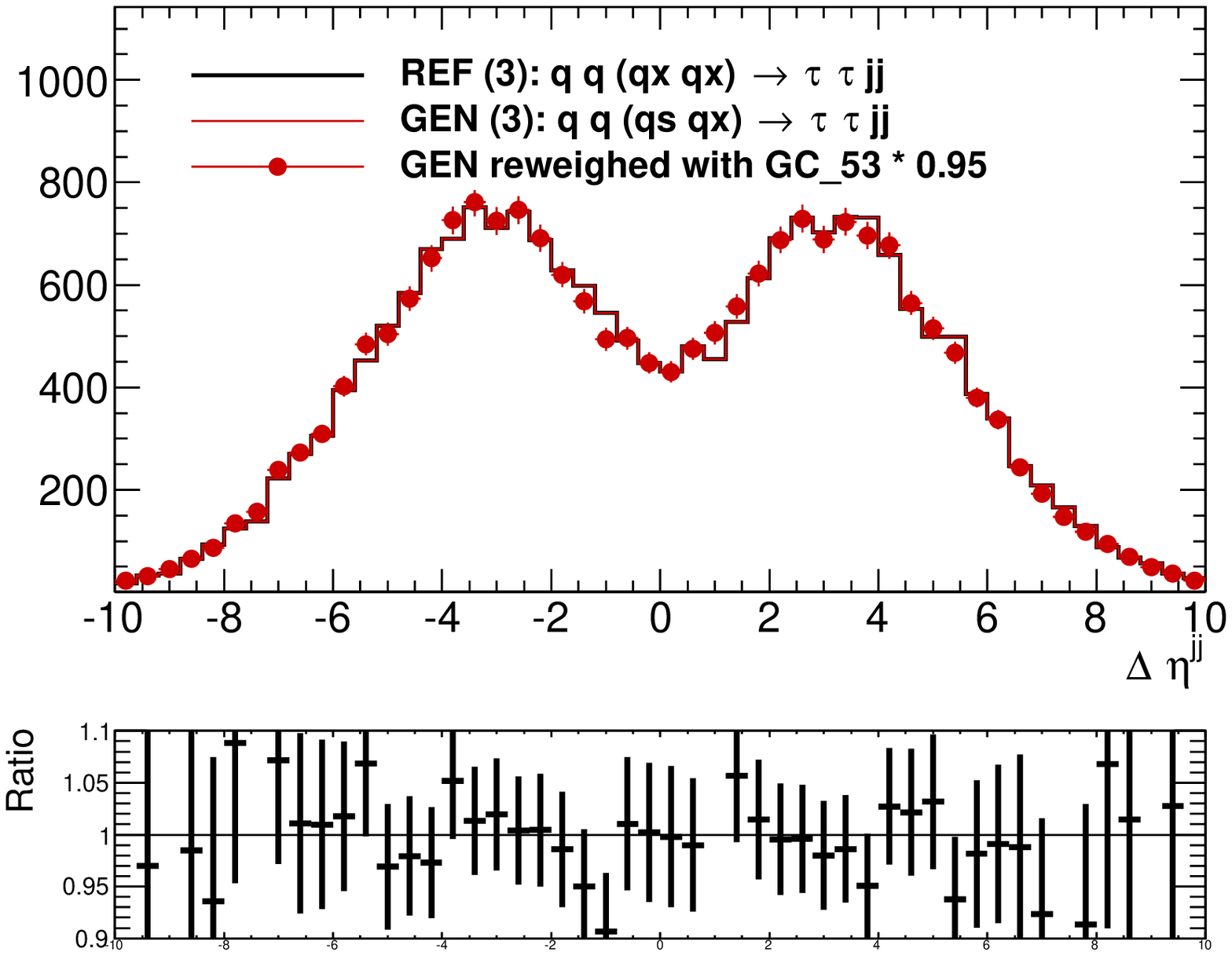}
   \includegraphics[width=7.5cm,angle=0]{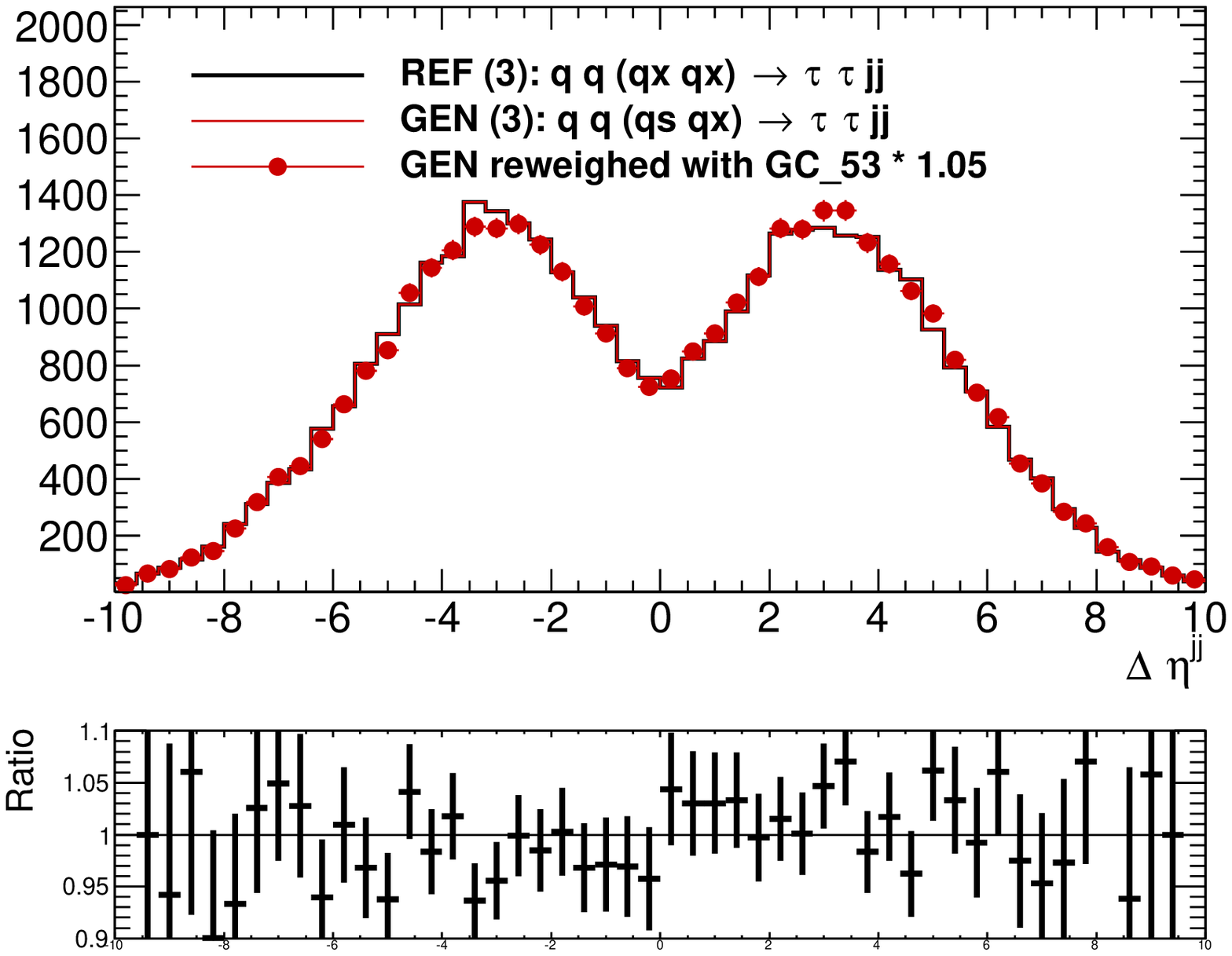}
}
{
   \includegraphics[width=7.5cm,angle=0]{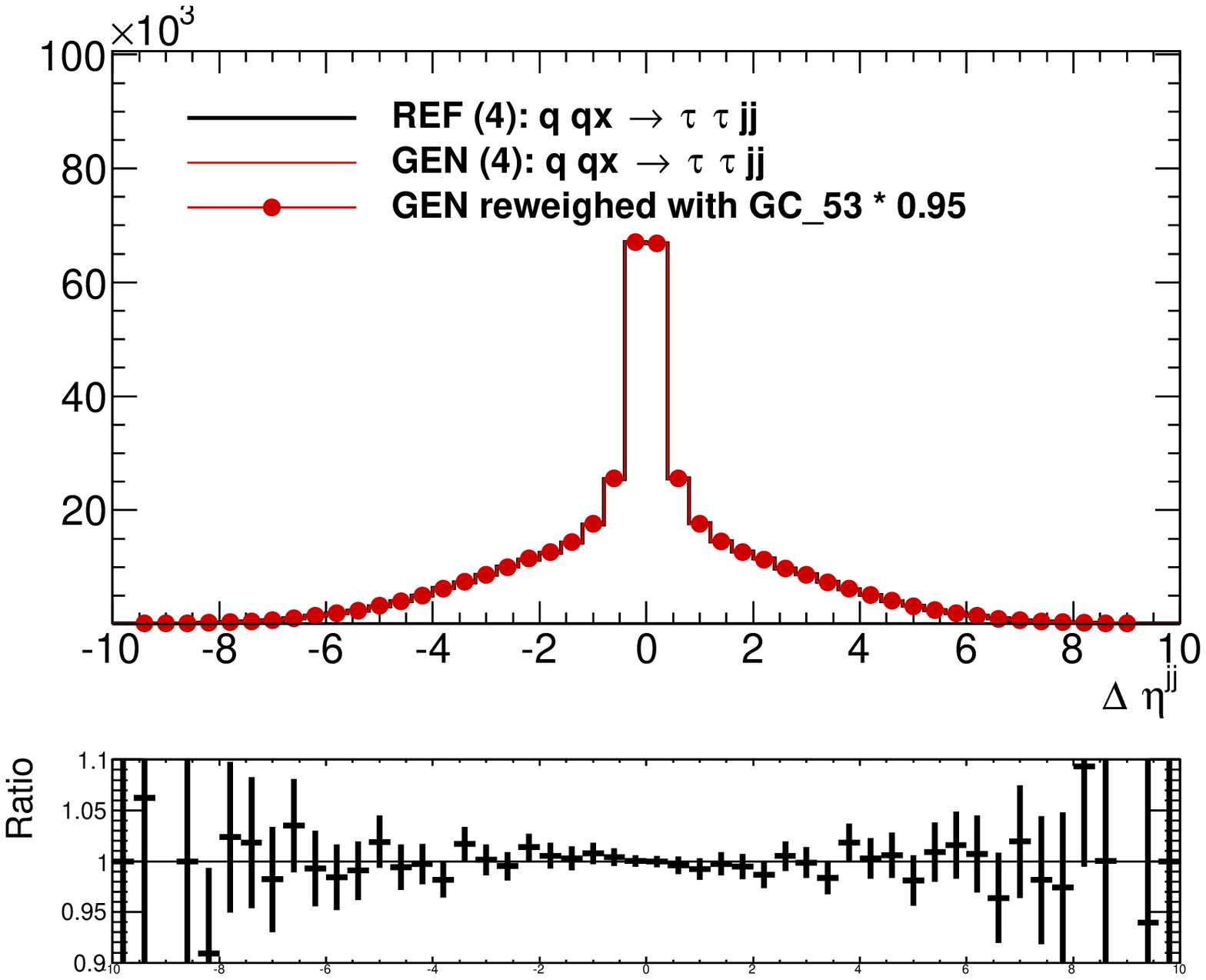}
   \includegraphics[width=7.5cm,angle=0]{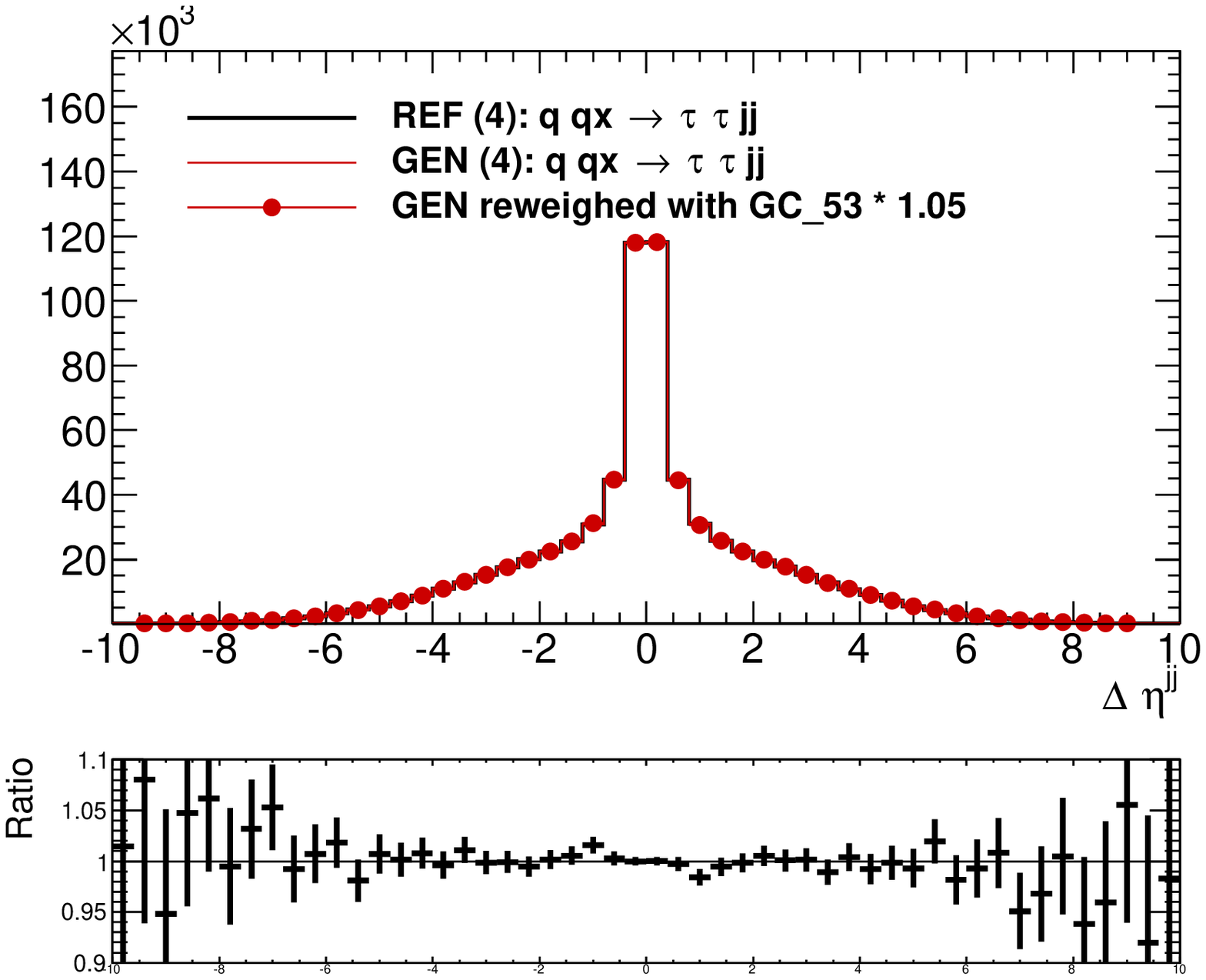}
}
\end{center}
\caption{Distribution of $\Delta jj$, reweighted to the one corresponding to $WWZ$ coupling
(internal {\tt MadGraph5} notation $GC\_53$) multiplied by 
factor 0.95 (right) and 1.05 (left), shown for $q\ q, \bar q\ \bar q$ (top) and $q\ \bar q$ (bottom) Drell-Yan processes.
\label{fig:EWModif} }
\end{figure}

For ($2 \to 2$) process, resummation of higher order effects into effective couplings is well established.
In  ($2 \to 4$) case, care is necessary, one may destroy gauge cancellations where matching of $Z$ emissions from
quark lines with the ones of the $t$-channel W must be preserved.
In Fig.~\ref{fig:EWModif} we demonstrate results, where effective $\sin^2 \theta_W$ is used in otherwise $G_F$ scheme, 
one can see that varying arbitrarily of $WWZ$ coupling by $\pm 0.05$ bring marginal effects only, even for the 
$q\ q, \bar q\ \bar q$ processes, chosen to maximize the relative effect of $WWZ$ coupling mismatch. 
The effect is negligible for the shown, most sensitive kinematical distribution studied. The estimate of the average polarisation 
remains unchanged. This is an expected result as for our amplitudes condition $\sin^2 \theta_W = 1 - M_W^2/M_Z^2$ 
is in principle not needed for gauge cancelation. 

\section{Numerical results}
\label{sec:MEpheno}

Once we have completed our technical tests, and gained confidence in 
the functioning of the $2 \to 4$ extension of {\tt TauSpinner} algorithms, let us turn to presentation of  
numerical results. 
In spite of a limited scope of the present version, like lack of the loop-induced gluon coupling to the Higgs boson, 
or subprocesses with b-quarks, {\tt TauSpinner} can already be used as a tool to obtain  numerical results of interest for 
phenomenology. Note that b-quarks as final jets can be tagged, and should thus be treated separately, 
while the contribution from the b-quark PDFs is rather small. Possible applications of {\tt TauSpinner} are presented below.


\subsection{Average $\tau$ lepton polarisation}

For calculation of weights earlier versions of {\tt TauSpinner}  used  the elementary  $(2 \to 2)$ parton level
$q  \bar q (gg) \to Z/\gamma/(H) \to \tau^- \tau^+$ amplitudes  
factorized out from the complex event processes. 
This approach can now be verified  with the explicitly implemented $(2 \to 4)$ matrix elements when two
hard jets are present in the calculation of the amplitudes. 
The physics of interest is the measurement of the Standard Model Higgs boson properties 
in decays to the $\tau$ leptons and its separation from the Drell-Yan background of $\tau$-pair production. 

We start by confirming the overall consistency of the calculations, comparing results from  $(2 \to 2)$ 
and  $(2 \to 4)$ calculations on inclusive $\tau \tau jj$ events, with $\tau$-pair around the Z-boson mass 
peak, but with very loose requirements on the accompanying jets, $p_T^{jet} > 1$ GeV.   
In Tables~\ref{tab:ew_gmu} and~\ref{tab:ew_gmu_above}, the estimated average polarisation is shown using  matrix elements for 
$(2 \to 2)$ and $(2 \to 4)$ for four categories of hard processes and for cuts selecting events at the $Z$ peak 
or above. For the $(2 \to 4)$ implementation
shown is also the difference when estimating  polarisation using an  average of all hard processes, 
or for only specific category.

\begin{table}
\caption{Comparison of the $\tau$-lepton polarisation in $\tau \tau jj$ events, calculated using {\tt TauSpinner}
weight $wt_{spin}$  of $(2 \to 2)$ and ($ 2 \to 4$) processes and $G_{F}$ EW schemes with $\sin^2 \theta_W$ = 0.22222. 
Required invariant mass of the $\tau$ pair of $m_{Z} \pm 10$~GeV and low threshold on outgoing partons
transverse momenta, $p_T > 1$~GeV. Rows of the Table correspond to different subsets of events generated with
{\tt MadGraph5}, selected accordingly to flavours of incoming partons. {\tt TauSpinner} algorithm is not using this 
information and the average of all possible configurations is used. In case of the last collumn, we restrict the average
to the ones actually used for the selected subset of events.} 
\label{tab:ew_gmu}
 \vspace{2mm}
  \begin{center}
  \begin{tabular}{|l|r|r|r|r|r|}
  \hline\hline 
 Process                                           & Fraction  &  Polarisation         & Polarisation             & Polarisation \\ 
                                                   & of events &   $(2 \to 2)$         & $(2 \to 4)$              &  $(2 \to 4)$ \\ 
                                                   &           &   Average             &  Average                 & Process specific \\ 
  \hline\hline
  All processes                                    &           & -0.2142 $\pm$ 0.0003  &  -0.2140 $\pm$ 0.0003    & -0.2135 $\pm$ 0.0003  \\
  \hline\hline
  $g\ g \to \tau\ \tau\ jj$                        &    3.1\%  & -0.2085 $\pm$ 0.0018  &  -0.2094 $\pm$ 0.0018    & -0.2122 $\pm$ 0.0018 \\
  \hline
  $g\ q\ , g\ \bar q\to \tau\ \tau\ jj$            &    59.3\% & -0.2132 $\pm$ 0.0004  &  -0.2133 $\pm$ 0.0004    & -0.2130 $\pm$ 0.0004 \\
  \hline
  $q\ q\ , \bar q \ \bar q\to \tau\ \tau\ jj$      &    1.8\%  & -0.2151 $\pm$ 0.0024  &  -0.2167 $\pm$ 0.0024    & -0.2146 $\pm$ 0.0024 \\
  \hline
  $q\ \bar q \to \tau\ \tau\ jj$                   &   35.7\%  & -0.2163 $\pm$ 0.0005  &  -0.2156 $\pm$ 0.0005    & -0.2140 $\pm$ 0.0005 \\
\hline
\end{tabular}
\end{center}
\caption{Comparison of the $\tau$-lepton polarisation in $\tau \tau jj$ events, calculated using {\tt TauSpinner}
weight $wt_{spin}$   of ($2 \to 2$) and ($ 2 \to 4$) processes and $G_{F}$ EW schemes with $\sin^2 \theta_W$ = 0.22222. 
Required invariant mass of the $\tau$ pair of $100-130$~GeV and low threshold on outgoing partons
transverse momenta, $p_T = 1$~GeV. The collumns are organised as in Table~\ref{tab:ew_gmu}.} 
\label{tab:ew_gmu_above}
 \vspace{2mm}
  \begin{center}
  \begin{tabular}{|l|r|r|r|r|r|}
  \hline\hline 
 Process                                           & Fraction  &  Polarisation         & Polarisation             & Polarisation \\ 
                                                   & of events &   $(2 \to 2)$   &   $(2 \to 4)$        &   $(2 \to 4)$ \\ 
                                                   &           &   Average             &  Average                 & Process specific \\ 
  \hline\hline
  All processes                                    &           & -0.4837 $\pm$ 0.0028  &  -0.4852 $\pm$ 0.0028    & -0.4864  $\pm$ 0.0028      \\
  \hline\hline
  $g\ g \to \tau\ \tau\ jj$                        &    2.6\%  & -0.4939 $\pm$ 0.0175  &  -0.5023 $\pm$ 0.0174    & -0.4864 $\pm$ 0.0176 \\
  \hline
  $g\ q\ , g\ \bar q\to \tau\ \tau\ jj$            &    56.0\% & -0.4815 $\pm$ 0.0038  &  -0.4838 $\pm$ 0.0038    & -0.4864 $\pm$ 0.0038 \\
  \hline
  $q\ q\ , \bar q \ \bar q\to \tau\ \tau\ jj$      &    1.7\%  & -0.4902 $\pm$ 0.0118  &  -0.4727 $\pm$ 0.0119    & -0.4770 $\pm$ 0.0119 \\
  \hline
  $q\ \bar q \to \tau\ \tau\ jj$                   &   39.8\%  & -0.4863 $\pm$ 0.0045  &  -0.4857 $\pm$ 0.0045    & -0.4857 $\pm$ 0.0045 \\
\hline
\end{tabular}
\end{center}
\end{table}

\begin{figure}
  \begin{center}                               
{
   \includegraphics[width=7.5cm,angle=0]{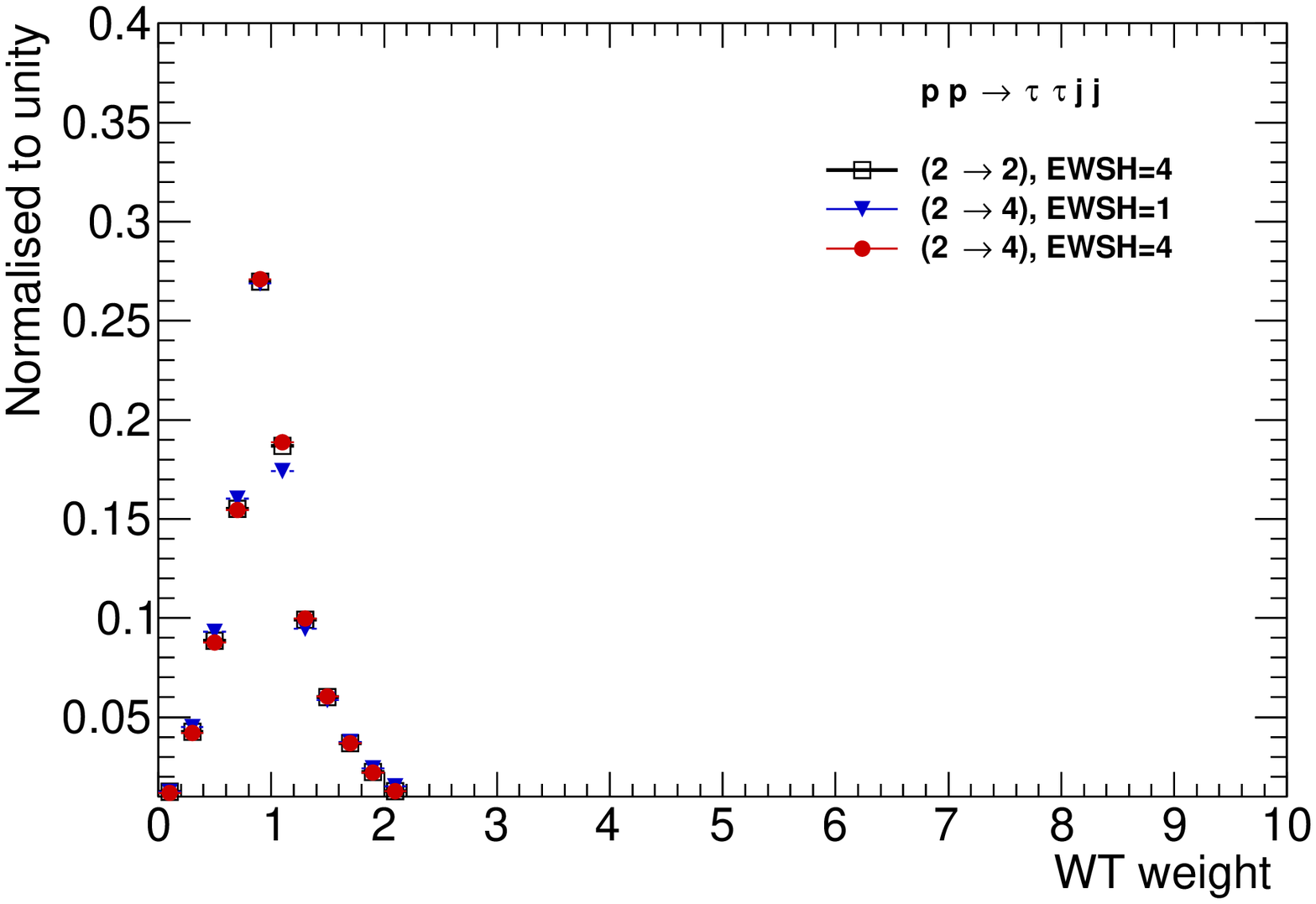}\\
   \includegraphics[width=7.5cm,angle=0]{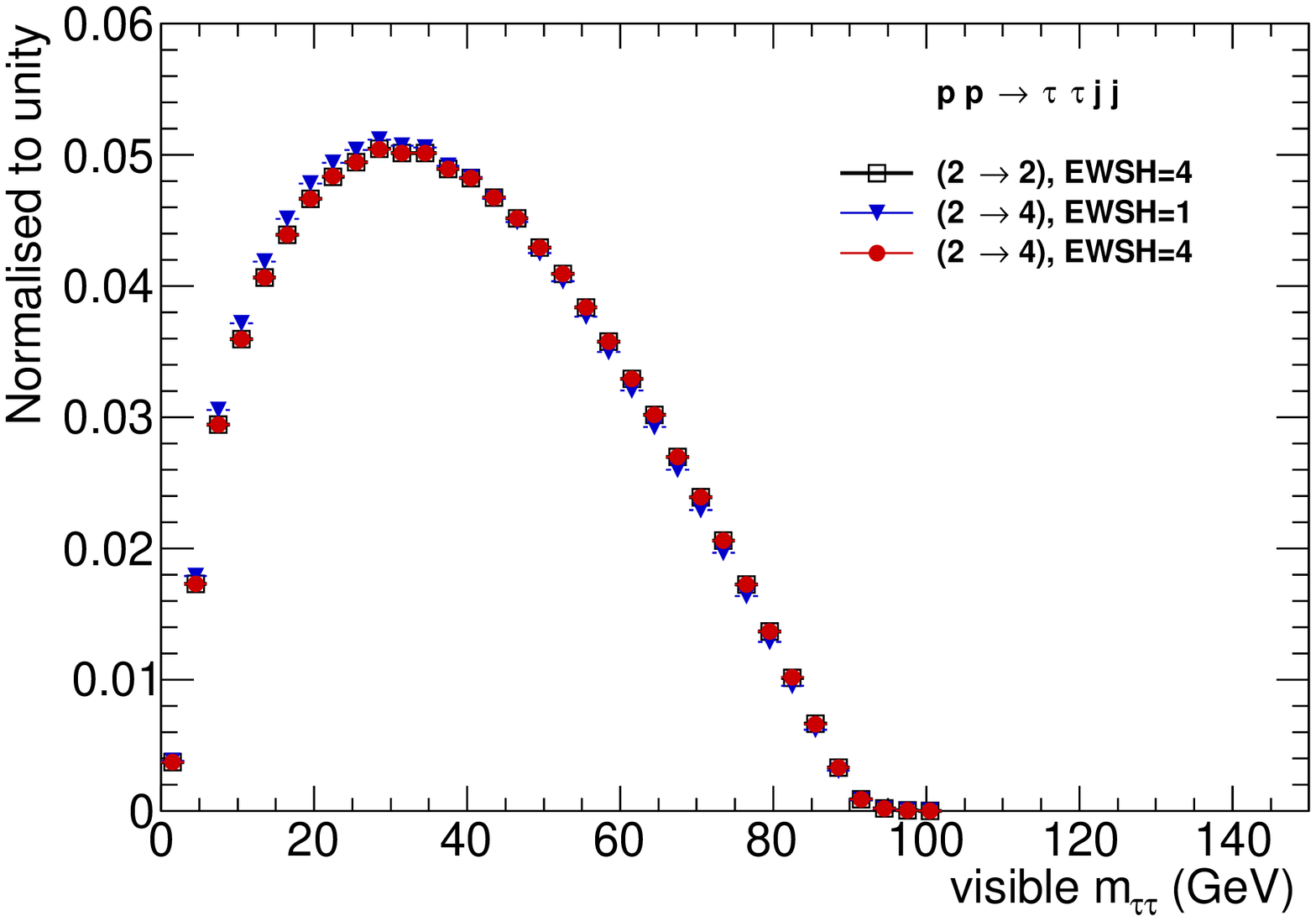}
   \includegraphics[width=7.5cm,angle=0]{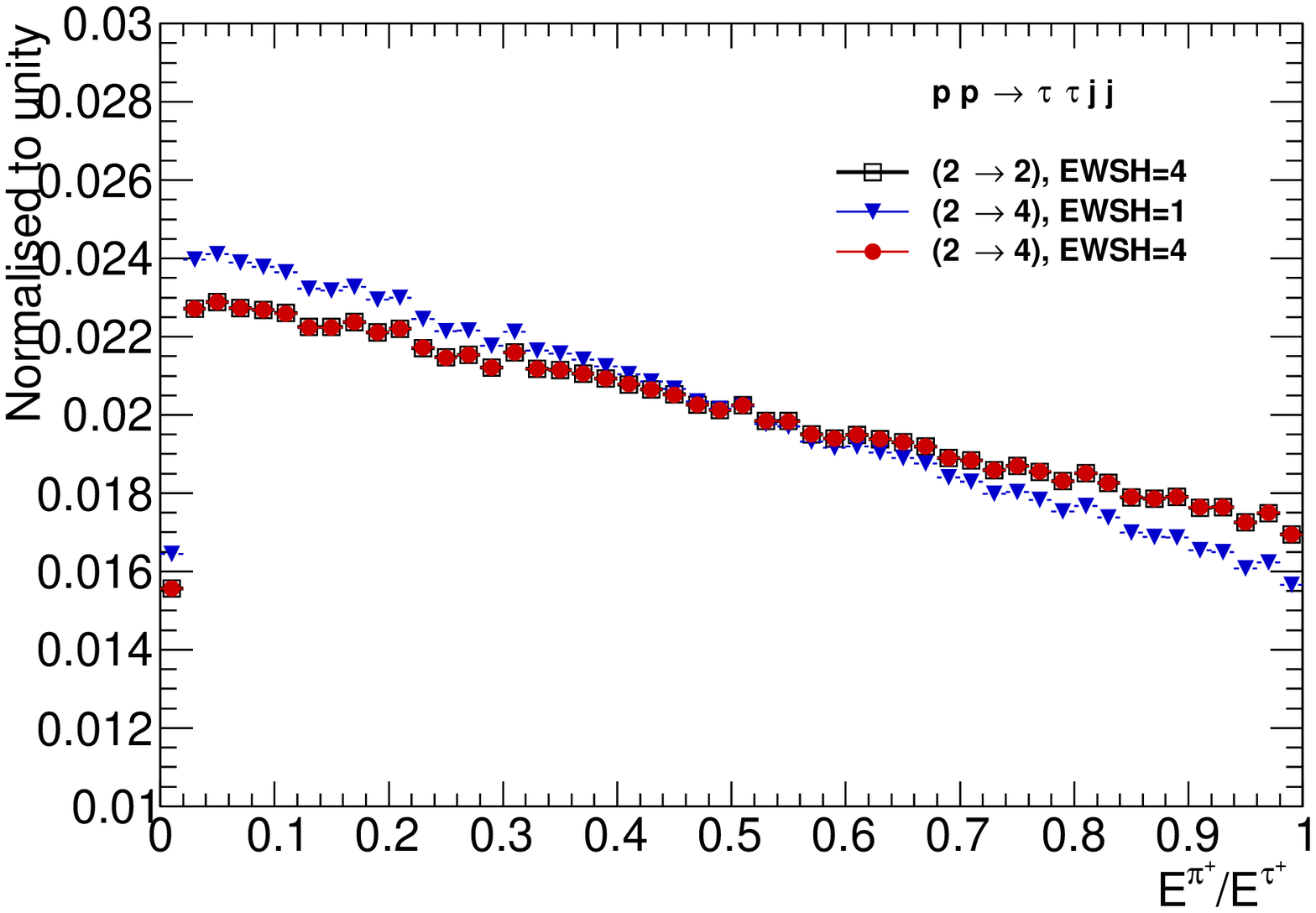}
}
\end{center}
\caption{ Distribution of the spin weight (top), of the invariant mass of visible decay products of $\tau$-pairs 
(bottom-left) and the energy fraction of the decaying $\tau$ lepton carried by 
$\pi^{\pm}$ in $\tau \to \pi^{\pm} \nu$ (bottom-right),
weighted with  ($2 \to 2$)  and  ($2 \to 4$) matrix elements  and for different EW schemes.
\label{fig:polar} }
\end{figure}

To verify that not only the calculation of spin averaged amplitudes, but the
contributions from specific helicity configurations are properly matched 
between ($2 \to 2$) and ($2 \to 4$), we have checked the  $E_\pi/E_\tau$ spectra
in the $\tau^\pm \to \pi^\pm \nu$ decays. This variable is sensitive to the polarisation
of the $\tau \tau$ system and longitudinal spin correlations.
To introduce spin effects to the sample, otherwise featuring non-polarized $\tau$ decays, 
we have used weights calculated by {\tt TauSpinner}. 
The spin weight distribution, the visible mass of $\tau$'s decay products combined 
and the energy fraction carried by the $\pi^{\pm}$ in $\tau \to \pi \nu$ decays
are compared for two different EW schemes in Fig.~\ref{fig:polar}. 

To emphasize possible differences between using the $(2 \to 2)$ or  $(2 \to 4)$  matrix elements for 
calculating spin weights for $\tau \tau j j$ events, we have applied simplified kinematic selection
inspired by the analysis of ~\cite{AtlasHtautau:2015}, called in the following {VBF-like} selection:
transverse  momenta of outgoing jets above 50 GeV; 
pseudorapidity gap between jets, $|\Delta \eta^{j j}| > 3.0$;
Transverse momenta of outgoing $\tau$ leptons of 35 GeV and  30 GeV, respectively 
and pseudorapidity $|\eta^{\tau}| < 2.5$.
It is also required that the invariant mass of the $\tau$-lepton pairs and jj pair is above the 
$Z$-boson peak. Results for the average polarization, are shown in Table \ref{tab:polar_vbf}. 

The { VBF-like} selection  enhances contributions from $qq, \bar q \bar q \to \tau \tau jj$ processes to about 25\% of the total
cross section. The highest discrepancy found  between the predicted $\tau$ lepton polarisation with
($2 \to 2$) and ($2 \to 4$) matrix element is at the level of 4\% in absolute value, being relative 10\% of the polarisation.
Using the average ($2 \to 4$) matrix element i.e.\ assuming that the initial state is not known, 
reduces the discrepancy by factor 2. For the polarisation averaged over all production processes the difference 
between  ($2 \to 2$) and ($2 \to 4$) matrix element is at the level of 1.0 - 1.5\% in absolute value,  which is
only 2\% relative effect. 

The above results indicate strongly that 
{\tt TauSpinner} in the  ($2 \to 2$) mode is sufficient for the evaluation of spin effects
observable in $\tau$ decays. The  ($2 \to 4$) mode is useful mainly for validations or systematic studies.

Please note, that results were obtained with the $G_F$-on-shell scheme {\tt EWSH=1}, thus are different from {\it physically} expected
values.
Let us continue now with the discussion of typical initializations used in calculations of matrix elements for ($2 \to 4$) processes.

\begin{table}
 \caption{Polarisation of the $\tau$-lepton in $\tau \tau jj$ events, calculated using {\tt TauSpinner}
weight $wt_{spin}$ and  ($2 \to 2$) and ($ 2 \to 4$) processes and {\tt EWSH=1} scheme with $\sin^2 \theta_W$ = 0.22222. 
For this comparison the initialisation of $(2 \to 2)$ process was also adopted to {\tt EWSH=1} scheme.
Required is the invariant mass of the $\tau$-pair and $jj$-pair above 120~GeV and VBF-like selection (see text).} 
\label{tab:polar_vbf}
 \vspace{2mm}
  \begin{center}
  \begin{tabular}{|l|r|r|r|r|r|}
  \hline\hline 
 Process                                           & Fraction   &  Polarisation          & Polarisation            & Polarisation \\ 
                                                   & of events  &  $(2 \to 2)$           &  $(2 \to 4)$            & $(2 \to 4)$ \\ 
                                                   &            &   Average              &  Average                & Process specific \\ 
 \hline \hline
  All processes                                    &            & -0.5026 $\pm$ 0.0019   & -0.5184  $\pm$ 0.0018   & -0.5110  $\pm$ 0.0018 \\
  \hline\hline
  $g\ g \to \tau\ \tau\ jj$                        &     3.2\%  & -0.5046 $\pm$ 0.0092   &  -0.5126 $\pm$ 0.0092   & -0.5027 $\pm$ 0.0092 \\
  \hline
  $g\ q\ , g\ \bar q\to \tau\ \tau\ jj$            &    54.3\%  & -0.5041 $\pm$ 0.0025   &  -0.5156 $\pm$ 0.0025   & -0.5013 $\pm$ 0.0025 \\
  \hline
  $q\ q\ , \bar q \ \bar q\to \tau\ \tau\ jj$      &    24.9\%  & -0.4989 $\pm$ 0.0037   &  -0.5253 $\pm$ 0.0037   & -0.5396 $\pm$ 0.0037 \\
  \hline
  $q\ \bar q \to \tau\ \tau\ jj$                   &    16.7\%  & -0.5026 $\pm$ 0.0045   &  -0.5188 $\pm$ 0.0045   & -0.5290 $\pm$ 0.0045 \\
\hline
\end{tabular}
\end{center}
\end{table}

\subsection{EW scheme dependence}

In initialization of programs like {\tt MadGraph5}, the tree level formula for weak mixing angle,
$\sin^2 \theta_W =  1 -M_W^2/M_Z^2 = 0.222246$, is often used following the {\tt EWSH=1} or {\tt EWSH=3} schemes described previously.
This theoretically motivated choice is quite distant from the  $\sin^2 \theta_W^{eff} =0.23147$ 
describing the ratio of vector to axial vector couplings of $Z$-boson to fermions and which is used in the  {\tt EWSH=2} scheme.
The LO approximation used in {\tt MadGraph5} initialization and in our tests so far,  can not be used for the program default initialization.
We are constrained by the  measured values of the $M_W$,  $M_Z$ and $\sin^2 \theta_W^{eff} $ and that is why the  {\tt EWSH=4} scheme is chosen as a default.

One must  keep in mind, that the $\tau$-lepton polarisation in Z-boson decays is very sensitive
to the scheme used for the electroweak sector. Table~\ref{tab:ew} gives numbers for the average polarisation
in the case of $G_{F}$ and effective EW schemes. Results for using $(2 \to 2)$ and $(2 \to 4)$ matrix elements 
coincide within statistical error, for event sample with rather loose kinematical cuts.
On the contrary, results of calculations strictly following the $G_F$ scheme are off by {\it 50 \%} with respect to 
experimentally measured value, -0.1415 $\pm$ 0.0059, see Table \ref{tab:ew}. This  must be taken into account if results are compared
with the data, as it was done in the LEP times \cite{ALEPH:TauPol}.

\begin{table}
 \caption{Polarisation of the $\tau$-lepton in $\tau \tau jj$ events, calculated using {\tt TauSpinner}
weight $wt_{spin}$  of $(2 \to 2)$ and $(2 \to 4)$ processes and different EW schemes.
Required is the invariant mass of the $\tau$ pair of $m_{Z} \pm 10$~GeV and 
low threshold on gluon transverse momenta of $p_T > 1$~GeV. }
\label{tab:ew}
 \vspace{2mm}
  \begin{center}
  \begin{tabular}{|l|c|r|r|}
  \hline\hline 
 EW parameter                & EW scheme  & Polarisation           & Polarisation \\ 
 (sensitive)                 &            &  $(2 \to 2)$         &  $(2 \to 4)$ \\ 
  \hline\hline
$\sin^2 \theta_W$ = 0.222246  & {\tt EWSH=1}   & -0.2140 $\pm$ 0.0004   & -0.2134 $\pm$ 0.0004  \\
\hline
$\sin^2 \theta_W$ = 0.231470  & {\tt EWSH=2}   & -0.1488 $\pm$ 0.0008   & -0.1487 $\pm$ 0.0008  \\
\hline
$\sin^2 \theta_W$ = 0.222246  & {\tt EWSH=3}   & -0.2140 $\pm$ 0.0008   & -0.2144 $\pm$ 0.0008  \\
\hline
$\sin^2 \theta_W$ = 0.231470  & {\tt EWSH=4}   & -0.1488 $\pm$ 0.0008   & -0.1486 $\pm$ 0.0008  \\
\hline
\end{tabular}
\end{center}
\end{table}






\section{Summary and outlook} \label{sec:summary}

In this paper, new developments of the 
{\tt TauSpinner} program for calculation of spin and matrix-element weights for the previously generated events 
have been presented. The extension of the program enables the calculation
of spin and matrix-element weights with the help  of ($2\to 4$) amplitudes
convoluted with parton distribution functions. Required only are kinematical configurations of the outgoing $\tau$ leptons, 
their decay products and two accompanying jets.

The comparisons of results of the new version of {\tt TauSpinner},  where 
matrix elements feature additional jets, and the previous one where  the Born-level ($2 \to 2$) matrix element is used, 
offer the possibility to evaluate systematic errors due to the neglect of transverse 
momentum of jets in calculating spin weights. We have found that for observables 
sensitive to spin, the bias was not exceeding 0.01 for sufficiently inclusive observables with tagged jets.
 
Numerical tests and technical details on how the new option of the program can 
be used were discussed. Special emphasis was put on spin effects sensitive to variants for 
SM electroweak schemes used in the generation of samples and available in initialization 
of {\tt TauSpinner}. The effect of using different electroweak schemes can be as big as 50\% of the spin effect 
and can be even larger for angular distribution of outgoing $\tau$ leptons.
For the configurations of final states with a pair of jets close to the $W$ mass the effect can be also high, 
up to 40\%. For applications of {\tt TauSpinner}, we recommend the 
effective scheme leading to results on $\tau$ polarisation, $m_Z$ and $m_W$  close to measurements.

The phenomena of $\tau$ decay and production are separated by the $\tau$ lifetime. This simplifying feature
is used in organizing the programs. As a consequence 
for  the generated Monte Carlo sample, different variant of the electroweak initialization
may be  used for generation of $\tau$ lepton momenta and later, for implementation of spin effects in the $\tau$'s decays.
Such a flexibility of the code may be a desirable feature: {\tt TauSpinner} weight calculation can be also adjusted 
to situation when the matrix element weight and spin weight have to be calculated with distinct initializations. 

Numerical results in the paper  were obtained with the help of weights. Not only spin weight, but also the production
weight has been used to effectively replace the  matrix element of the generation. 
This feature, introduced and explained in Ref.~\cite{Banerjee:2012ez}, was targeting an implementation of anomalous contributions. 
However, its use can easily be  adopted for studies of the electroweak sector initialization.  
This helps to get results quicker thanks to  correlated sample method. 
It provides technical advantage for the future, namely the possibility for the use of externally  provided matrix elements 
or initialisation of EW schemes. 

In tests discussed in this  paper we have used {\tt MadGraph5} generated events for the 
$pp  \to \tau^+ \tau^- jj$ process. The incoming partons were distributed according to PDFs, 
but in most cases neither the transverse momentum of incoming state nor additional initial state jets
were allowed. We will return to this point in the future, with greater attention.
We may also be able to extend, with the help of the program developed for the present paper, 
the work on factorization of the effective Born of ($2 \to 2 $) configuration.
This, in turn, will help to check the factorization of additional $p_T$ activity of our 
{\it parton parton $\to \tau^+ \tau^-$ j j } hard processes from the $pp$ collision. 
The special case of processes with a single hard jet in the final state and its corresponding matrix elements will 
be also useful for such tests and we plan to return to  such topic in the near future.

The program is now ready for studies with matrix elements featuring extensions 
of SM amplitudes. Systematic errors have been discussed. Let 
us stress, that results of such studies depend on the definition of observable  
and need to be repeated whenever new observables or selection cuts are introduced. 
In the evaluation of impact of new physics or variation of SM parameters 
on experimentally accessible distribution it is necessary to compare results 
of calculations which differ by such changes. The Monte Carlo simulations 
are used, whenever detector acceptance and other  effects are to be taken 
into account.

\vskip 1 cm
\centerline{\bf \Large Acknowledgments}
\vskip 0.5 cm

We thank Tomasz Przedzi\'nski for work on the early versions of  {\tt TauSpinner} ($2\to4$) implementation
and help with documentation of technical aspects. 

This project was supported in part from funds of Polish National Science
Centre under decisions UMO-2014/15/ST2/00049 and by PLGrid Infrastructure of the Academic 
Computer Centre CYFRONET AGH in Krakow, Poland, where  majority of numerical calculations were performed.
JK, ERW and ZW were supported in part by the Research Executive  Agency (REA) of the European Union under the Grant 
Agreement PITNGA2012316704 (HiggsTools). WK was supported in part by the German DFG grant STO 876/4-1.

\providecommand{\href}[2]{#2}
\clearpage
\appendix
\section{Comments on the code organization and how to use it.} 
\label{app:HowToUse}


In this Section, we collect information on how to use the  ($2 \to 4$) option of the {\tt TauSpinner} program. 
We will concentrate on aspects, which are 
important to demonstrate the general scheme and organization of the new functionality of the code.
We assume that the reader is already familiar with previous versions of {\tt TauSpinner} or   
Refs.~\cite{Kaczmarska:2014eoa, Przedzinski:2014pla}. 

\subsection{Technical implementation} 
\label{sec:MEtechnical}

The general strategy of the reweighting technique of {\tt TauSpinner} for the case of configurations with 
$\tau \tau j j$ final states does not differ much from the  previous one in which 
only four-momenta of outgoing $\tau$ leptons and their decay products have been used. Nevertheless 
a few extensions with respect to  
 Refs.~\cite{Kaczmarska:2014eoa, Przedzinski:2014pla} have been introduced as explained below. 

\begin{enumerate}
\item 
For calculation of $\tau$ polarimetric vectors from their decay products  
and for the definition of boost routines from $\tau$-lepton's rest frames to the laboratory frame the same 
algorithms as explained in Refs. \cite{Kaczmarska:2014eoa, Przedzinski:2014pla} are used.
\item Before evaluation of production matrix elements and numerical values of PDF functions, one  has to 
reconstruct the four momenta of the incoming partons. For that purpose 
the following assumptions are made:
  \begin{enumerate}
   \item  For calculation of the hard process  virtuality $\bar Q^2$ and $p_z$, the 
          four-momenta of $\tau$ leptons and jets are summed to  a four-momentum vector $\bar Q^\mu$.
   \item The $\bar Q^\mu$ determined from experimental data, or from events generated by another Monte Carlo program, may have sizable transverse 
   momentum which has to be taken into account when 
         the directions $\hat e^{j}_z$ of the two incoming partons $j=1,2$ are constructed. To this end,
         versors of the beam directions in the laboratory frame are  boosted to the rest 
         frame of $\bar Q^\mu$. 
         The time-like components of boosted versors are dropped and the remaining space-like 
         part is normalized to unity. Note that  $\hat e^{1,2}_z$  obtained in this way  do not 
         need to remain back-to-back.  
\item Four-momenta of $\tau$ leptons and of 
         accompanying jets/partons are forced to be on mass-shell to eliminate all possible 
         effects of the rounding errors. This is necessary, to assure 
         the numerical stability of spin amplitude calculations.
\item The four-momenta for  the incoming partons are constructed 
using\footnote{We take
 weighted average of the two $\hat e^{1,2}_z$ directions, see in the code of {\tt vbfdistr.cxx} method {\tt getME2VBF()}
definition of {\tt P[6][4]}.} 
the direction of the versors $\hat e^{1,2}_z$ and enforcing four-momentum 
conservation.  
  \end{enumerate}
 This exhausts list of steps and changes to the  components for production and $\tau$ decay matrix elements
of the {\tt TauSpinner} in ($2 \to 4$) mode with respect to ($2 \to 2$) one. Only in steps (a) and (b) there are differences with respect 
to the original ($2 \to 2$) case.
\item The new source code for the matrix elements library
and interfaces is stored in {\tt TAUOLA/TauSpinner/src/VBF}
\item 
An exemplary code {\tt example-VBF.cxx} showing how to use {\tt TauSpinner} with ($2 \to 4$) matrix elements 
can be found in directory {\tt TAUOLA/TauSpinner/examples/example-VBF}. 
The extract of this code is given in Subsection~\ref{sect:MainProgram}.
In the same directory the 
code {\tt read\_particles\_for\_VBF.cxx,  read\_particles\_for\_VBF.h}   for reading the events from 
the file in the  {\tt HepMC} format,   as well as
a file {\tt events-VBF.dat} with a sample of 100 events,  are also stored.
Some further technical details can be found in the {\tt README} file of that directory.
\item
At the initialization step, basic information on the input sample like the center of mass energy for 
the $pp$ collisions or the set of parton density functions, PDF's, should be configured.
This part of the configuration has not changed since previous version of {\tt TauSpinneR}.
See Section~\ref{sect:Initialisation} and~\ref{sect:MainProgram} 
\item
The  {\tt example-VBF.cxx} provides also a prototype for implementation of the user code to
replace the default $(2 \to 4)$ matrix element of {\tt TauSpinner}.
\item
The spin weight {\tt WT}  (denoted in this paper as $wt_{spin}$) is calculated using ($2 \to 4$) matrix element by invoking method \\
{\tt WT = calculateWeightFromParticlesVBF(p3,p4,X,tau1,tau2,tau1\_daughters,tau2\_daughters);}\\
Note that  only  final state four-vectors of $\tau$'s, their decay products and outgoing jets,
are passed  to calculate  the weight.  
\item  
The method {\tt getME2VBF(p3,p4,X,tau1,tau2, W,KEY)} returns a double-precision table
{\tt  W[2][2]}, which contains partonic-level 
cross sections  for $\tau^+\tau^-$ helicity states $(1,-1)$, $(-1,1)$, $(-1,-1)$, $(1,1)$, respectively. 
They are obtained by summing matrix element squared over all parton flavour configurations and convoluted
 with the corresponding PDFs. Direct use of this method is optional. 
It is invoked internally by {\tt TauSpinner} though.
\item 
Several scenarios (models) of the hard process for calculating corresponding spin weight (note that  at the same 
time the weight for the production matrix elements is calculated), are possible.
At the initialisation step the choice is made and 
the technical internal parameter {\tt  KEY} is set. 
We give below  some  details: 
\begin{itemize}
\item
{\tt  KEY=1}  for the Standard Model Higgs process, matrix elements explained in Section~\ref{sub:incorpo} is used.
\item {\tt KEY=0} for 
non-Higgs Drell-Yan--like processes, matrix elements explained in Section~\ref{sub:incorpo} is used.
\item 
{\tt KEY}$>1$  is reserved for non-standard calculations, that is when {\tt nonSM=true} 
option is used\footnote{See later Appendix~\ref{sect:Initialisation}, the first and the second bullets.}, and 
 the  matrix element calculations  are modified  with the routines provided by the user.
Provisions with 
 {\tt  KEY=3} have been prepared  for the non-standard Higgs-like production process 
 and {\tt  KEY=2} for the Drell-Yan--like.
The required choice  is made  implicitly, at the initialisation step when setting the pointer to the 
user provided  function {\tt vbfdistrModif} and selecting initialization variable {\tt nonSM2=1}
which will set internal global  variable {\tt nonSM=true}.
The   result of default calculations will be passed to 
{\tt vbfdistrModif} function to be overwritten with user-driven modifications to the matrix elements,
 without the need of re-coding and recompiling standard {\tt TauSpinner} library.
\item
The internal parameter {\tt KEY} in general does not need explanation. However, 
as it is passed latter to the methods for calculating $\alpha_s$ or matrix elements, which may be replaced from  
user main program by re-setting the pointer, documentation was necessary. 
\end{itemize}
\item 
The method {\tt double getTauSpin()} returns helicities attributed to $\tau$ leptons on the statistical basis.
It is the same method as already implemented for the ($2 \to 2$) case. 

\item  The value which is returned by the method  {\tt double getWtNonSM()} depends on the configuration
of two flags: {\tt nonSM} and  {\tt  relWTnonSM}.
\begin{itemize}
\item
For {\tt  relWTnonSM=true}: in case of {\tt nonSM=true} the method returns the weight obtained from Eq.~(\ref{eq:WT}),
for {\tt nonSM=false} the value of 1 is returned. 
\item
For {\tt  relWTnonSM=false}:  in case of {\tt nonSM=true} the method returns the numerator of  Eq.~(\ref{eq:WT}),
for {\tt nonSM=false} the value of Eq.~(\ref{eq:WT}) denominator is returned.
\end{itemize}
The above discussed weight features matrix elements squared and summed over spin degrees of freedom.
It has similar functionality as already implemented for the $(2 \to 2)$ case. 
It is supposed to supplement the spin weight {\tt WT} of {\tt TauSpinner}.  In general, spin weight
differs for the {\tt SM} and {\tt nonSM}  calculation.  The ratio
of these  two has to be used for modifying decay product kinematic distributions.
Finally let us point out, that also the helicity of $\tau$'s will be attributed 
at the {\tt nonSM} step of the calculation, corresponding to the chosen {\tt nonSM} model.
\end{enumerate}

Let us bring some further points on the details of the use of the example.

\begin{itemize}
\item
This example has been prepared to read events in {\tt HepMC} format.
An additional tool {\tt lhe-to-hepmc.exe} convertin {\tt LHE} event to {\tt HepMC}
has been provided as well.
\item
To read events form the data file, the  method  {\tt read\_particles\_for\_VBF} stored in file \\
{\tt TAUOLA/TauSpinner/examples/example-VBF/read\_particles\_for\_VBF.cxx } is used. It is invoked as follows: 
{\tt int status = \\
read\_particles\_for\_VBF(input\_file,p1,p2,X,p3,p4,tau1,tau2,tau1\_daughters,tau2\_daughters);}\\
which reads consecutive  event and retrieves the following information:
four-momenta of incoming and outgoing partons, denoted as  {\tt p1,p2}, and {\tt p3,p4}, respectively;
four-momenta of outgoing $\tau$ leptons, {\tt tau1, tau2} and \\ {\tt tau1\_daughters,tau2\_daughters}
which stand for lists of decay products (their four-momenta and {\tt PDG-id's}) and, 
if available, also the four-momentum of an intermediate resonance $X$ and its {\tt PDG-id}.
It returns  {\tt status=1} if no event to read was found (this is specific to the method 
chosen for reading the events and is used in the user program only).

 Let us stress, that the above interface to read event record is {\it not} a part of {\tt TauSpinner} library. It is used 
in the  demonstration 
program and is adopted to the particular conditions. It is expected to be replaced by the user
 with the  customized one. 
Implementation of such a method must match conventions for the format and 
information on  different particles in the stored event.  
For example, for  formats (eg. {\tt HepMC}~\cite{Dobbs:2001ck} or {\tt lhe}~\cite{Alwall:2006yp}) 
distinct conventions are used in Monte Carlo generators:
i.e. for relations among  particles and intermediate states (resonances) 
which may be  explicitly written into event record or omitted. The same is true for  
production/decay vertices, status codes, etc. 
With a variety of conventions used, it can be highly error-prone and lead to necessity 
of non trivial implementations in the code,  like can be seen  in {\tt read\_particles\_for\_VBF} 
method provided in the distribution tar-ball.
See eg.~\cite{Davidson:2010rw,Davidson:2010ew} for discussion of similar difficulties in other projects. 
In fact, 
for {\tt TauSpinner} algorithms this is less an issue, as  the information on intermediate and 
incoming states is not used,
even though it can be very useful for testing purposes.

The information required by {\tt TauSpinner} algorithms for $(2 \to 4)$ processes 
which must be read in,  is limited to: 
four-momenta of outgoing jets and of $\tau^\pm$ including all their decay products and  
 $PDG$ identifiers 
of $\tau^\pm$ and their decay products. 
The four momenta of incoming partons and the intermediate resonance are not needed but 
can be used for tests.
The imminent next step is to exploit also, if available, information on the  
four-momentum of the intermediate
$Z/\gamma/H$ state (or any other non-standard resonance) which decays to $\tau^\pm$ lepton pair 
as well, as they can be used to tackle the effect of 
the QED bremsstrahlung in its decay, similar as it was done for the {\tt TauSpinner} algorithms in $(2 \to 2)$ case. 
\end{itemize}

\subsection{Initialisation methods}
\label{sect:Initialisation}

\begin{itemize}
\item {\bf Matrix elements:} \\
The {\tt TauSpinner} library includes codes for calculation of matrix elements squared for all 
(first two families) parton level cross sections of $(2 \to 4)$ processes. Use of this library
functions can be over-loaded, with the user's own matrix elements implementation 
by providing respective function 
{\tt vbfdistrModif(...)}. Its usage is activated at initialization with command
{\tt TauSpinner::set\_vbfdistrModif(vbfdistrModif)}, which sets the pointer to {\tt vbfdistrModif(...)}. 
A skeleton function,  {\tt vbfdistrModif(...)}, for user provided calculation of matrix elements squared 
is included in \\
{\tt /TAUOLA/TauSpinner/examples/example-VBF/example-VBF.cxx}.

The invocation inside {\tt TauSpinner} library  of the  function \\
{\tt vbfdistrModif(int I1,int I2,int I3,int I4,int H1,int H2,double P[6][4],int KEY,double result)}\\
includes among its arguments the  {\tt result} of the default {\tt TauSpinner} Standard Model $(2 \to 4)$ 
calculation. \\
The {\tt vbfdistrModif(...)} of the demonstration  program returns directly 
the {\tt result}\footnote{This enables possibility to obtain weights for different setting of the 
electroweak initialization, but calculated otherwise with the same matrix elements.}. \\
The following  arguments are passed to this function, and this must be obeyed in its declaration:
\begin{itemize}
\item
The first four arguments {\tt I1,I2,I3,I4} denote PDG   identifiers \cite{PDG:1998} 
of incoming and outgoing partons (for gluon {\tt ID=21}) .
\item
 The following two
{\tt H1,H2= $\pm$ 1} denote helicities of outgoing $\tau^+$ and   $\tau^-$.
\item
Matrix {\tt P[6][4]} encapsulates four-momenta of all incoming/outgoing partons and $\tau^\pm$ leptons.
They are for massless partons and for massive $\tau$ leptons. Energy momentum 
conservation is required at the double precision level. The partons are not expected to 
be in the phase-space regions close to the 
collinear/soft boundaries.
\item
The parameter {\tt KEY=0} is reserved for the SM  default
processes of Drell-Yan--type (all Feynman diagrams included, but the ones with 
$H\to\tau^+\tau^-$), while {\tt KEY=1}  for SM processes with the  Higgs production 
and its decay to $\tau$-lepton pair. In these two cases {\tt vbfdistrModif} is
not activated. For {\tt KEY=2,3} the SM calculation (again respectively for Drell-Yan and Higgs processes) is performed 
first and the result     
is passed into {\tt vbfdistrModif()} where it can be just modified, before being used 
for final weight calculations. The user may choose to modify the value of the default calculations for all
or only for subset of processes involved. This is why complete information of the initial and final state 
configurations is exposed.
If a completely new calculation is to be performed using the above method, then 
it is advised to use options {\tt KEY=4,5}, 
so that the Standard Model calculation will be avoided (to save CPU) and {\tt result=0} 
will be passed to {\tt vbfdistrModif()}. The {\tt KEY=4,5} is reserved for optional use of {\tt vbfdistrModif()}.
\end{itemize}
\item {\bf Electroweak schemes:}\\
In Table \ref{tab:EWschemes}  options of initialization for the EW schemes 
implemented in {\tt TauSpinner} are explained. The particular choice can be made with  
{\tt vbfinit\_(\&ref,\&variant)} as follows:
\begin{verbatim}
    int EWSH_ref=4;      // EW scheme to be used for the default 2 -> 4 calculation.
    int EWSH_variant =5; // EW scheme to be used for non-standard  2 -> 4 calculation.
    vbfinit_(&ref,&variant);
\end{verbatim}
The {\tt EWSH\_ref} will set initialization as used for the default
calculation, and {\tt EWSH\_variant} for the reweighting with modified amplitudes.
The  choices 1, 2, 3, 4 correspond to {\tt EWSH=1, EWSH=2, EWSH=3, EWSH=4} respectively. 
The default  {\tt EWSH=4}, as explained in the main text,  leads to correct $\tau$ lepton polarisations and angular distributions.  
As it causes at  tree-level inconsistencies
in the calculation of the $WWZ$ coupling, we provide an additional option, {\tt EWSH=5},
for which parameter setting as for {\tt EWSH=4} is used, but with the $WWZ$ coupling 
modified by 5\%.   It can be 
used for testing sensitivity of the analysed distributions to the missed higher order corrections
to the $WWZ$ coupling. For more discussion, see Section~\ref{sec:EWdetails}. 
\item {\bf PDFs and $\alpha_s$:} \\

Any PDF set from \texttt{LHAPDF5} library~\cite{Whalley:2005nh} can be used for calculating spin weight.
The choice can be configured by setting,  \\
\texttt{string name="cteq6ll.LHpdf";}\\
\texttt{LHAPDF::initPDFSetByName(name);}

The choice of renormalization and factorization scales (imposed is case of $\mu_F=\mu_R$) can be set with the help of the following command:

\begin{verbatim}
    int QCDdefault=1; // QCD scheme to be used for default 2 ->4 calculation.
    int QCDvariant=1; // QCD scheme to be used in optional matrix element reweighting (nonSM2=1).
    setPDFOpt(QCDdefault,QCDvariant);
\end{verbatim}
The choice can be different for the default (SM) calculation and the variant one ({\tt nonSM=true}), see Appendix~\ref{sec:MEtechnical}, point 9.
The $Q^2$ evolution and starting value of $\alpha_s$ used in PDF's is internally defined by the  \texttt{LHAPDF5} library.
For the matrix element calculations we do not impose consistent definition of $\alpha_s$ but it can be enforced 
by the user, see next point. As a default, we fix  starting point at $\alpha_s(M_Z)=0.1180$ value
and evolve it with $Q^2$ with a simple formula of Eq.~(\ref{eq:alfas}).

\item {\bf User own $\alpha_s$ in matrix element calculation}:\\
User can supersede the simple, leading logarithmic 
function provided by us for $\alpha_s(Q^2)$ used in   the matrix element calculation (Eq.~(\ref{eq:alfas})) 
 with his preferred one, and pass it to the program.
The function calculating $\alpha_s$ has to have the following arguments: \\
\\
{\tt  alphasModif(double Q2,int scalePDFOpt, int KEY)} \\
\\
In {\tt alphasModif} one can also use directly a method 
{\tt  LHAPDF::alphasPDF(sqrt(Q2))} of {\tt LHAPDF5} library~\cite{Whalley:2005nh}, the same assuring consistency
between value of $\alpha_s$ in the matrix element and the structure functions.
Such function can be used by executing {\tt set\_alphasModif(alphasModif); }.
An example of such setup has been provided in {\tt example-VBF.cxx} program.
\end{itemize}

\subsection{Random number initialization}
In most of the calculations the {\tt TauSpinner} algorithms are not using random numbers. However, there are  
two exceptions. 
In both cases random generators from {\tt TAUOLA} are used, see Appendix C.12 
of Ref.~\cite{Davidson:2010rw}.
\begin{itemize}
\item
The helicity states attribution uses {\tt Tauola::RandomDouble}. It should be replaced by 
the user, with the help of \\
{\tt Tauola::setRandomGenerator(double (*gen)())} method and then properly initialized, with distinct seed 
for each parallel run. In our example program the actual command is 
{\tt Tauola::setRandomGenerator( randomik );}
\item If the {\tt read\_particles\_for\_VBF.cxx} code is required to generate
$\tau$ decays, then a second random generator, coded in {\tt FORTRAN}  has to be also initialized with distinct seed for 
each individual parallel 
run: \\ {\tt Tauola::setSeed(int ijklin, int ntotin, int ntot2n)}.
\end{itemize}

\subsection{Main program -- an example}
\label{sect:MainProgram}

The following files are prepared for the user prototype program
in the  {\tt TAUOLA/TauSpinner/examples/example-VBF} directory
\begin{itemize}
\item
The  user example program {\tt  example-VBF.cxx}.
\item
The  prototype method {\tt read\_particles\_for\_VBF.cxx } to read in events stored in {\tt 
HepMC} format  is  prepared specifically for {\tt MadGraph5}
generated events.
\item
The  separate program {\tt lhef-to-hepmc.cxx} for translating 
{\tt MadGraph5} events from {\tt lhe}~\cite{Alwall:2006yp} to {\tt HepMC}~\cite{Dobbs:2001ck} format.
\item
The {\tt README } file which contains  auxiliary  information.
\end{itemize}
Only the program {\tt  example-VBF.cxx} is generic, and does not depend on the specific environment 
for event generation.  This is why we provide an extract from this code below.
For the {\tt TauSpinner} library to work, the $\tau$ decay products must be present 
in the event. In case they are absent, like e.g.\ in events generated with $\tau$'s as final states in  {\tt MadGraph5},  we prepared 
settings for their decays in {\tt Tauola} library using the mode of  not-polarised $\tau$ decays
and {\tt Tauola universal interface}. 
Such additional processing is implemented in {\tt read\_particles\_for\_VBF.cxx} code. 

In our demonstration program for {\tt TauSpinner} spin correlations between $\tau$ leptons are then introduced, 
using $(2 \to 4)$ matrix elements and calculating respective spin weight. 
The purpose of the example is to demonstrate the default initialisation of the {\tt TauSpinner} program 
and a flow of the main event loop.

\begin{center}
\bf{Extract from an example for main  user program, {\tt example-VBF.cxx} file. }
\end{center}
{\scriptsize
\begin{verbatim}
//-----------------------------------------------------------------
//replacement of default (not best quality) random number generator
// #include <TRandom.h>
// TRandom gen;
// double randomik(){
// return gen.Rndm();
// }
//-----------------------------------------------------------------
int main(int argc, char **argv) {
     // Initialize Tauola
    Tauola::initialize();
    Tauola::spin_correlation.setAll(false);
    // Initialize random numbers:
    // ##1##
    // Important when you re-decay taus: set seed fortauola-fortran random number generator RANMAR
    // int ijklin=..., int ntotin=..., int ntot2n=...; /
    // Tauola::setSeed(ijklin,ntotin,ntot2n);
    // Tauola::setSeed(time(NULL), 0, 0);
    // ##2##
    // Important when you use attributed by TauSpinner  helicities
    // Replace C++ Tauola Random generator with your own (take care of seeds). Prepared method: 
    // gen.SetSeed(time(NULL));
    // Tauola::setRandomGenerator( randomik );
    // Initialize LHAPDF
    // string name="MSTW2008nnlo90cl.LHgrid";
    string name="cteq6ll.LHpdf";
    // choice used for events-VBF.lhe which is tiny, thus it is not 
    // string name="MSTW2008nlo68cl.LHgrid"; //      statistically important
    LHAPDF::initPDFSetByName(name);

    double CMSENE = 13000.0;  // 14000.0;
    bool   Ipp    = true;
    int    Ipol   = 1;
    int    nonSM2 = 0;
    int    nonSMN = 0;
    // Initialize TauSpinner
    initialize_spinner(Ipp, Ipol, nonSM2, nonSMN,  CMSENE);

    int ref=4;      // EW scheme to be used for default vbf calculation.
    int variant =4; // EW scheme to be used in optional matrix element reweighting (nonSM2=1). Then
                    //   for vbf calculation, declared above prototype method vbfdistrModif (or user function)
                    //   will be used. At its disposal result of calculation with variant of  EW scheme will be available.
    vbfinit_(&ref,&variant);
    int QCDdefault=1; // QCD scheme to be used for default vbf calculation.
    int QCDvariant=1; // QCD scheme to be used in optional matrix element reweighting (nonSM2=1).
    setPDFOpt(QCDdefault,QCDvariant);

    // Set function that modifies/replaces Matrix Element calculation of vbfdistr
    // TauSpinner::set_vbfdistrModif(vbfdistrModif);

    // Set function that modifies/replaces alpha_s calculation of vbfdistr
    // TauSpinner::set_alphasModif(alphasModif);
    // Open I/O files  (in our example events are taken from "events.dat")
    HepMC::IO_GenEvent input_file(input_filename,std::ios::in);

    int events_read  = 0;
    int events_count = 0;
    double wt_sum    = 0.0;

     //- Event loop --------------------------------------------------------------
     while( !input_file.rdstate() ) {
        double    WT      = 1.0;
        double    W[2][2] = { { 0.0 } };
        SimpleParticle p1, p2, X, p3, p4, tau1, tau2;
        vector<SimpleParticle> tau1_daughters, tau2_daughters;
        int status = read_particles_for_VBF(input_file,p1,p2,X,p3,p4,tau1,tau2,tau1_daughters,tau2_daughters);
        ++events_read;
        WT = calculateWeightFromParticlesVBF(p3, p4, X, tau1, tau2, tau1_daughters, tau2_daughters);
        wt_sum += WT;
        ++events_count;
        if( events_limit && events_count >= events_limit ) break;
    }
    cout<<endl<<"No of events read from the file: "<<events_read<<endl;
    cout<<endl<<"No of events processed for spin weight: "<<events_count<<endl;
    cout<<      "WT average for these processed events: "<<wt_sum/events_count<<endl;
}
\end{verbatim}
}

\subsection{New option for the $ (2 \to 2)$ case}

To synchronize the old code with the equivalent method implemented now for ($2 \to 4$) process\\  
{\tt TauSpinner::set\_vbfdistrModif(vbfdistrModif)} which enables introduction of the user-defined function for matrix
elements which are sensitive to flavours of incoming partons, we provide  such an option for  the 
($2 \to 2$) variant of {\tt TauSpinner} as well. Just from now on the first argument of user-defined function {\tt nonSM\_adopt}
denotes the incoming parton flavour {\tt PDGid} and is respectively treated when calculating matrix element for ($2 \to 2$) process. 
 
The necessary  changes were introduced, and from now on the first argument {\tt ID}
passed by {\tt TauSpinner} library to  the user-defined function {\tt nonSM\_adopt}
activated by the pointer: \\
{\tt set\_nonSM\_born( nonSM\_adopt )}  \\
of {\tt /TAUOLA/TauSpinner/examples/tau-reweight-test.cxx} \\
denotes the incoming parton flavour {\tt PDGid}.
In constrast, in earlier version of {\tt TauSpinner} library~\cite{Banerjee:2012ez}, it was possible to invoke
 user-defined function {\tt nonSM\_adopt} activated by the pointer in {\tt set\_nonSM\_born( nonSM\_adopt )}  
The first argument of this method was passing to the user function the
information if incoming parton was up- or down-type quark only, without specifyingits family affiliation.

\section{Tests of reweighting the differential cross-sections for Drell-Yan--like processes}
\label{app:XsecTestsDY}

\begin{figure}
  \begin{center}                               
{
   \includegraphics[width=7.5cm,angle=0]{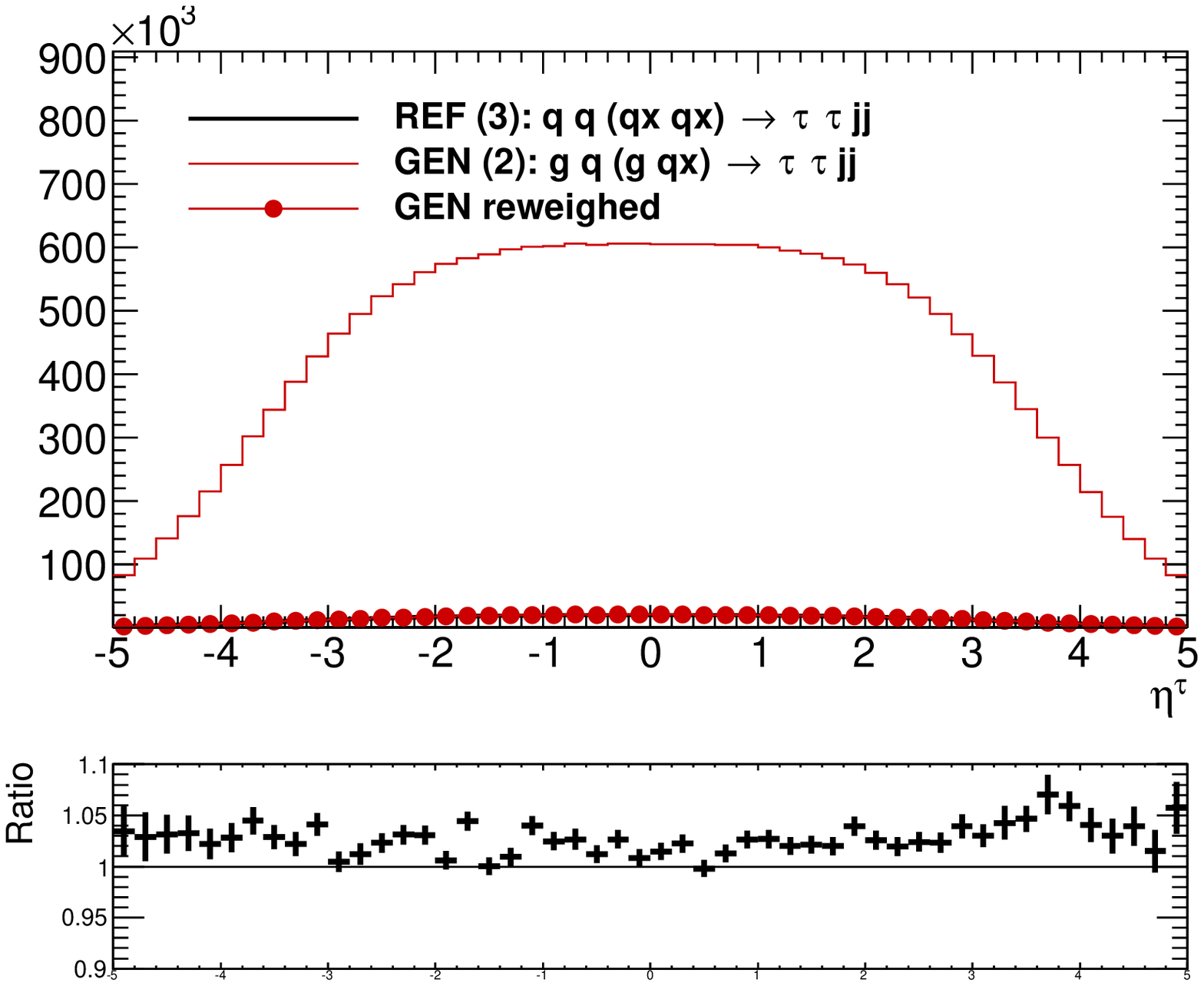}
   \includegraphics[width=7.5cm,angle=0]{IDPROD2_wto_IDPROD3_hist31006505.eps}
   \includegraphics[width=7.5cm,angle=0]{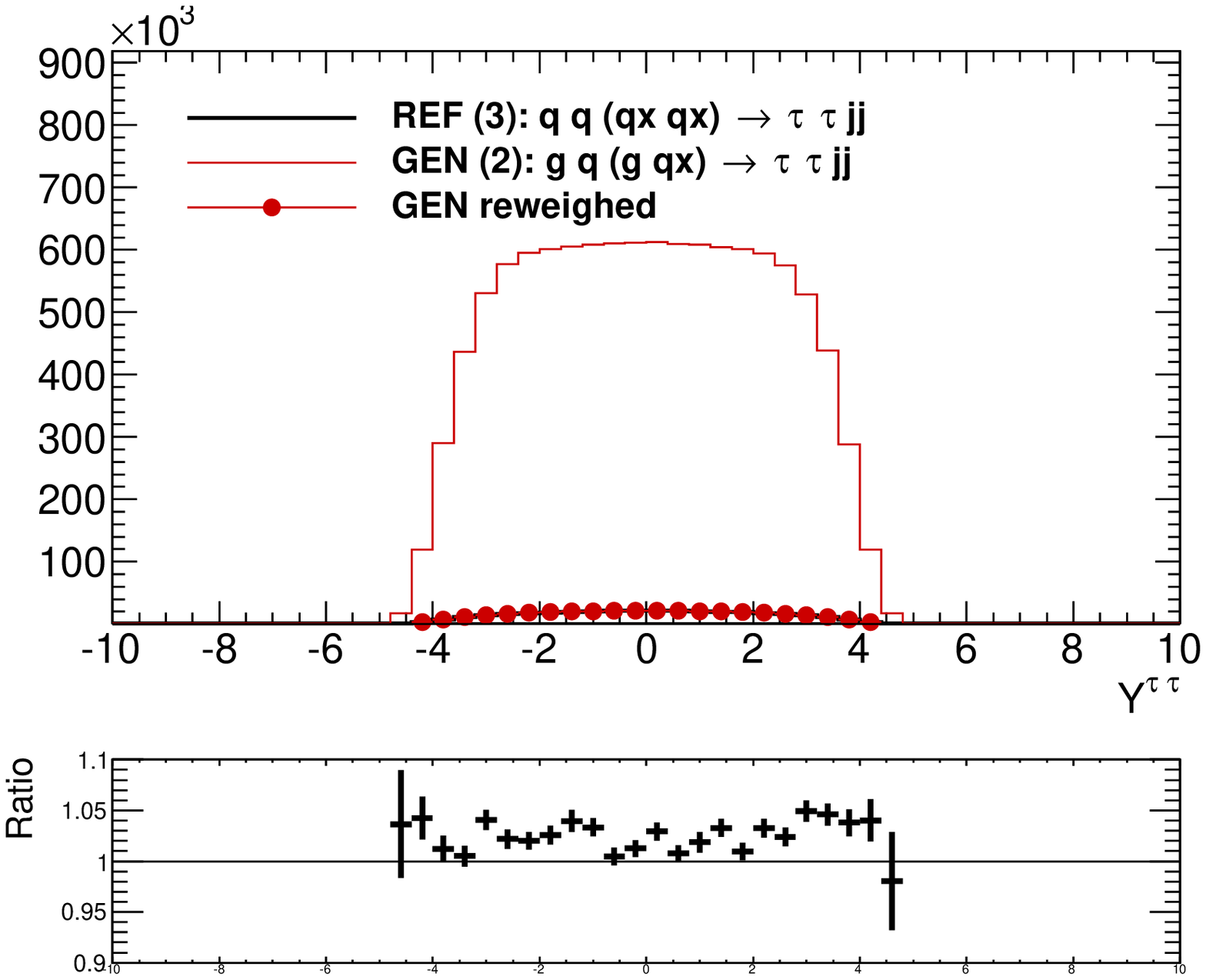}
   \includegraphics[width=7.5cm,angle=0]{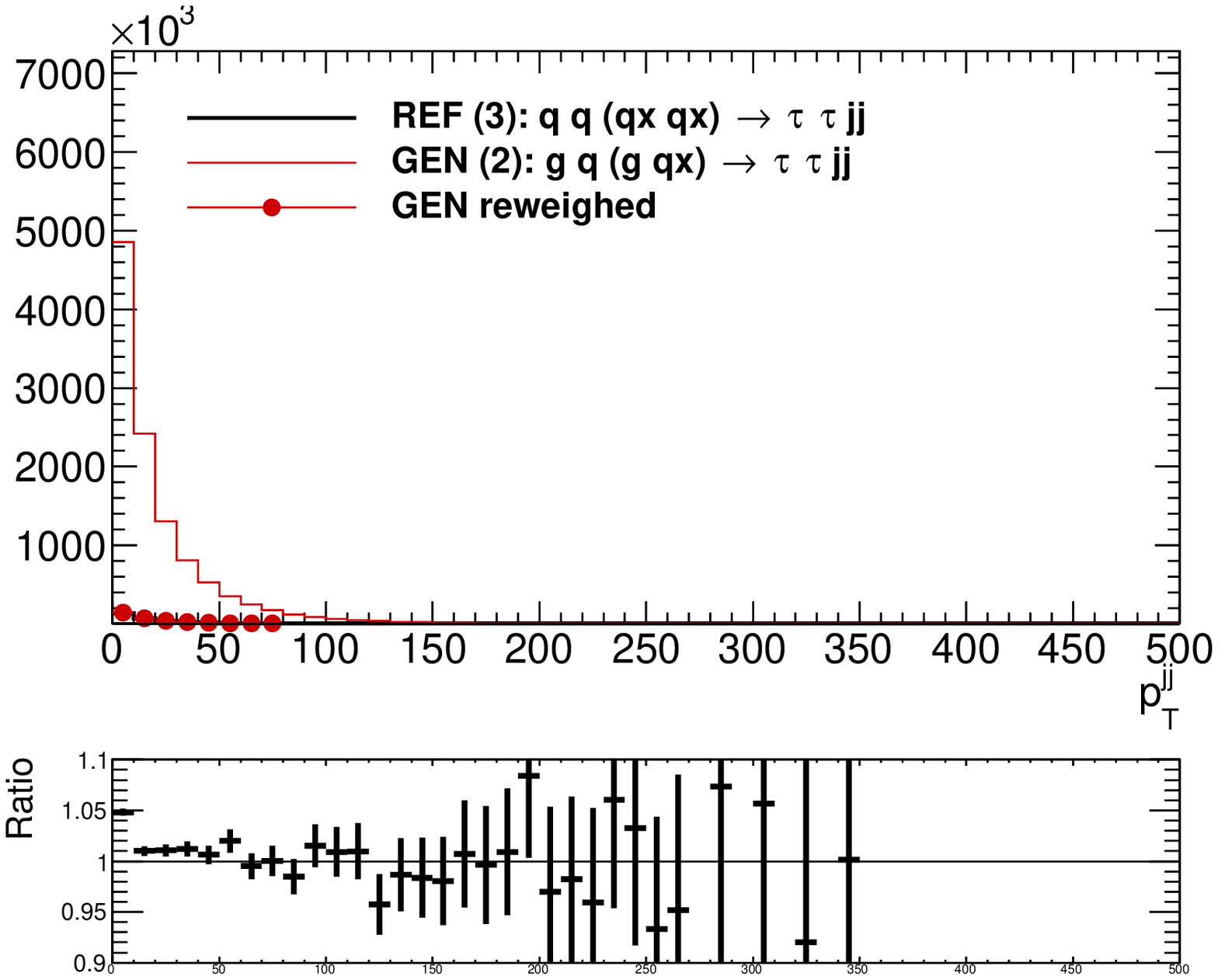}
   \includegraphics[width=7.5cm,angle=0]{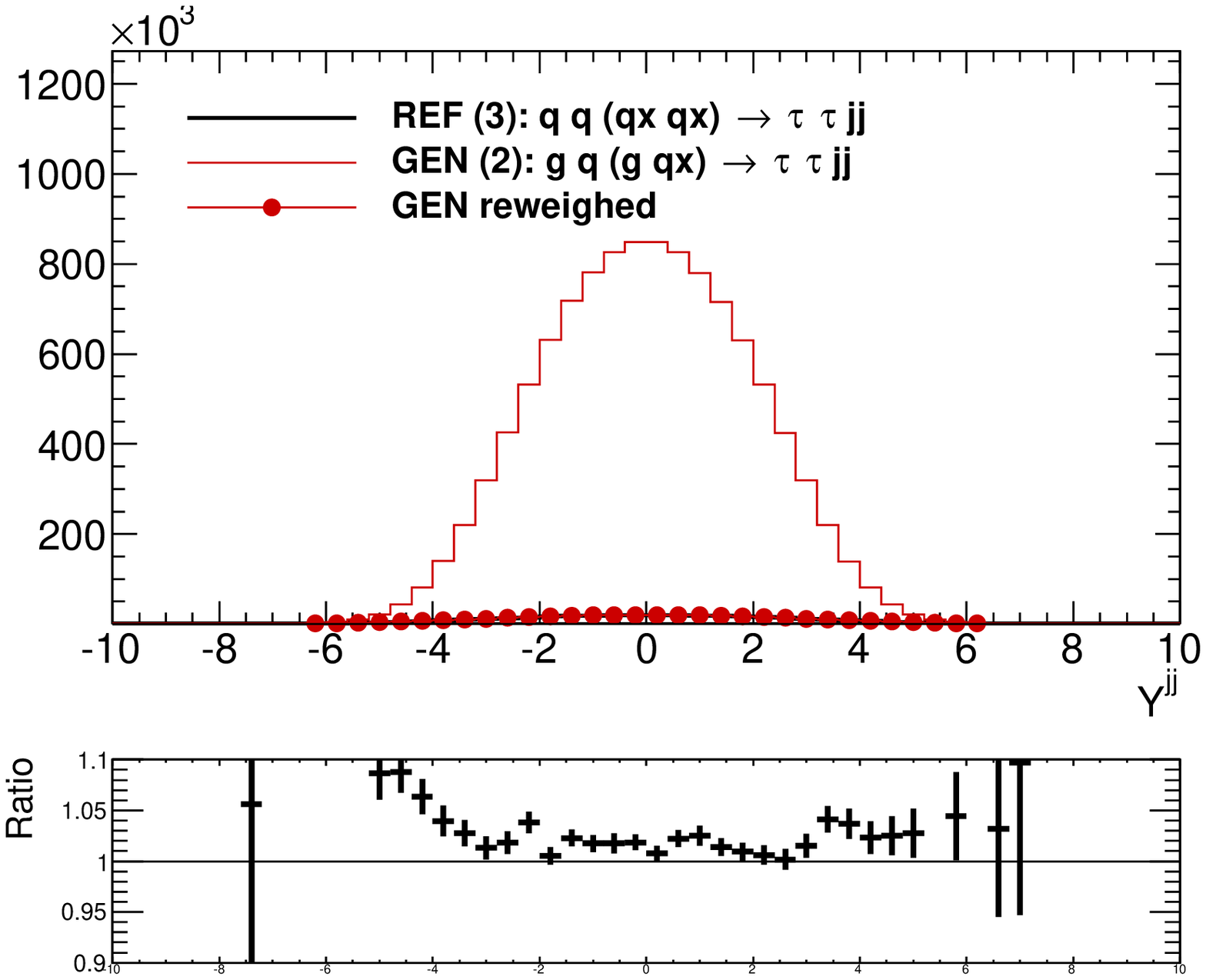}
   \includegraphics[width=7.5cm,angle=0]{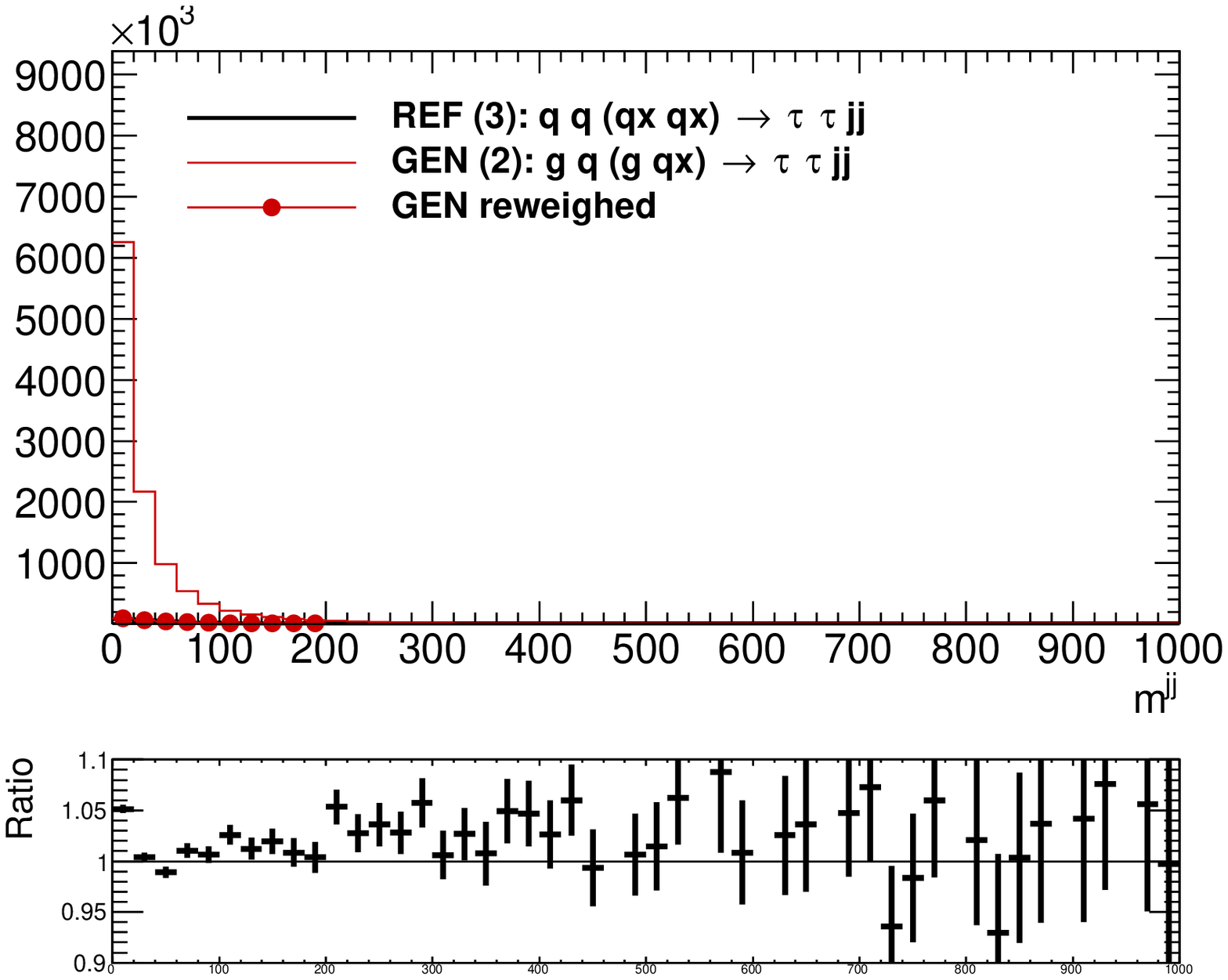}
}
\end{center}
\caption{Shown generated   $g\ q \to \tau \tau jj$ (thin red line) after reweighing 
to  $ q\ q\ (qx\ qx) \to \tau \tau jj$  (red points). Reference  $ q\ q\ (qx\ qx) \to \tau \tau jj$  distribution shown with black line.
\label{fig:A1} }
\end{figure}

\begin{figure}
  \begin{center}                               
{
   \includegraphics[width=7.5cm,angle=0]{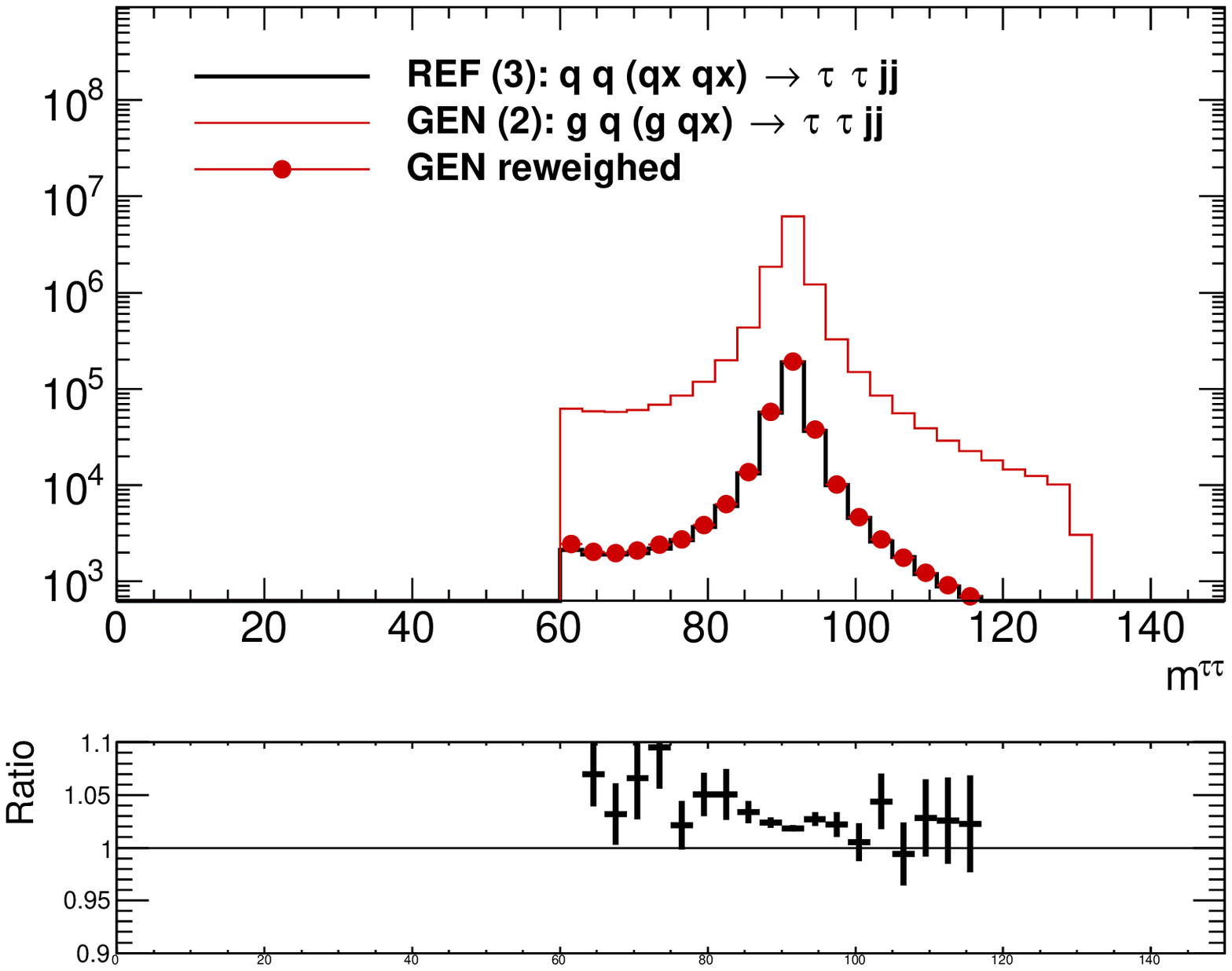}
   \includegraphics[width=7.5cm,angle=0]{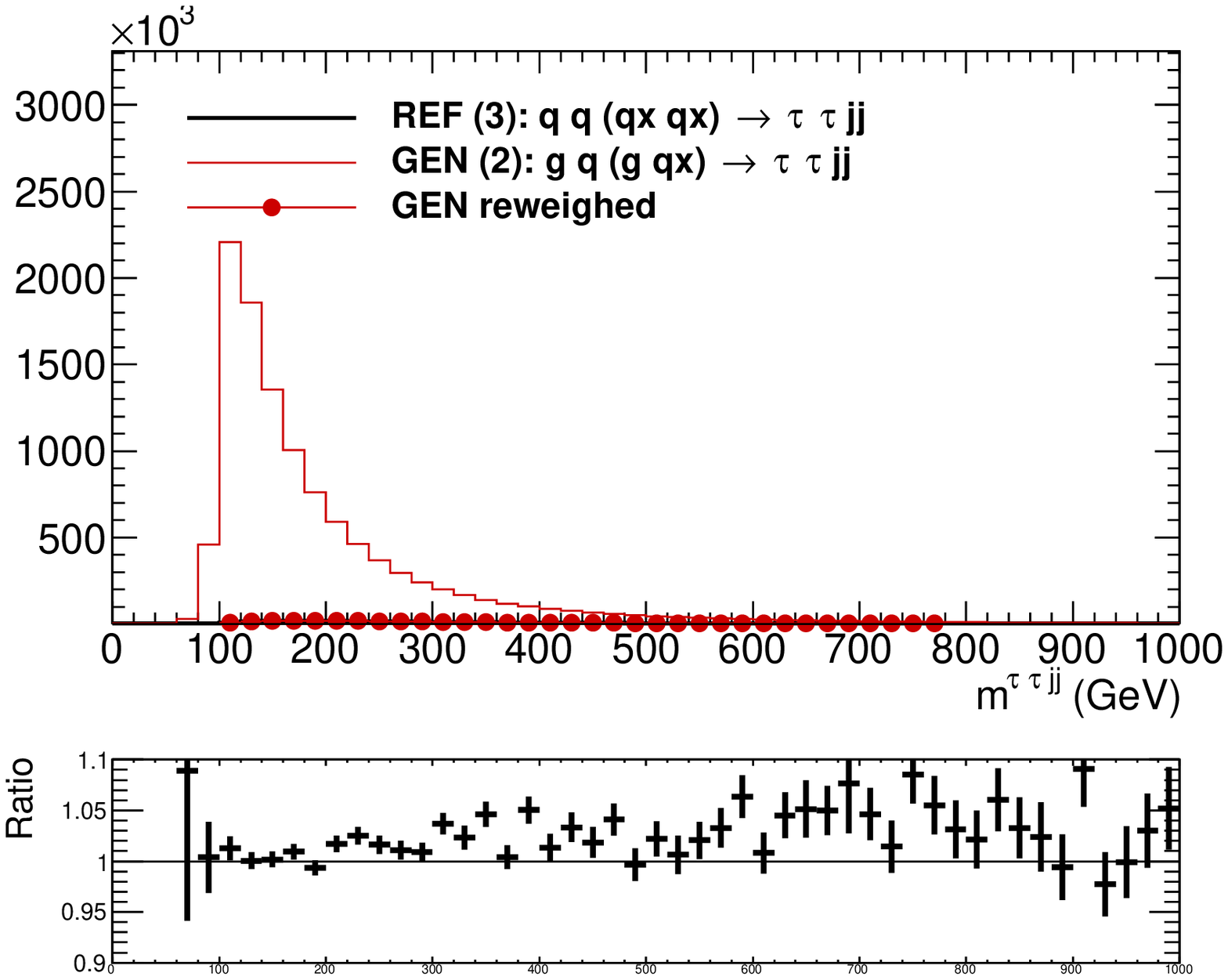}
   \includegraphics[width=7.5cm,angle=0]{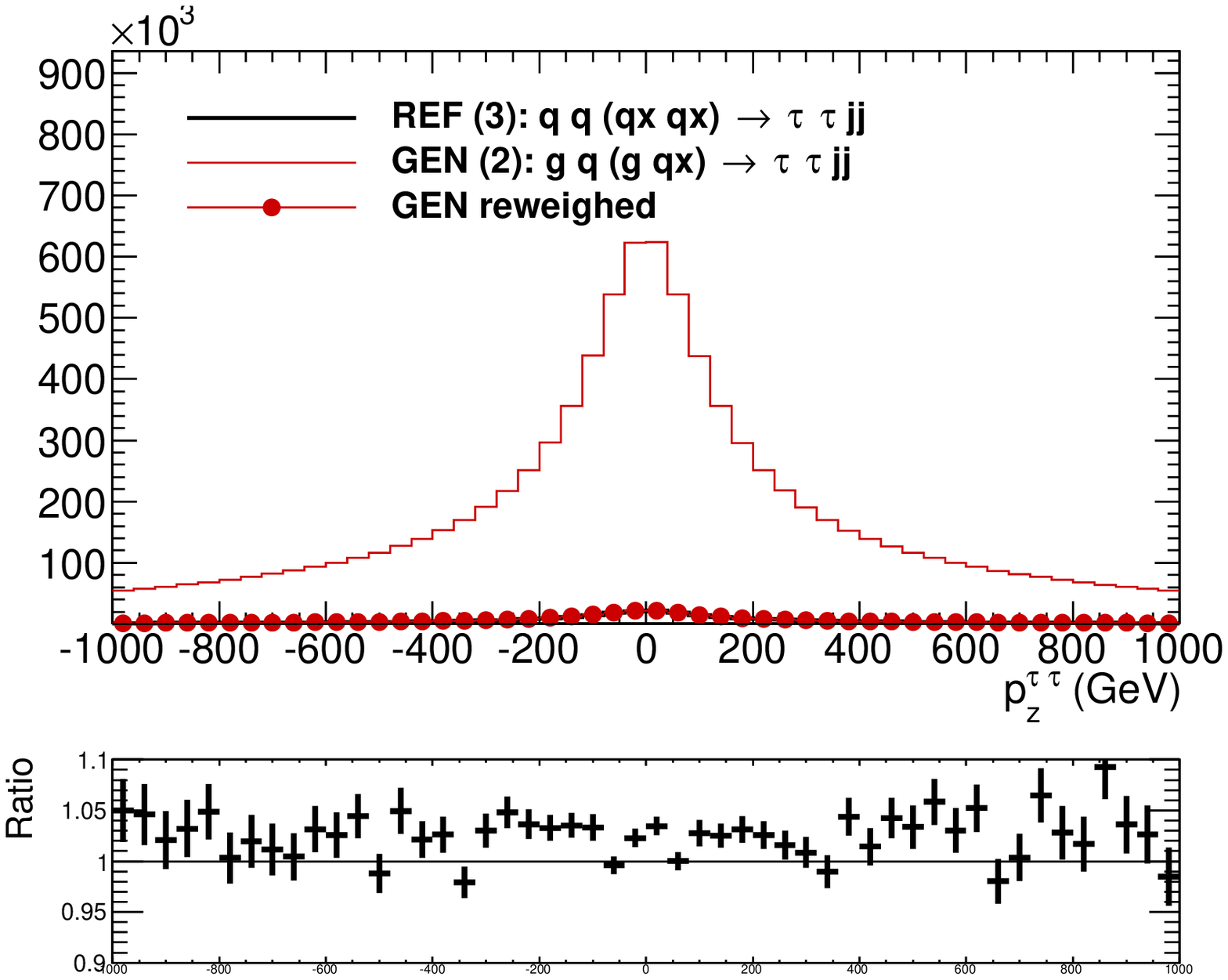}
   \includegraphics[width=7.5cm,angle=0]{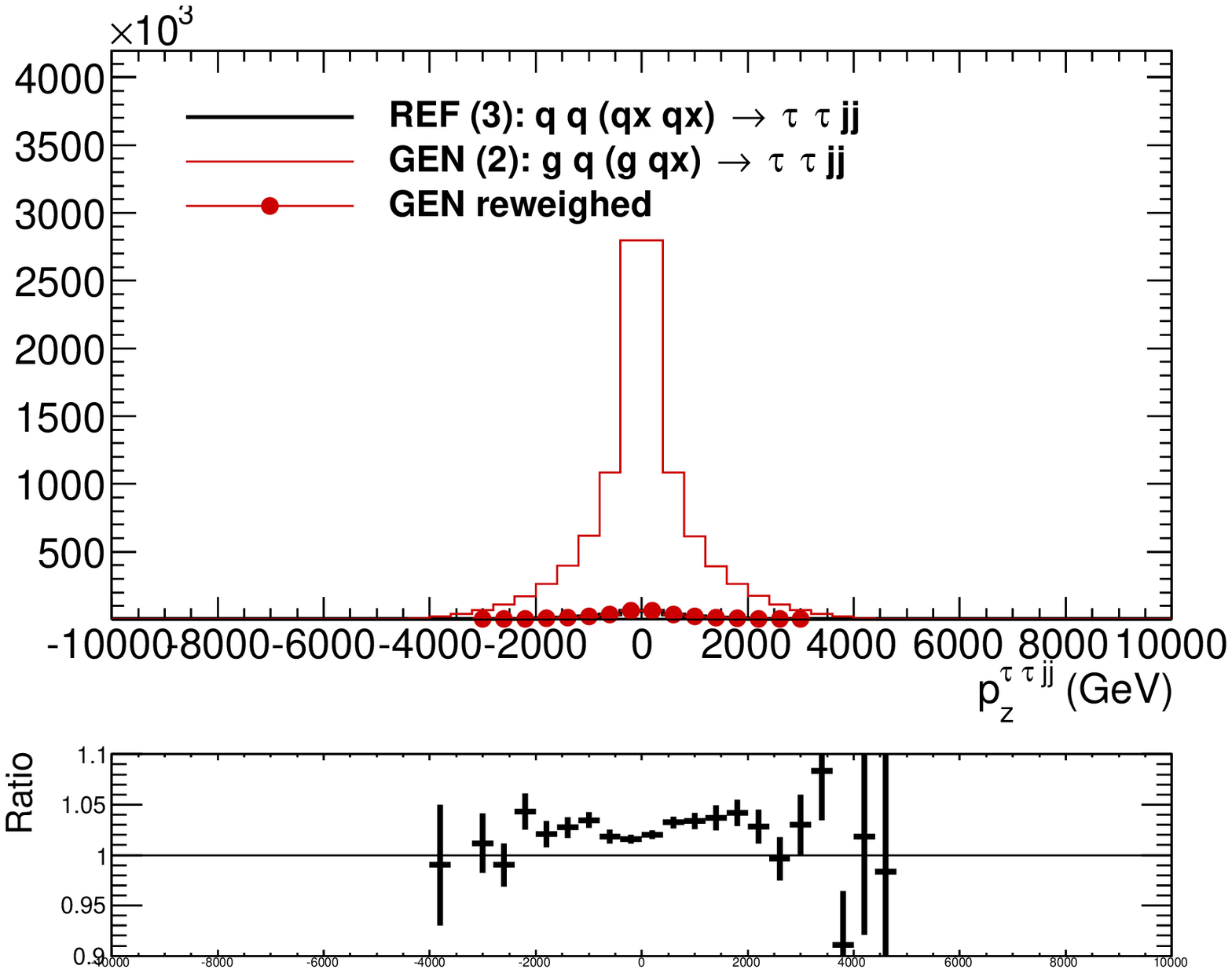}
   \includegraphics[width=7.5cm,angle=0]{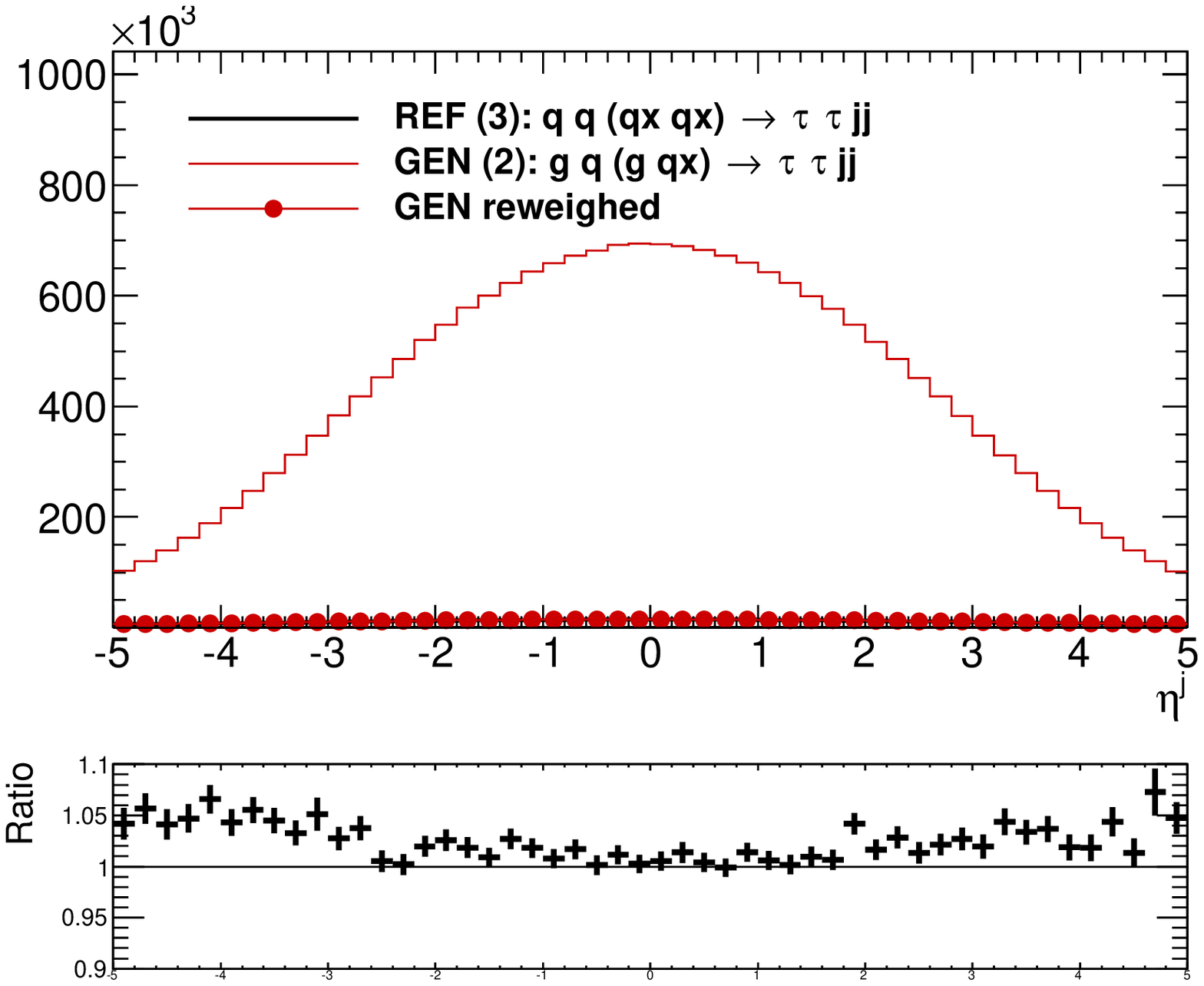}
   \includegraphics[width=7.5cm,angle=0]{IDPROD2_wto_IDPROD3_hist31006514.eps}
}
\end{center}
\caption{Shown generated  $g\ q \to \tau \tau jj$ (thin red line) after reweighting 
to $ q\ q\ (qx\ qx) \to \tau \tau jj$  (red points). Reference  $ q\ q\ (qx\ qx) \to \tau \tau jj$ distribution shown with black line.
\label{fig:A2} }
\end{figure}

In this Appendix we show in Figs.~\ref{fig:A1} and ~\ref{fig:A2} a complete set of 
kinematic distributions validating implementation of  ($2 \to 4$) non-Higgs Drell-Yan--like processes.
We split $pp \to \tau \tau jj$ events into four groups, depending on the initial partons,
see Table~\ref{tab:processes} for definition of parton level processes. We use the  implemented
($2 \to 4$) matrix elements to calculate per event a weight, $wt_{prod}^{C \to D} = d\sigma_D/d\sigma_C$, see Eq. (\ref{wtxsec}), 
defined as a ratio of the cross-sections for events of groups $C$ and $D$. 
The expression is similar to  Eq.~(\ref{eq:WT}) except that the sum is over subprocesses which
belong to the chosen groups $C$ or $W$.
We apply $wt_{prod}^{C \to D}$ to events from  the group $C$  and compare both the  absolute normalisations 
and shapes of the re-weighted distributions with the distributions of events from the group $D$.

These tests were done on the large statistics samples and have been repeated between 
each groups of processes and within groups between subgroups. 
The achieved agreement between the reference and re-weighted distributions validates the correctness
of the implemented matrix elements.

\section{Tests of reweighting differential cross-sections for Higgs boson production}
\label{app:XsecTestsH}

\begin{figure}
  \begin{center}                               
{
   \includegraphics[width=7.5cm,angle=0]{HIGGS_IDPROD3_wto_IDPROD4_hist31006102.eps}
   \includegraphics[width=7.5cm,angle=0]{HIGGS_IDPROD3_wto_IDPROD4_hist31006105.eps}
   \includegraphics[width=7.5cm,angle=0]{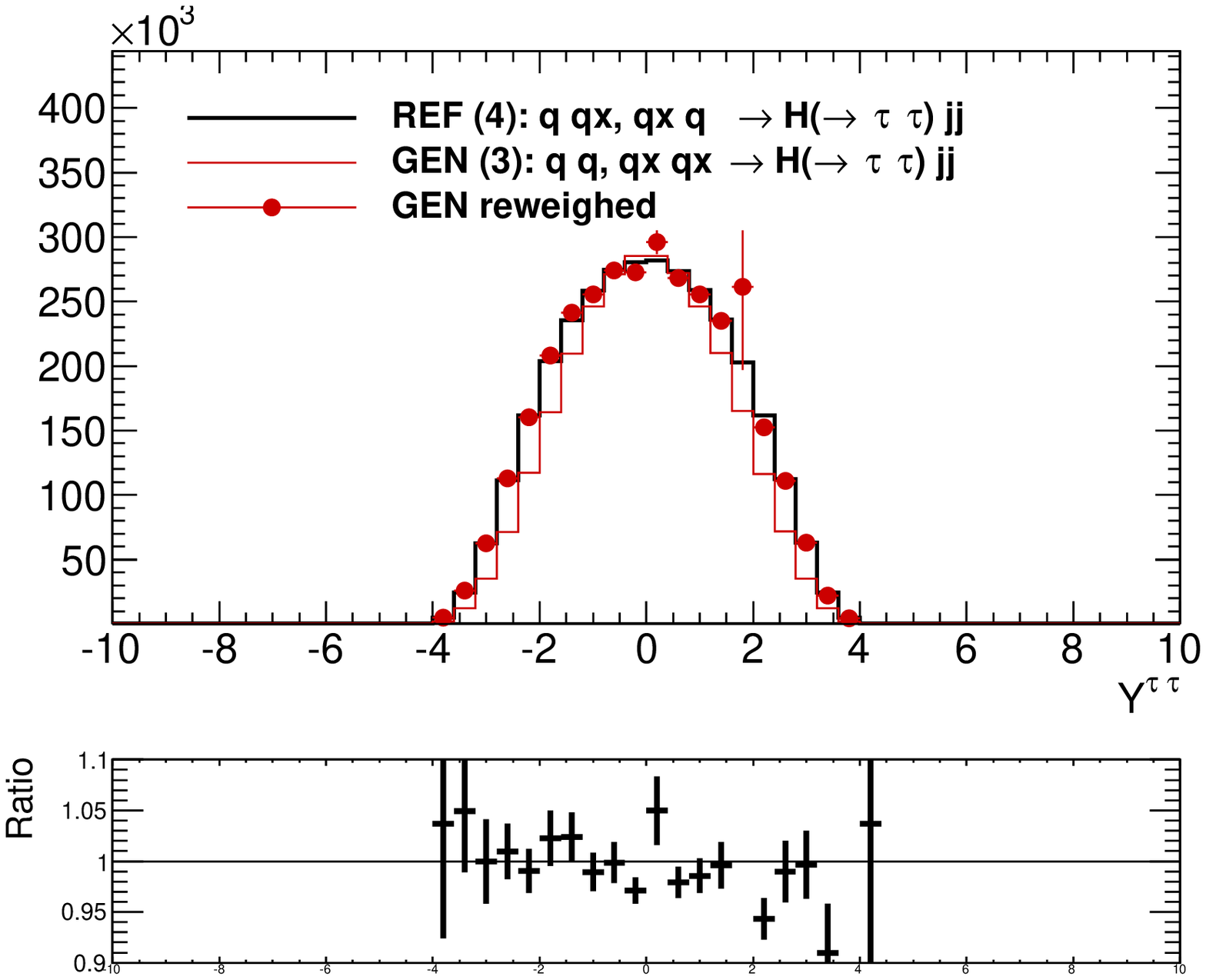}
   \includegraphics[width=7.5cm,angle=0]{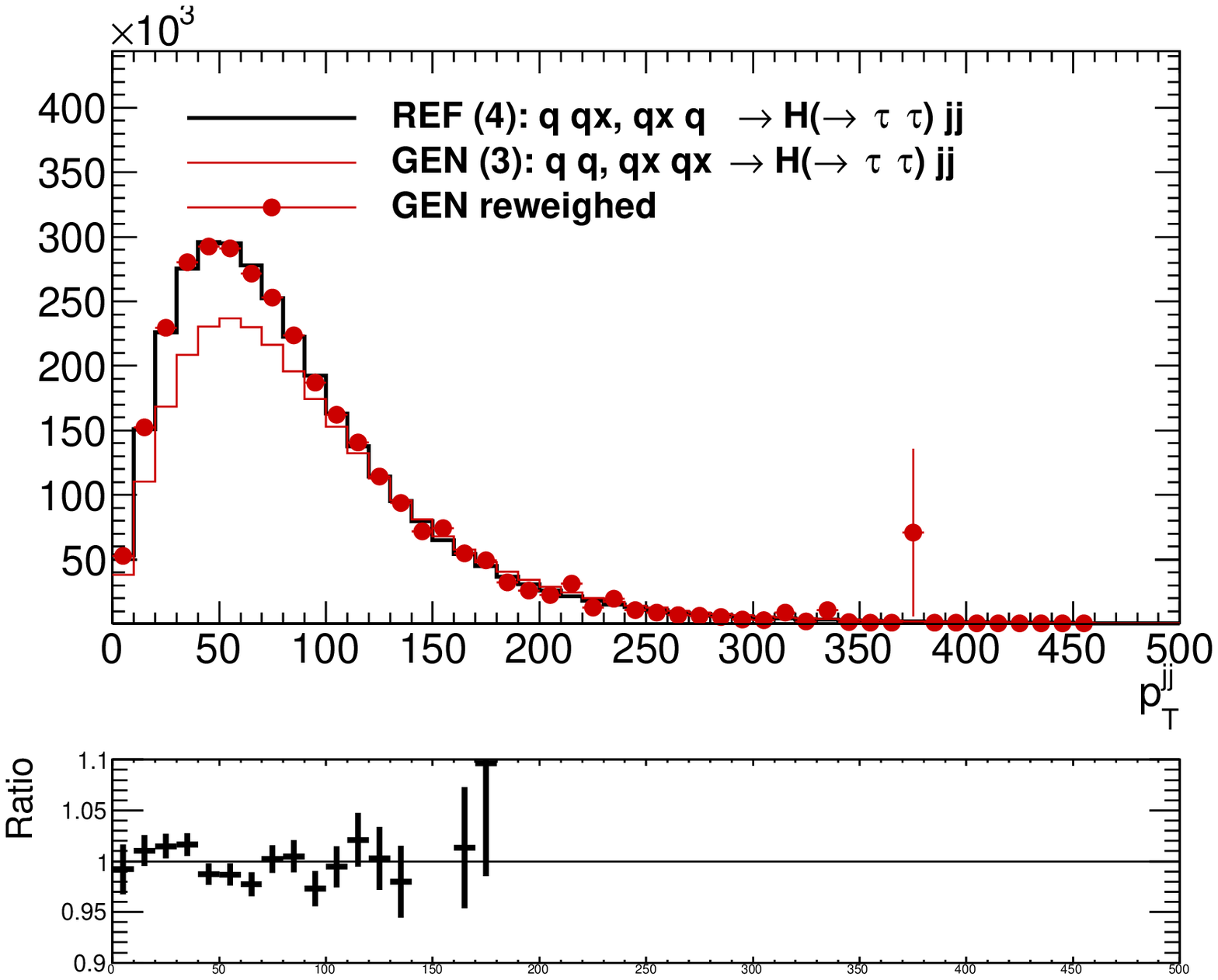}
   \includegraphics[width=7.5cm,angle=0]{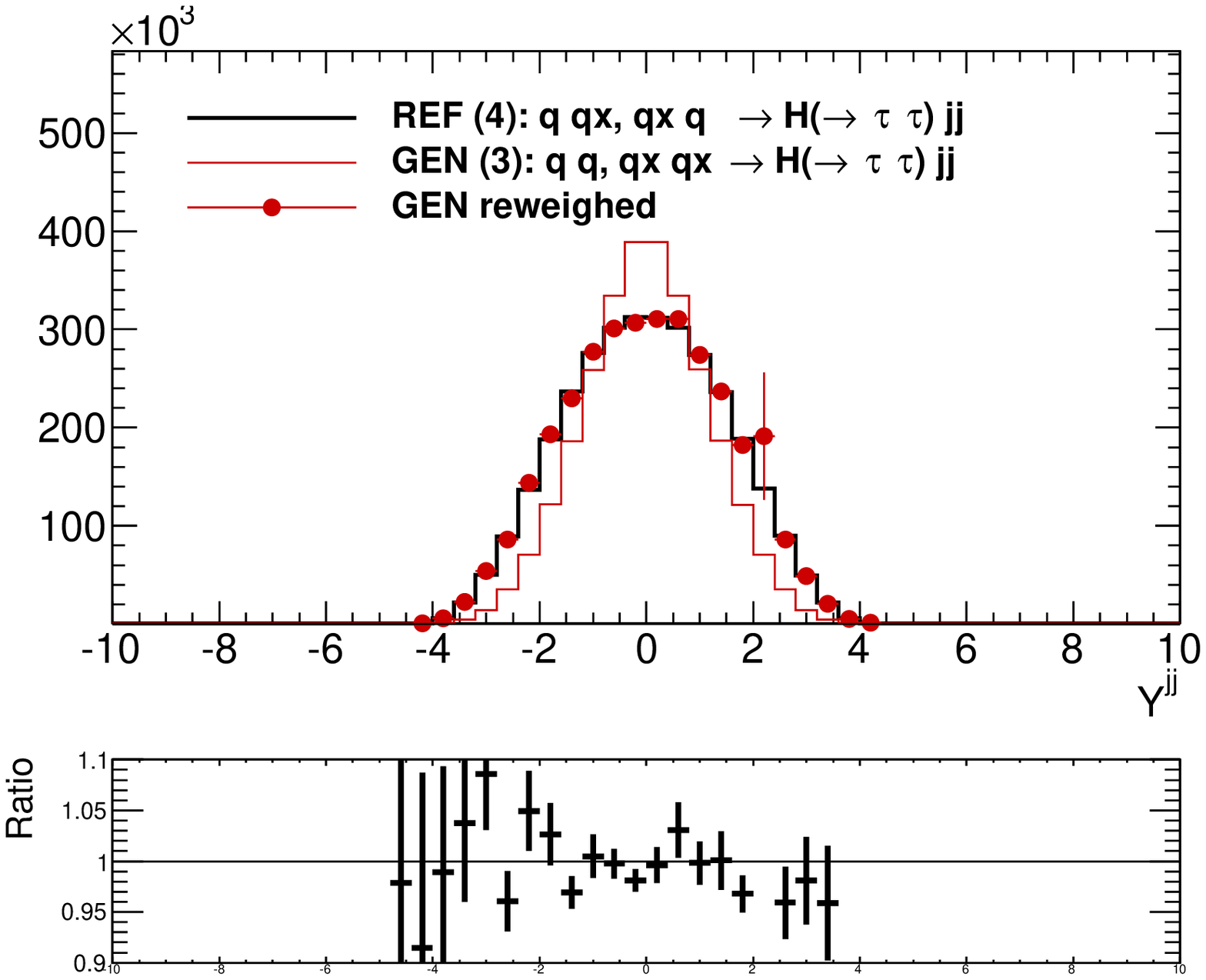}
   \includegraphics[width=7.5cm,angle=0]{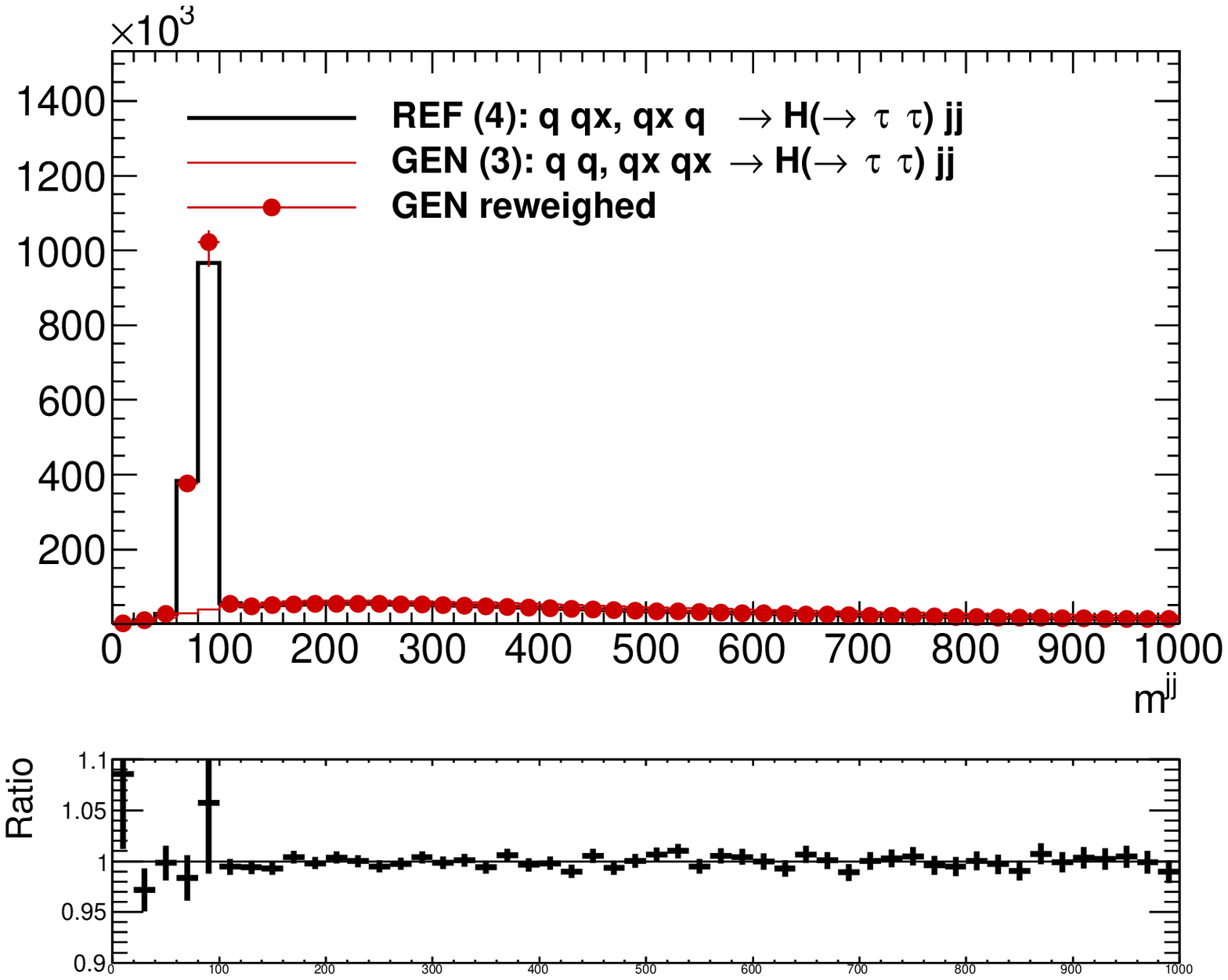}
}
\end{center}
\caption{Shown example of tests distributions for generated process   $q\ q, qx\ qx \to H(\to \tau \tau) jj$ (thin red line) after reweighting 
to  $q\ qx \to  H(\to \tau \tau) jj$ process (red points). Reference  $q\ qx \to  H(\to \tau \tau) jj$  distribution shown with black line.
\label{fig:B1} }
\end{figure}

\begin{figure}
  \begin{center}                               
{
   \includegraphics[width=7.5cm,angle=0]{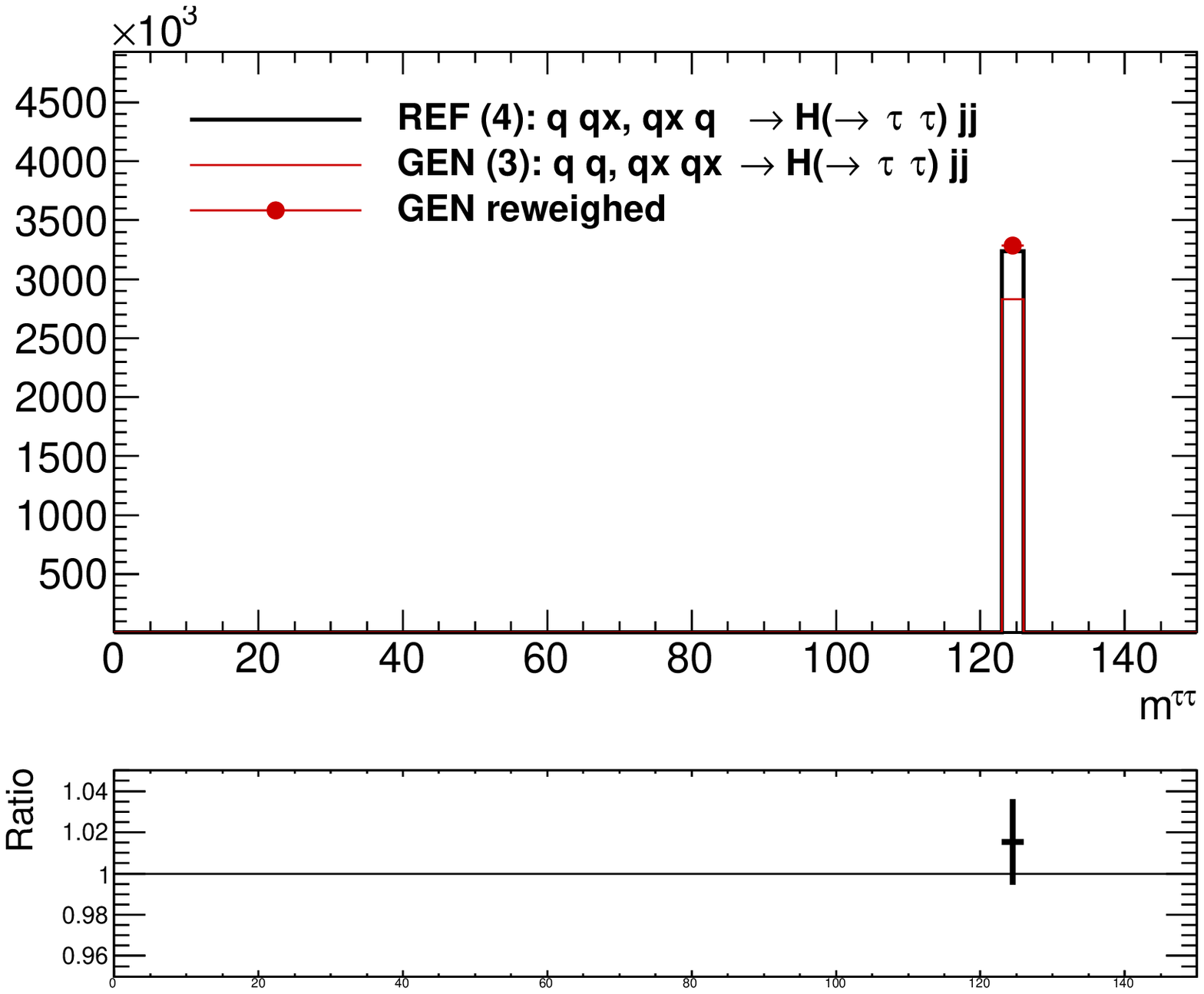}
   \includegraphics[width=7.5cm,angle=0]{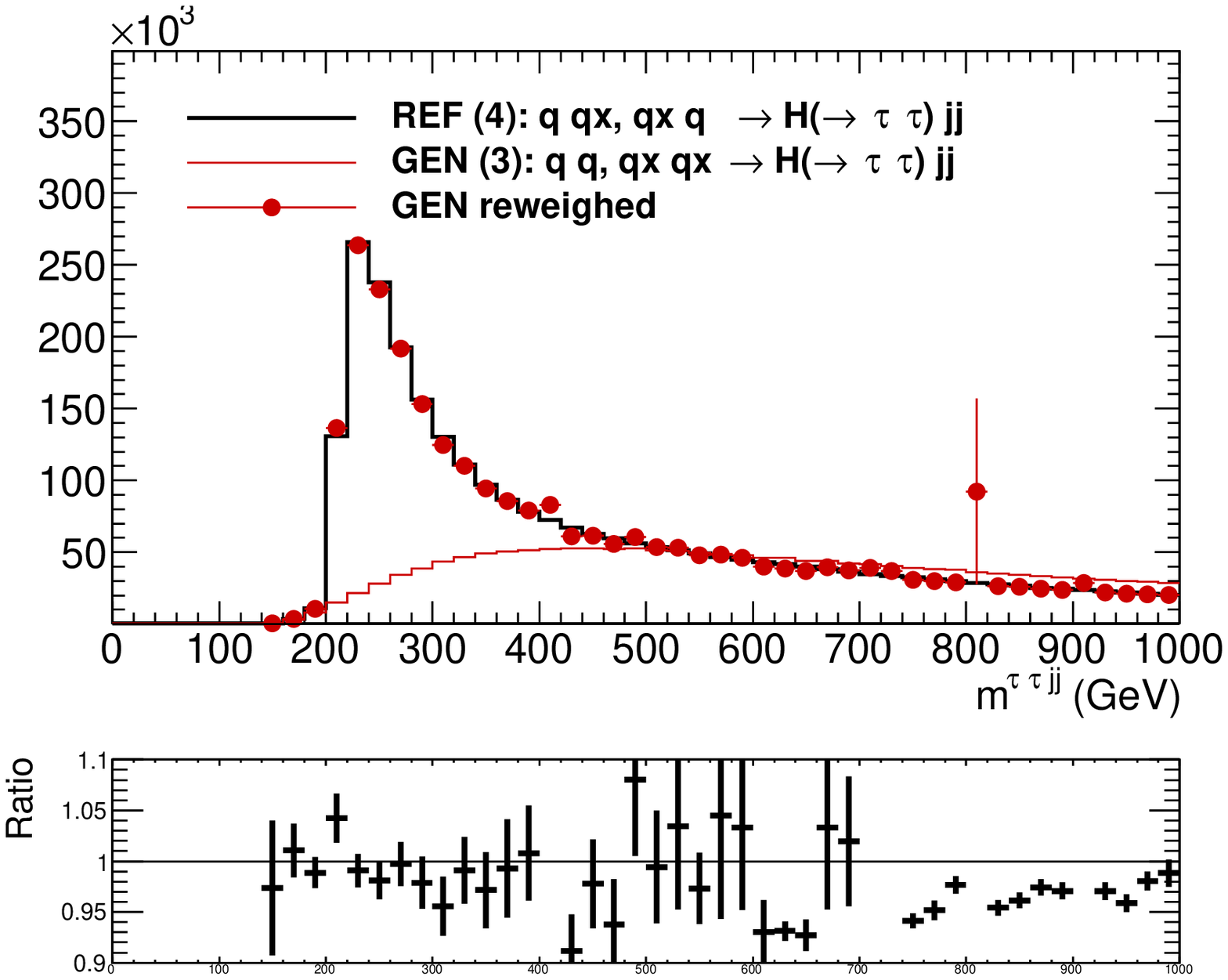}
   \includegraphics[width=7.5cm,angle=0]{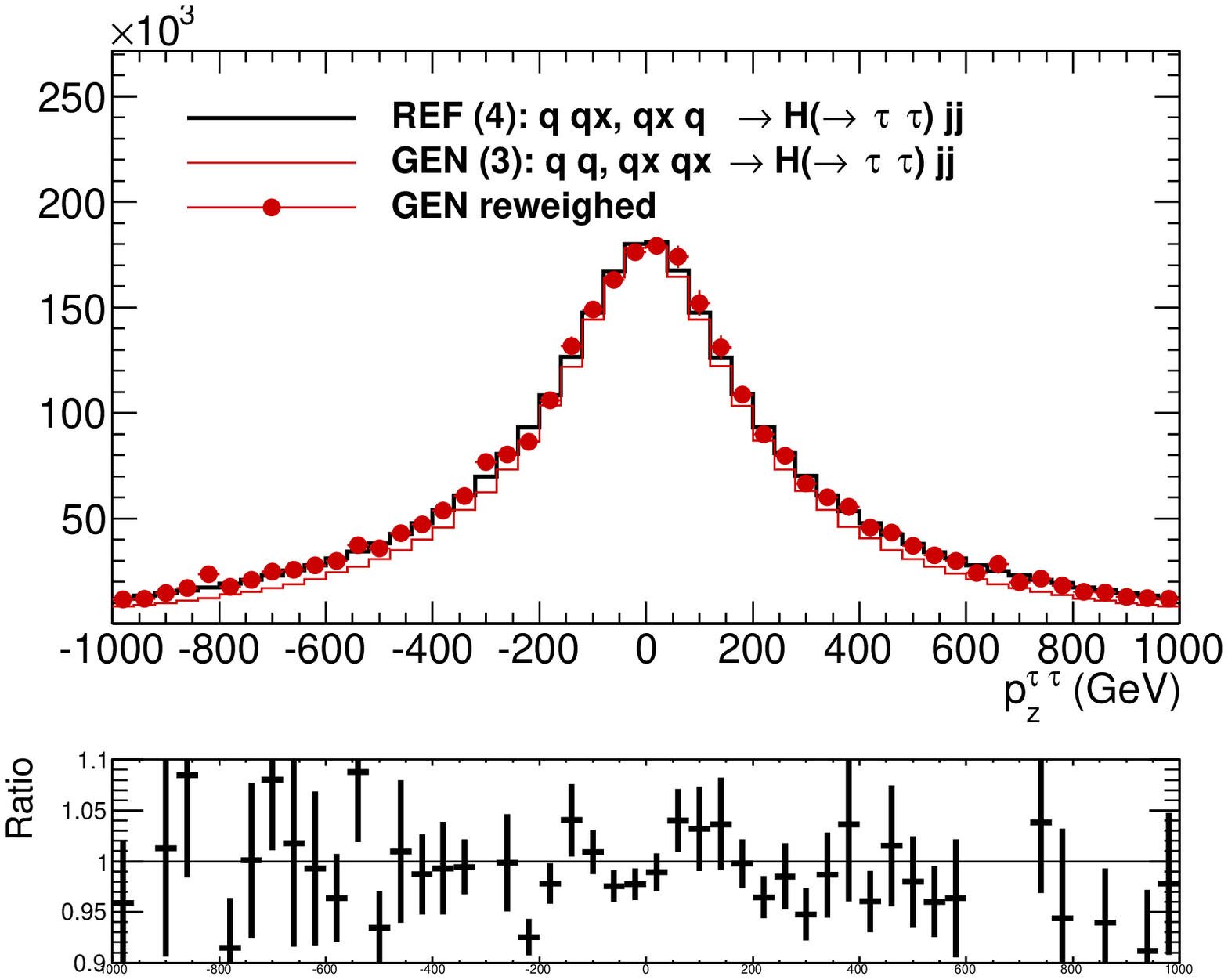}
   \includegraphics[width=7.5cm,angle=0]{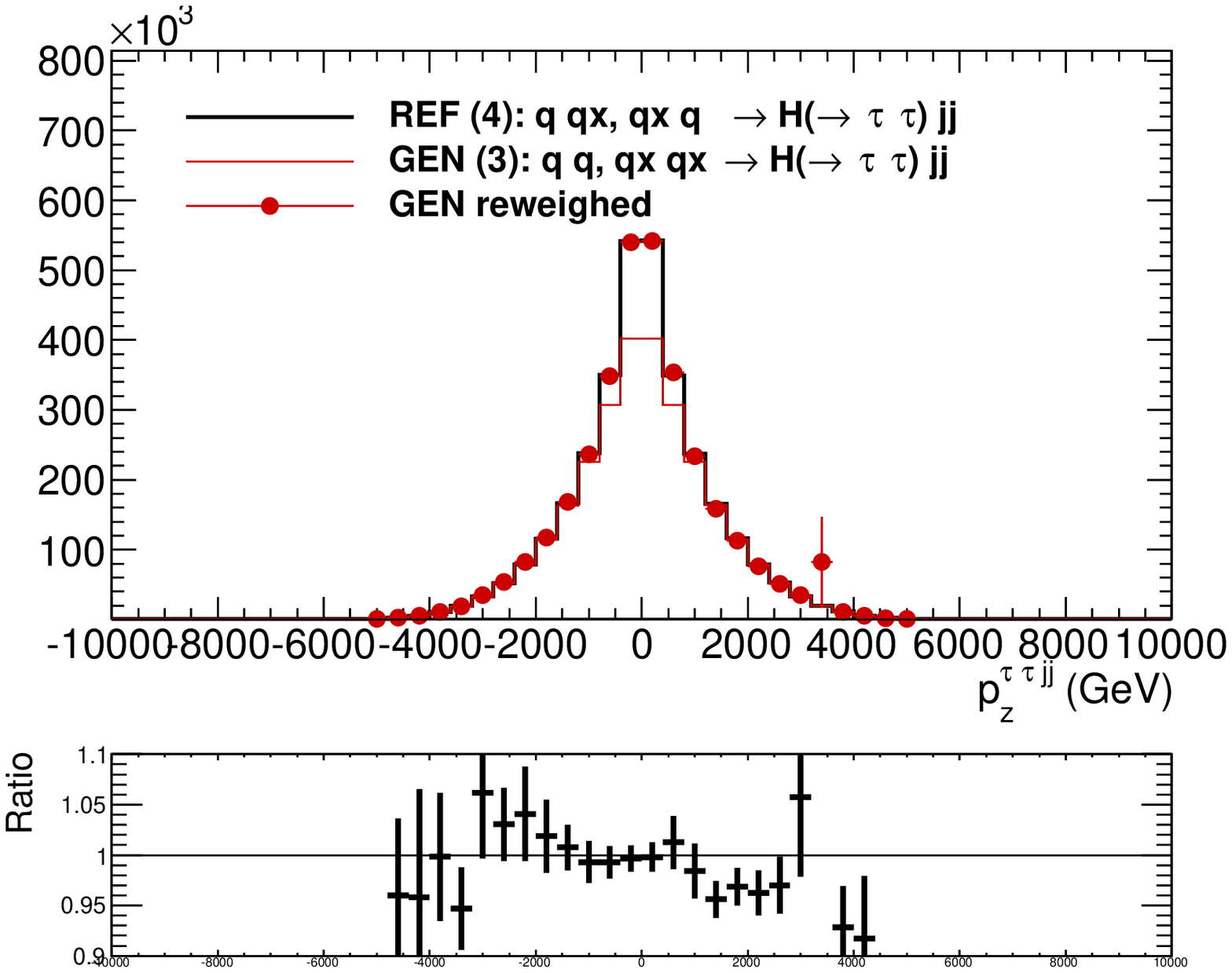}
   \includegraphics[width=7.5cm,angle=0]{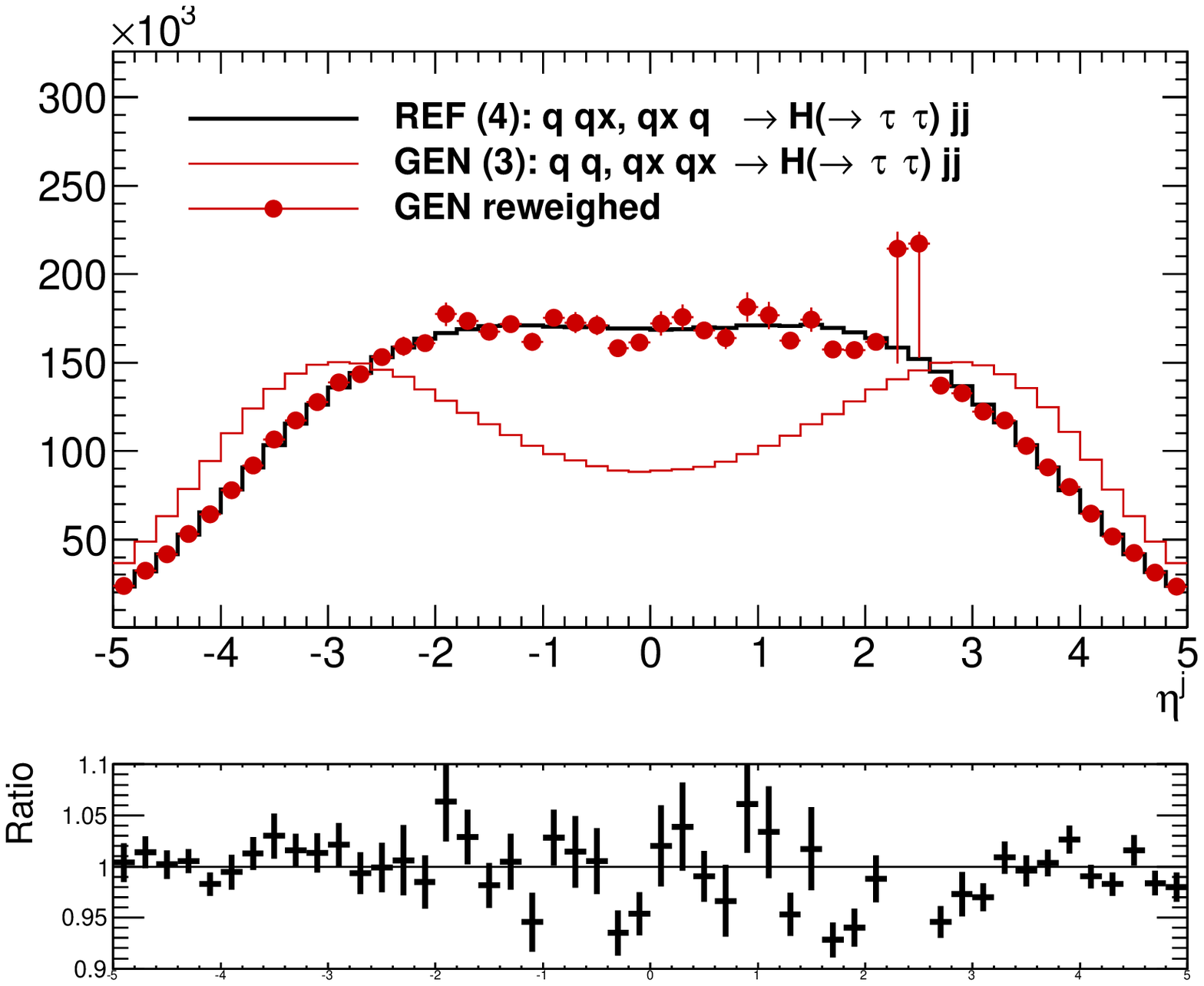}
   \includegraphics[width=7.5cm,angle=0]{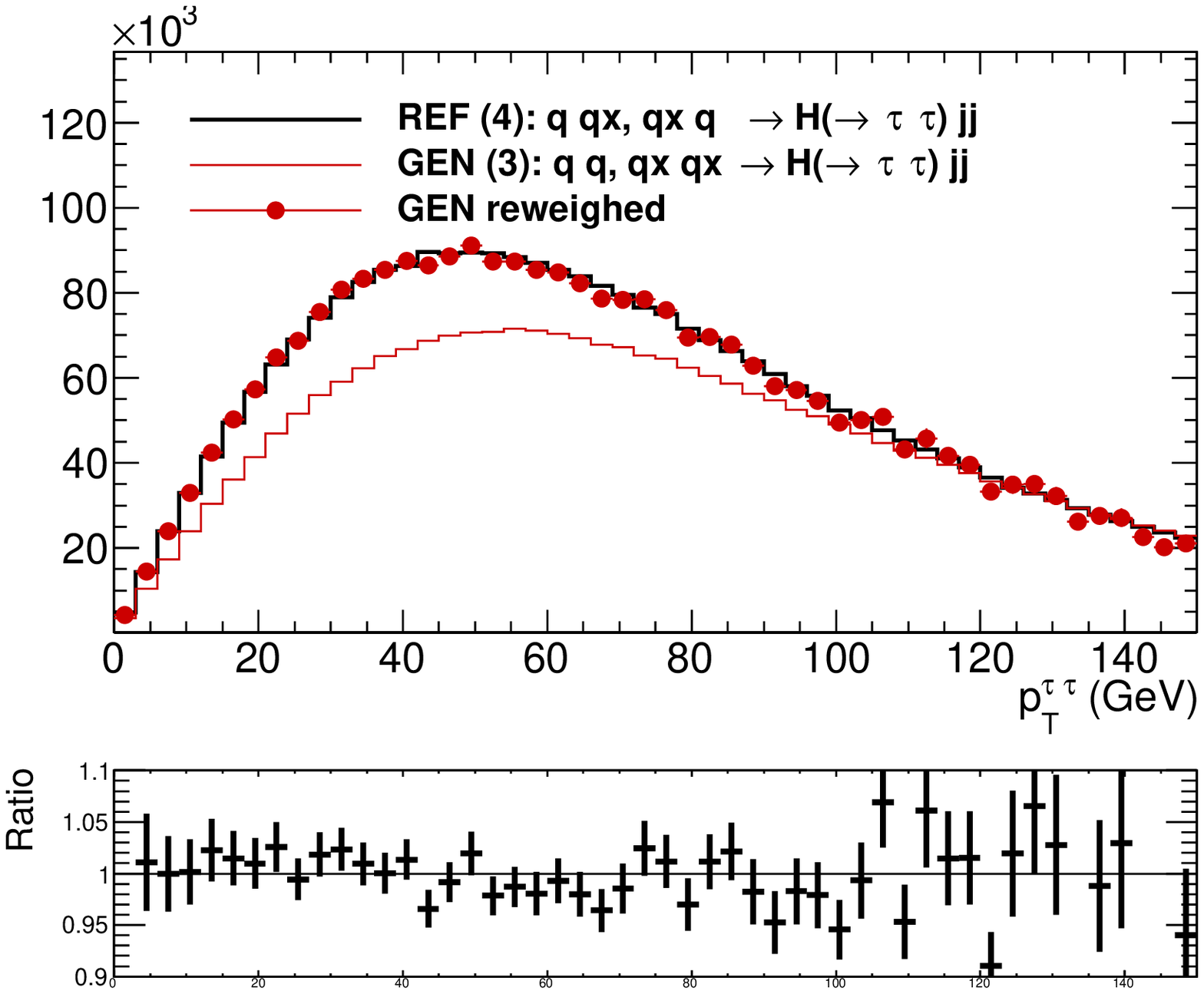}
}
\end{center}
\caption{Shown example of tests distributions for generated process $q\ q, qx\ qx \to  H(\to \tau \tau) jj$ (thin red line) after reweighting 
to $q \ qx \to  H(\to \tau \tau) jj$  (red points). Reference  $q \ qx \to  H(\to \tau \tau) jj$ distribution shown with black line.
\label{fig:B2} }
\end{figure}

\begin{figure}
  \begin{center}                               
{
   \includegraphics[width=7.5cm,angle=0]{HIGGS_IDPROD4_wto_IDPROD3_hist31006602.eps}
   \includegraphics[width=7.5cm,angle=0]{HIGGS_IDPROD4_wto_IDPROD3_hist31006605.eps}
   \includegraphics[width=7.5cm,angle=0]{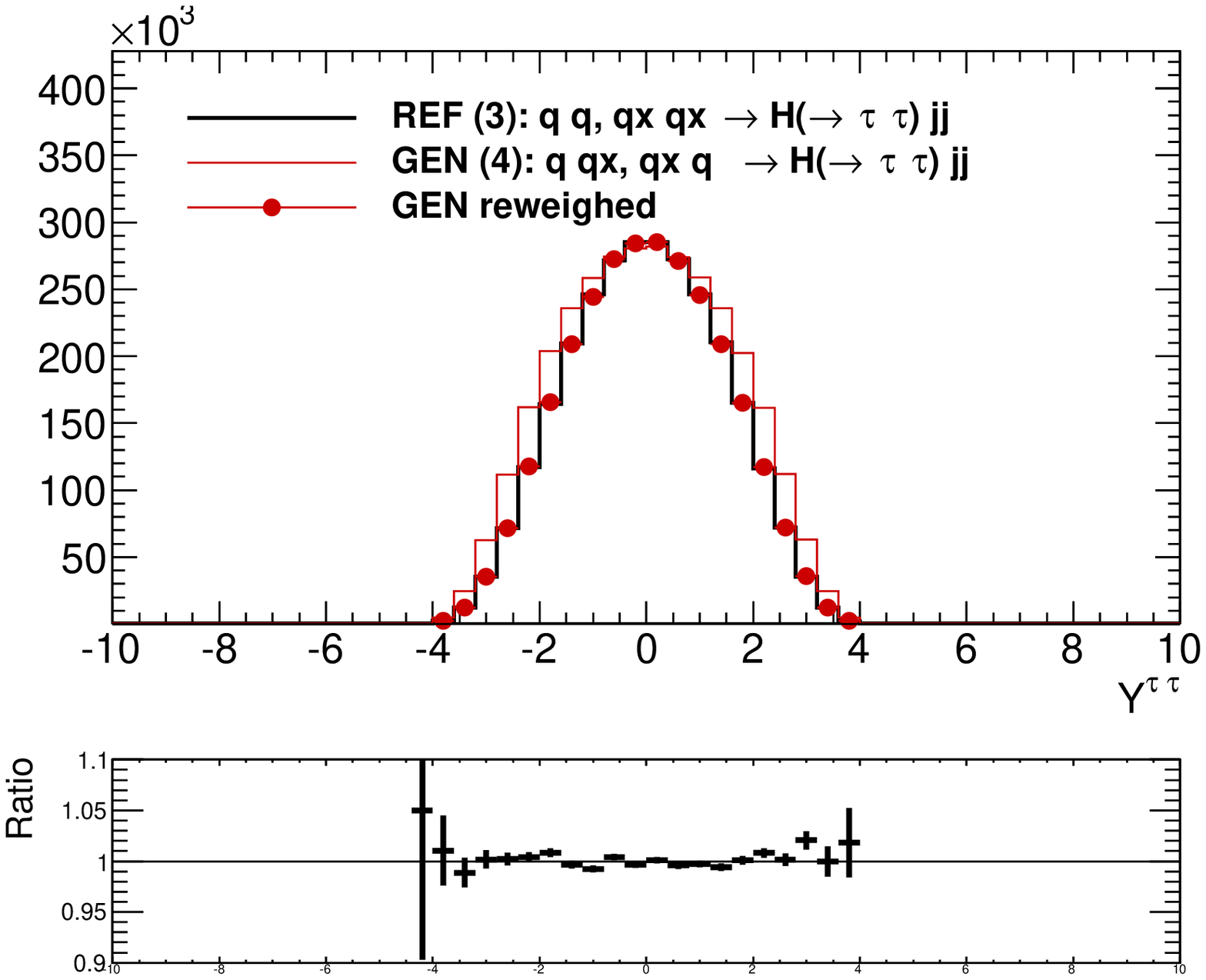}
   \includegraphics[width=7.5cm,angle=0]{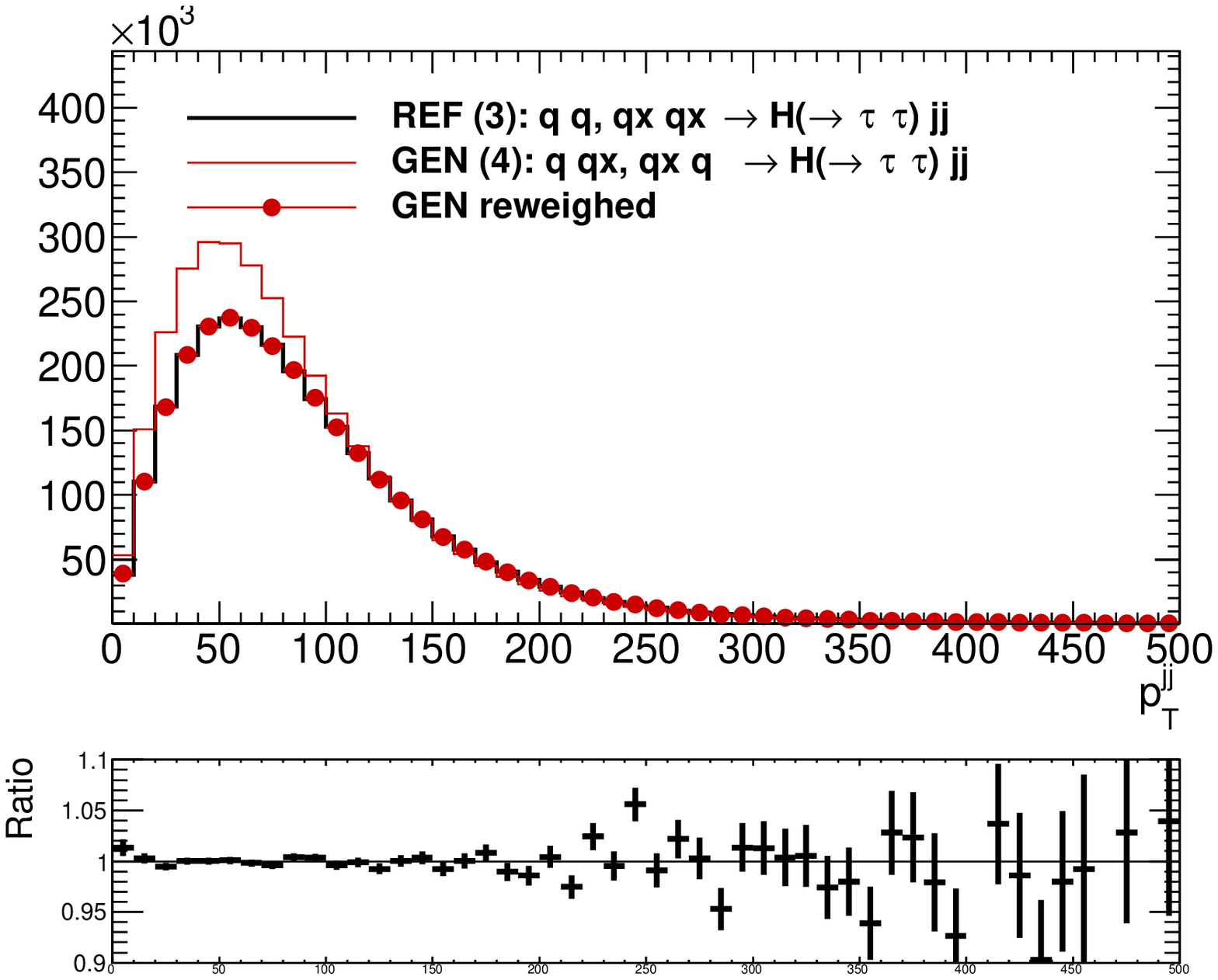}
   \includegraphics[width=7.5cm,angle=0]{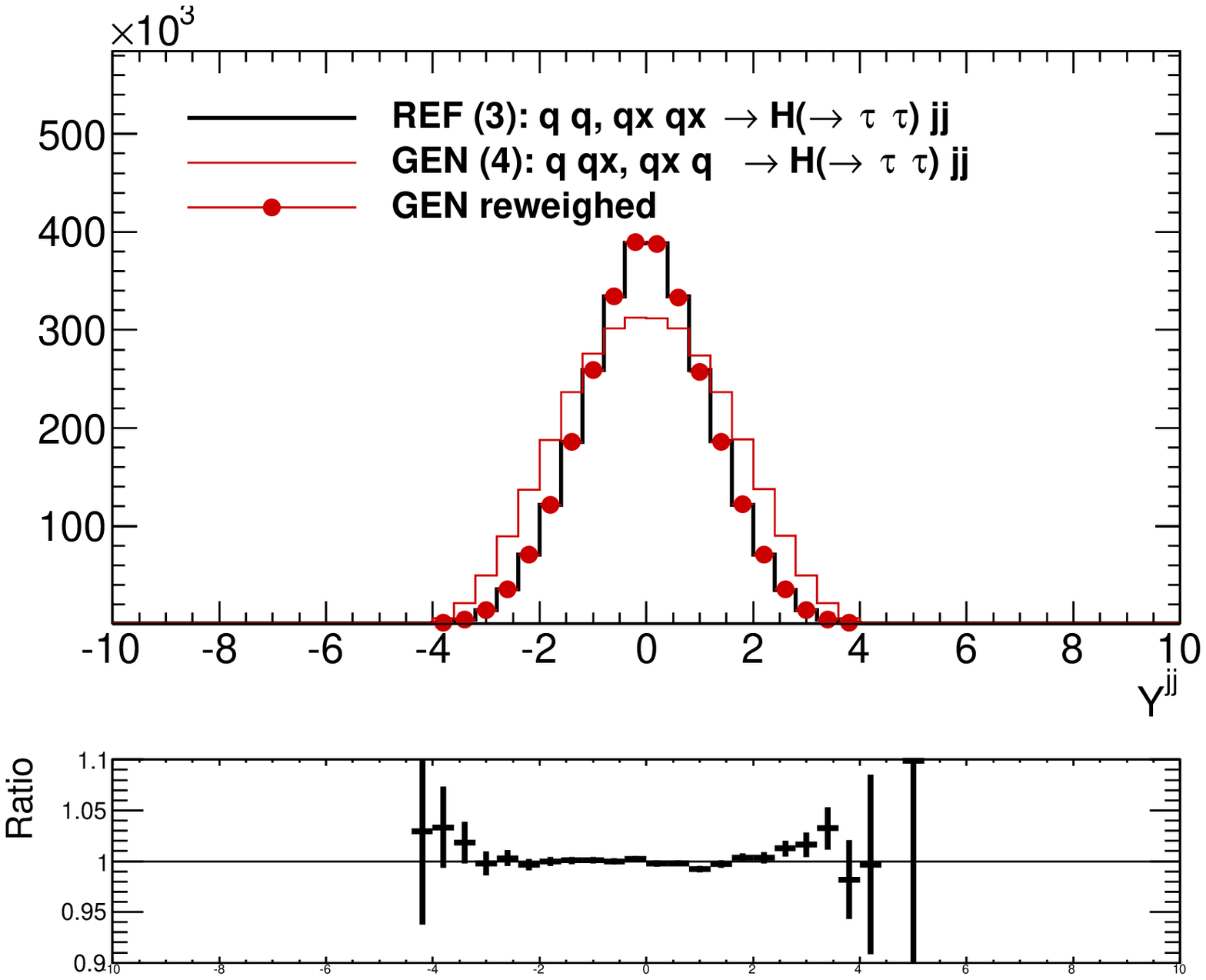}
   \includegraphics[width=7.5cm,angle=0]{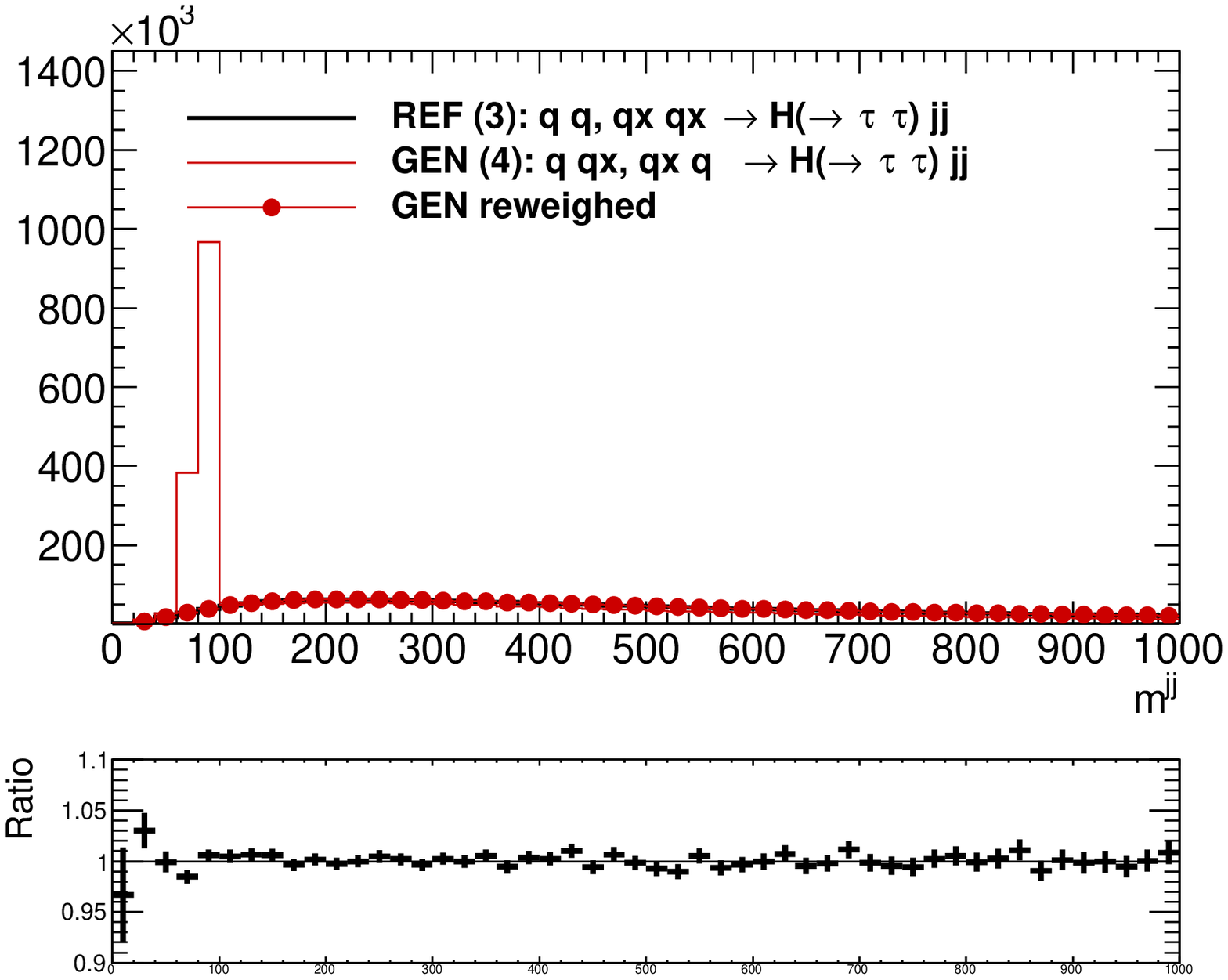}
}
\end{center}
\caption{Shown example of tests distributions for generated process   $q\ qx \to H(\to \tau \tau) jj$ (thin red line) after reweighting 
to  $q\ q, qx\ qx \to  H(\to \tau \tau) jj$ process (red points). Reference  $q\ q, qx\ qx \to  H(\to \tau \tau) jj$  
distribution shown with black line.
\label{fig:B3} }
\end{figure}

\begin{figure}
  \begin{center}                               
{
   \includegraphics[width=7.5cm,angle=0]{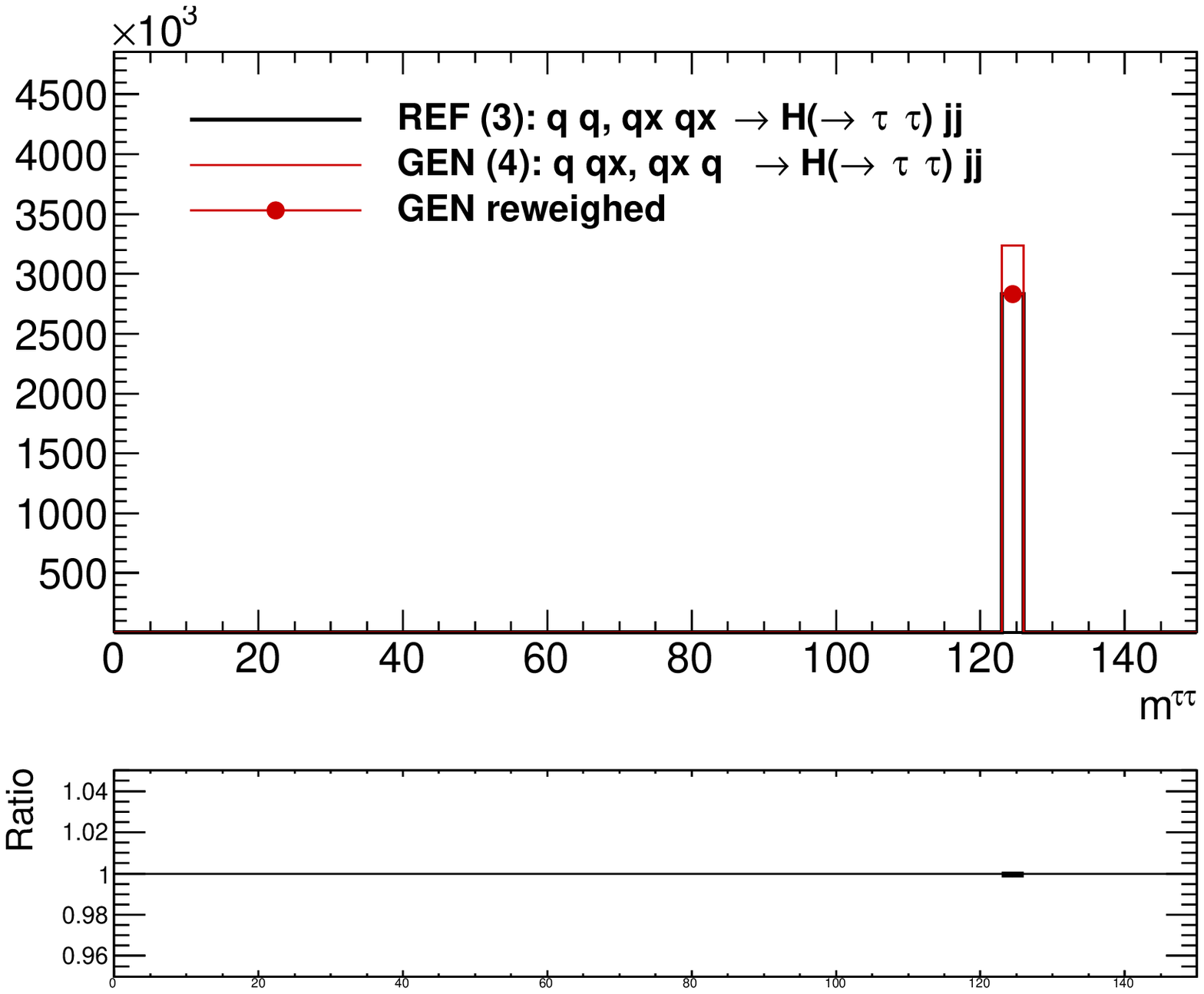}
   \includegraphics[width=7.5cm,angle=0]{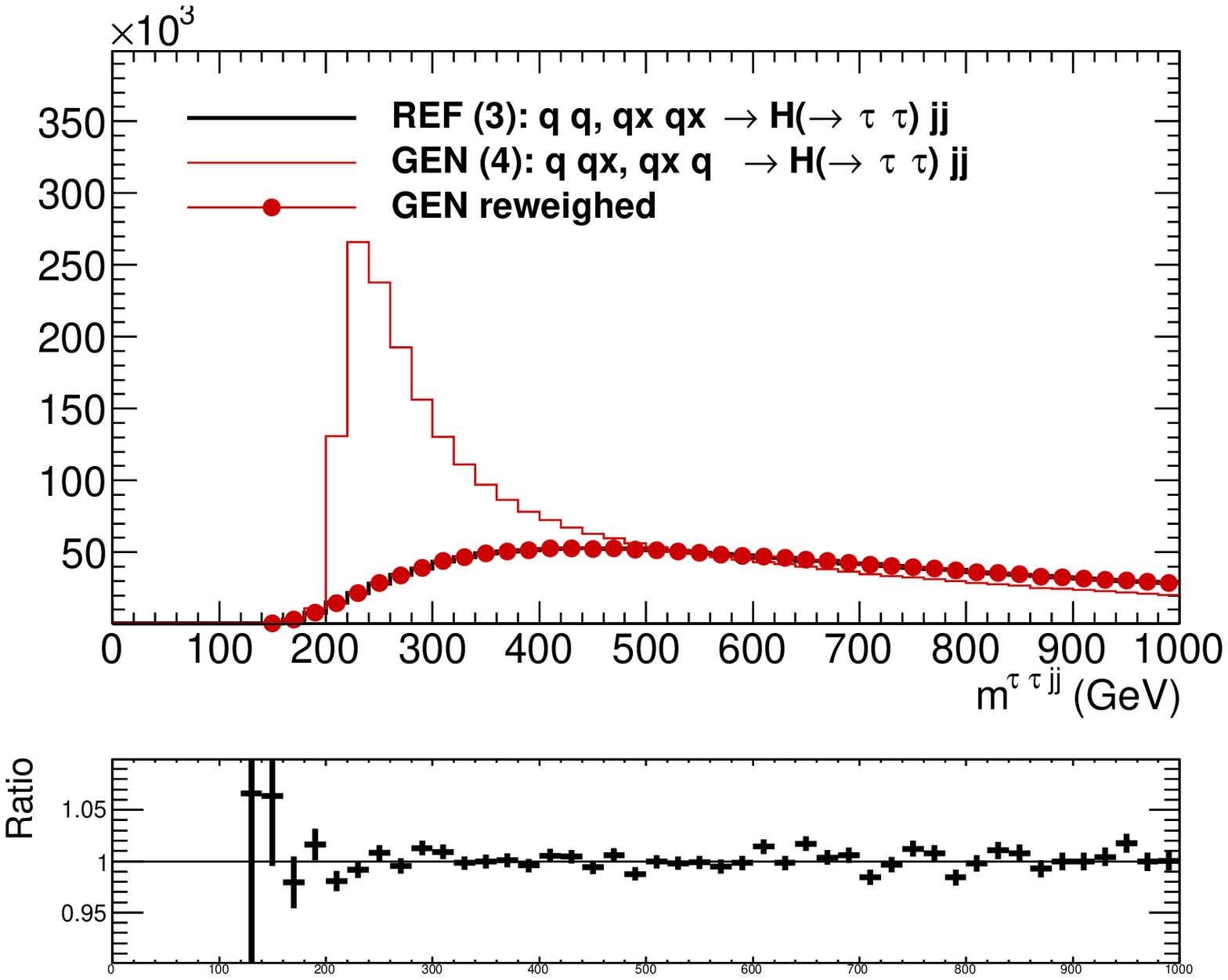}
   \includegraphics[width=7.5cm,angle=0]{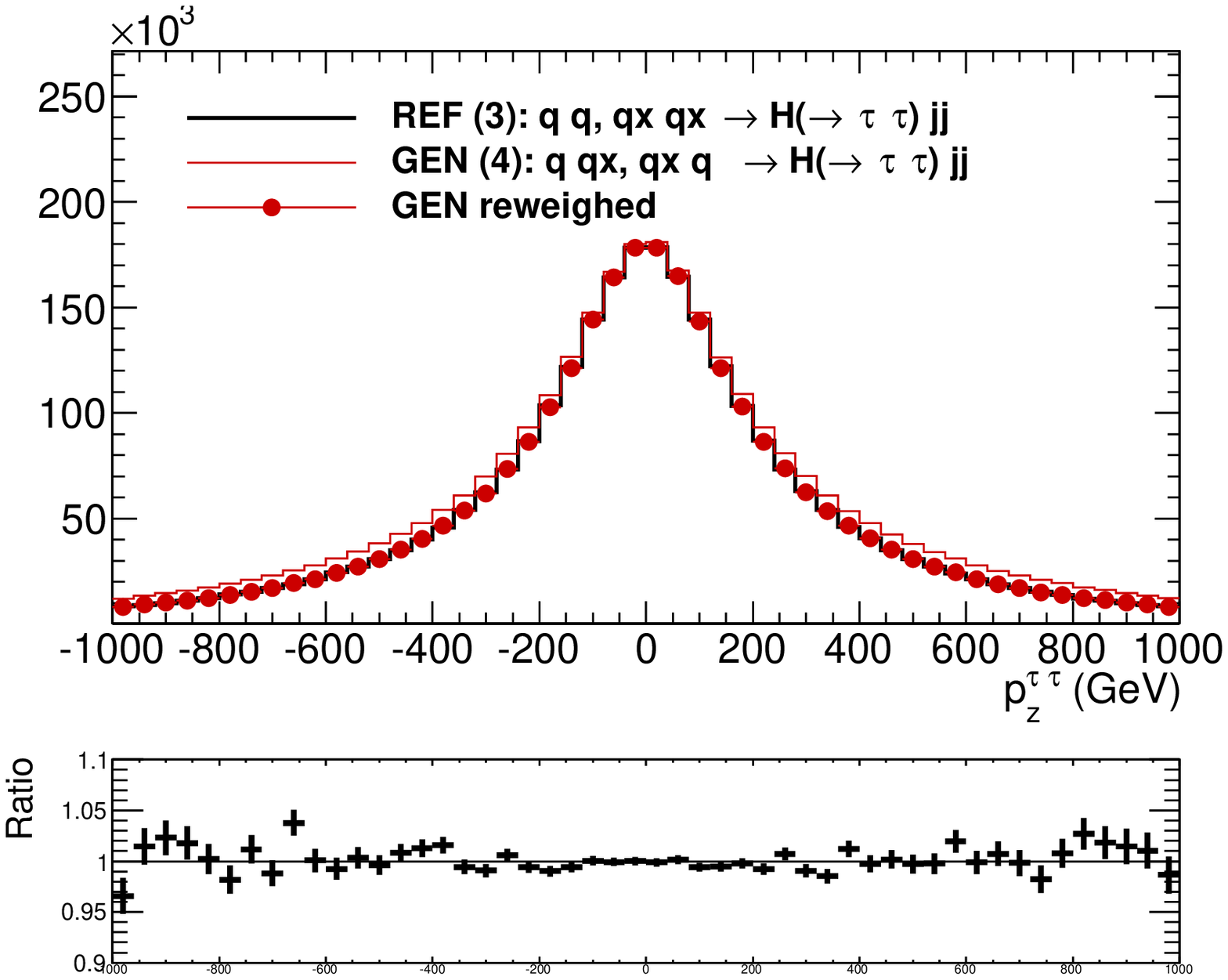}
   \includegraphics[width=7.5cm,angle=0]{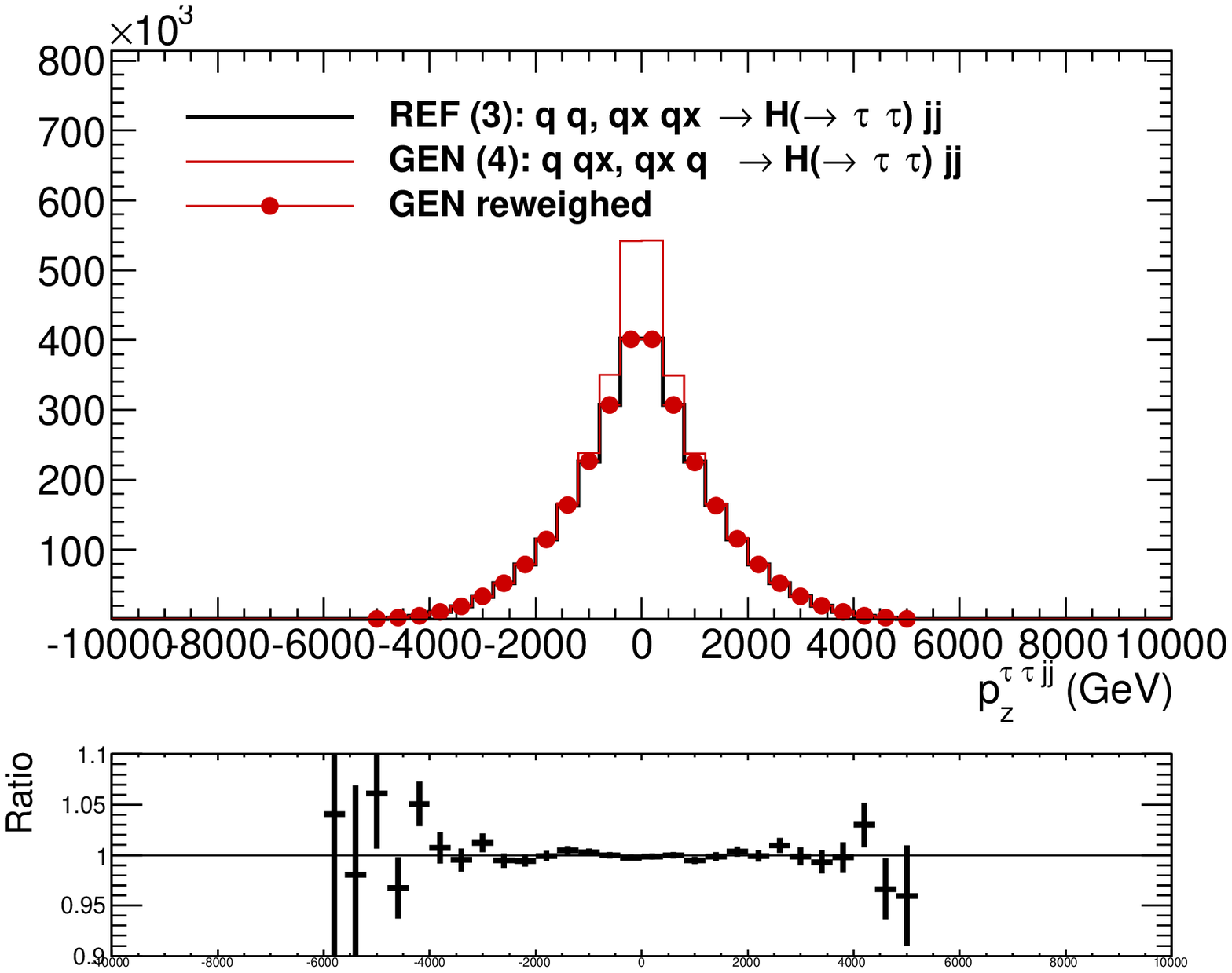}
   \includegraphics[width=7.5cm,angle=0]{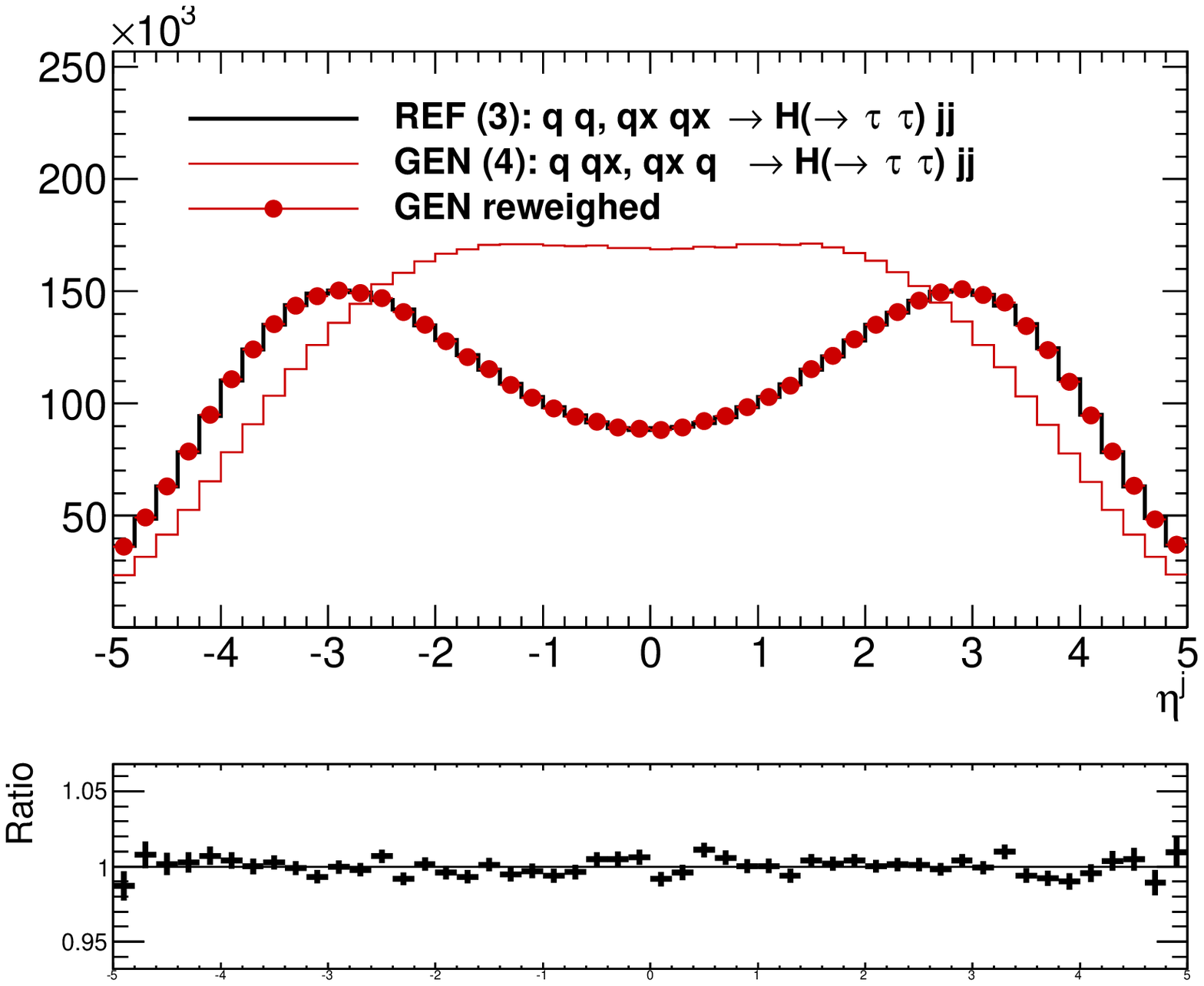}
   \includegraphics[width=7.5cm,angle=0]{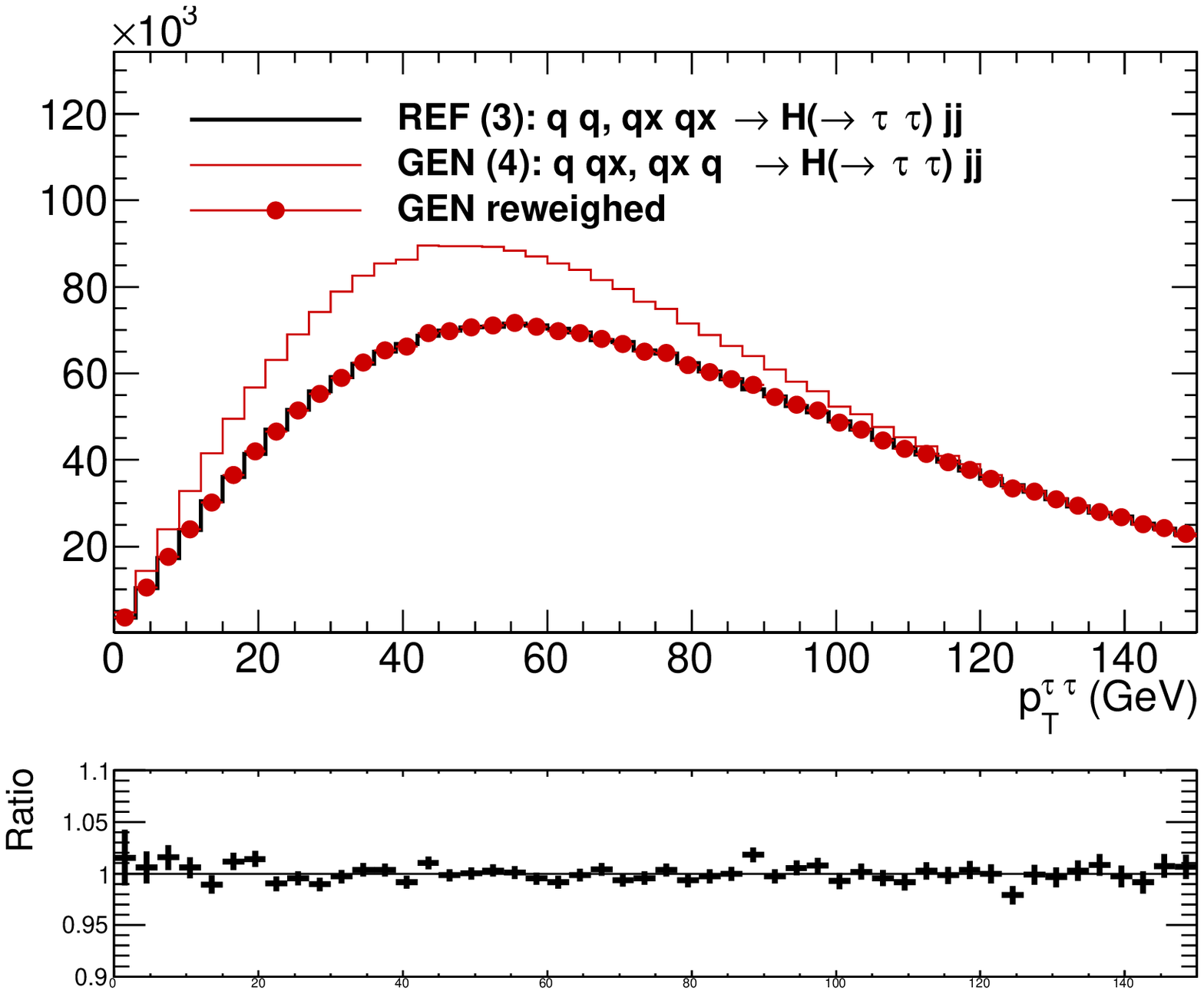}
}
\end{center}
\caption{Shown  example of tests distributions for  generated process $q\ qx \to  H(\to \tau \tau) jj$ (thin red line) after reweighting 
to $q \ q, qx\ qx \to  H(\to \tau \tau) jj$  (red points). Reference  $q\ q, qx\ qx \to  H(\to \tau \tau) jj$ distribution shown with black line.
\label{fig:B4} }
\end{figure}

Similar tests, as discussed in Appendix~\ref{app:XsecTestsDY}, have been repeated for the 
 $pp \to H(\to \tau \tau) jj$ processes. Results are shown in Figs. \ref{fig:B1} -~\ref{fig:B4}.
Very good agreement between the reference and re-weighted distributions is observed, 
both for shapes and relative normalisations.

\section{Optimalization of interface to Standard Model matrix element calculation}
\label{sect:CPUsave}

The steering function for calculating ($2 \to 4$) matrix elements squared\\
{\tt REAL*8 FUNCTION VBFDISTR(ID1,ID2,ID3,ID4,HH1,HH2,PP,KEYIN)}\\
is coded in {\tt FORTRAN} and stored in the  {\tt VBF\_distr.f} file. 
Before invoking calculation of particular  matrix element squared of the Standard Model, it performs several steps 
of filtering to speed up the CPU needed for numerical calculations by setting matrix element 
squared to zero without calculation for the cases when configuration of partonic PDG identifiers for 
incoming and outgoing partons  imply that it is the case.  
The following conditions are  consecutively checked (strictly in the given 
order\footnote{In the case of non-standard calculations, these checks are 
not performed, because  the function {\tt VBFDISTR} is not invoked.}).
Each condition must be passed to go to the next one, and finally to invoke 
the matrix 
element calculation.

\newpage
\vskip 2mm
\centerline{\bf Check if Matrix Element can be set to  zero}
\vskip 2mm

\begin{enumerate}
\item Two incoming (or two outgoing) parton identifiers correspond to gluons and the sum of the other two identifiers is
      zero, otherwise  {\tt ID1} $ \cdot$  {\tt  ID2} $\cdot $ {\tt ID3} $\cdot$ {\tt  ID4} must be positive, to pass 
      to the next step.
\item
 {\tt mod(ID1 +  ID2 + ID3 + ID4, 2) = 0 }, 
\item If both {\tt ID1,ID2} are negative or both   {\tt ID3,ID4 } are negative, and  at least one of the
other two  {\tt ID}'s is positive, then the result is zero. 
\item Charge conservation imposes that for processes without gluons the following  condition must be  fulfilled:\\
 {\tt mod(ID1,2)}$ \cdot$  {\tt sign(ID1)+ mod(ID2,2)}$ \cdot$  {\tt sign(ID2)=mod(ID3,2)}$ \cdot$  {\tt sign(ID3) +mod(ID4,2)} $\cdot$  {\tt sign(ID4)}. 
\item Number of gluons in the process must be zero or two. 
\item If there are two gluons, then for the process to give a non zero contribution it is required that\\ 
 {\tt ID1 +  ID2 = ID3 + ID4} or {\tt ID1 + ID2 =0} or {\tt ID3 +  ID4 =0}.
\end{enumerate}

For some configurations it is enough to change  the order of partons (arguments of  
{\tt  VBFDISTR} routine) or to 
apply  $CP$ symmetry, to avoid duplicating routines for  matrix element calculations. 
It is achieved by first copying 
kinematic variables into the local ones of  {\tt  VBFDISTR} routine and then 
performing  the following  permutations/modifications of the parton
positions and momenta:

\vskip 2mm
\centerline{\bf Reorder  arguments and apply  CP symmetry for convenient choice  of  {\tt ID1, ID2}}
\vskip 2mm

\begin{enumerate}
\item For incoming quark-quark pair, we interchange the order, if necessary,
 to assure {\tt |ID1|}$\ge$ {\tt |ID2|}.
\item If all {\tt ID}'s which do not correspond to gluons are negative, we change their signs. 
 At the same time we interchange 
positions of $\tau^+$ with $\tau^-$ and flip signs of helicities. Finally we change signs of all
3-momenta  to complete the $CP$ transformation.
\item For incoming quark-antiquark pair where at least one is non-first family, 
we require that 
{\tt |ID2|} $\le$ {\tt |ID1|}. For the first family  quarks we require that  {\tt ID1=-1 or ID1=2}
or {\tt |ID2|=|ID1|}. To achieve that goal, if condition is not fulfilled,
we change the signs of all {\tt ID}'s. At the same time we interchange 
positions of $\tau^+$ with $\tau^-$ their helicities  signs changed as well. 
Finally, we change signs of all
3-momenta to complete the $CP$ transformation. 
\item We enforce (by reordering) that the first parton is not an antiquark,  nor a gluon in the case of gluon-fermion initial state.  
\item If both  {\tt ID1, ID2 } are non gluon and positive, we enforce that {\tt ID1} $\ge$ {\tt ID2}
\end{enumerate}

That completes transformations triggered by the configuration of identifiers of  incoming partons. Note that  if the third family is to be  taken into account, also the sign of the $CP$ symmetry  breaking phase will have to be changed to complete the $CP$ transformation. 

\vskip 2mm
\centerline{\bf Reorder   arguments for convenient choice  of  {\tt ID3, ID4}}
\vskip 2mm

\begin{enumerate}
\item 
The {\tt ID3} can not be negative and {\tt ID4} can not be alone the gluon.
\item 
If both {\tt ID3} and {\tt ID4} are non-gluon and positive, then {\tt ID3} can be even 
and {\tt ID4} odd but not the other way.
\item If both {\tt ID3} and {\tt ID4} are odd non-gluon and  also {\tt ID3}$\cdot$ {\tt ID4}>0 ,  
then  {\tt ID4}  must be larger/equal  {\tt ID3}.
\item If {\tt ID3, ID4} are simultaneously even and also {\tt ID3}$\cdot$ {\tt ID4}>0,  then  {\tt ID4} 
must be larger/equal {\tt ID3}.
\end{enumerate}

Note that all of the above conditions are checked one after another. In particular,  all necessary transformations  
(flipping the  position of partons or invoking the $CP$ transformation) are performed
 in the order as listed above.
If all the above conditions are met, the matrix element is not set to zero 
and order of arguments is adjusted to available parton level routines, then  
numerical calculation of matrix element squared 
for a given helicity configuration, parton identifiers and momenta is performed.

\end{document}